\documentclass[11pt,a4paper]{article}
\usepackage{jheppub}
\usepackage{amsmath,multirow}
\usepackage[usenames,dvipsnames,table]{xcolor}
\usepackage{comment,mathrsfs}
\usepackage{graphicx}
\usepackage{soul}
\usepackage{blindtext}
\usepackage{slashed}
\usepackage{subcaption}
\usepackage[normalem]{ulem}

\usepackage{tikz}

\usepackage[compat=1.0.0]{tikz-feynman}

\def\tr{\textrm{tr}\,}
\def\det{\textrm{det}\,}
\def\be{\begin{equation}}
\def\ee{\end{equation}}
\def\lab{\label}
\def\6{\partial}
\def\m{\mu}
\def\n{\nu}
\def\a{\alpha}
\def\b{\beta}

\def\le{\left}
\def\ri{\right}
\def\nn{\nonumber}
\def\la{\langle}
\def\ra{\rangle}
\def\6{\partial} 
\def\tr{\textrm{Tr}}

    \newcommand{\bea}{\begin{eqnarray}}
   \newcommand{\eea}{\end{eqnarray}}

\title{Strings from interacting quantum fields}

\author[1,2]{Domingo Gallegos,} 
\author[1]{Umut G\"ursoy,}
\author[1]{Natale Zinnato}

\affiliation[1]{Institute for Theoretical Physics, Utrecht University,
Leuvenlaan 4, 3584 CE Utrecht, The Netherlands}
\affiliation[2]{Facultad de Ciencias, Universidad Nacional Aut\'onoma de M\'exico, Investigaci\'on Cient\'ifica C.U., 04510 Coyoacan, Ciudad de Mexico, Mexico }

\abstract{We generalize Gopakumar's microscopic derivation of Witten diagrams in large N free quantum field theory \cite{Gopakumar:2003ns} to interacting theories in perturbative expansion. For simplicity we consider a matrix scalar field with $\Phi^h$ interaction in d dimensions. Using Schwinger's proper time formulation and organizing the sum over Feynman diagrams by the number of loops $\ell$, we show that the  two-point function in the massless case 
 can be expressed as a sum over boundary-to-boundary propagators of massive bulk scalars in $AdS_{d+1}$ with masses determined by $\ell$. The two-point function of the massive theory has the same structure given by a sum over boundary-to-boundary propagators but on a geometry different than AdS. The coefficients in the sum contain information on the putative string geometry dual to the interacting QFT. We also consider the three-point function in the field theory and show that it can again be given as an infinite sum, this time over the products of three bulk-to-boundary propagators.  The issue of divergences and renormalization is discussed in detail. 
 
We also notice an intriguing similarity between field theory and string amplitudes. In particular we observe that, in the large-N limit, embedding function of string in the holographic direction corresponds to a continuum limit of Schwinger parameters of Feynman diagrams in the limit where $\ell$ diverges. This provides an interpretation of the holographic dimension emerging directly from field theory amplitudes.}

\begin{document}

\maketitle 

%\tableofcontents

%%%%%%%%%%%%%%%%%%%%%%%%%%%%%%%%%%%%%%%%%%%%%%%%%%%%%%%%%%%%%%%
%%%%%%%%%%%%%%%%%%%%%%%%%%%%%%%%%%%%%%%%%%%%%%%%%%%%%%%%%%%%%%%
%%%%%%%%%%%%%%%%%%%%%%%%%%%%%%%%%%%%%%%%%%%%%%%%%%%%%%%%%%%%%%%
%%%%%%%%%%%%%%%%%%%%%%%%%%%%%%%%%%%%%%%%%%%%%%%%%%%%%%%%%%%%%%%
\section{Introduction and summary}
\lab{sec::intro}
%%%%%%%%%%%%%%%%%%%%%%%%%%%%%%%%%%%%%%%%%%%%%%%%%%%%%%%%%%%%%%%
%%%%%%%%%%%%%%%%%%%%%%%%%%%%%%%%%%%%%%%%%%%%%%%%%%%%%%%%%%%%%%%
%%%%%%%%%%%%%%%%%%%%%%%%%%%%%%%%%%%%%%%%%%%%%%%%%%%%%%%%%%%%%%%
%%%%%%%%%%%%%%%%%%%%%%%%%%%%%%%%%%%%%%%%%%%%%%%%%%%%%%%%%%%%%%%
\noindent\\
Gauge-string correspondence \cite{Maldacena:1997re,Witten:1998qj,Gubser:1998bc} lacks a satisfactory microscopic derivation directly from quantum field theory, barring specific examples such as matrix quantum mechanics \cite{Kazakov:1985ea} (see \cite{Ginsparg:1993is} for a review), minimal model CFTs \cite{Gaberdiel:2010pz,Gaberdiel:2011zw,Gaberdiel:2012uj}, symmetric product CFTs \cite{Eberhardt:2018ouy}, free field theories \cite{Gaberdiel:2021jrv, Gaberdiel:2021qbb} and vector models \cite{Aharony:2020omh}. One fundamental question is how to reformulate holographic QFT correlation functions, in particular their expansion in terms of Feynman diagrams, so that, for example, emergence of gravity becomes manifest.  Various fundamental questions in this context are: how do propagators of dual gravitational space-time arise from field theory amplitudes;  how to determine which QFTs are holographic, which are not?;  given a holographic QFT and assuming a limit where the dual geometry is semi-classical, is there an algorithm to determine this dual background directly from QFT correlators? 

Among all different approaches in the literature, duality between open and closed string descriptions of D-branes \cite{Aharony:1999ti}, entanglement entropy \cite{Ryu:2006bv}, geometrization of RG flows \cite{Heemskerk:2010hk,deBoer:1999tgo,Lee:2013dln}, bulk reconstruction \cite{Harlow:2018fse}, quantum error correction \cite{Almheiri:2014lwa}, tensor networks \cite{Swingle:2009bg,Hayden:2016cfa}, etc., there is one which stands out as the most elementary: deriving dual gravity propagators directly from the QFT Feynman diagrams. For free field theories in $d>2$ this approach was pioneered by R. Gopakumar \cite{Gopakumar:2003ns}. The author considered free matrix field theories and studied $n$-point functions of composite single-trace operators in Schwinger's proper time formulation \cite{Schwinger:1951nm}. The key element, at least for the three-point function studied in \cite{Gopakumar:2003ns}, is a change of variables involving the moduli (Schwinger (or Feynman) parameters of a given graph) that is called the {\em star-triangle duality}\footnote{Earlier work relating matrix quantum mechanics and 2D non-critical string theory \cite{Kazakov:1985ea} involves a similar type of duality.}. The name derives from an analogous relation that involves electric circuits\footnote{See for example, \cite{Lam:1969xk} for a concise account of the map between Feynman diagrams and electric circuits.} which relates the total effective impedance of a triangle shaped electric circuit to that of a tri-star circuit, see fig. \ref{fig2}. In Schwinger's formulation the total proper time of the graph is related to the holographic direction of the dual gravity theory \cite{Gopakumar:2003ns} (see also \cite{Gopakumar:2004qb, Gopakumar:2005fx, Gopakumar:2004ys}) and the star-triangle relation becomes a clear manifestation of the gauge-string duality, or open-closed duality in string theory where the gauge theory three-point function is represented by the triangle and the corresponding Witten diagram \cite{Witten:1998qj} in the dual theory is represented by the tri-star, see Fig. \ref{fig2}. See \cite{Aharony:2020omh} for a more recent work, based on a different approach, that also derives dual gravity theory directly from field theory, in the case of vector models \cite{Gubser:2002tv}.   

In this note we suggest that a generalization of the star-triangle type duality of Feynman diagrams to interacting field theories might be a fundamental manifestation of the gauge-string duality and a key to generalize it beyond the known specific cases\footnote{E.g. based on D-brane descriptions \cite{Aharony:1999ti}, lower dimensional examples \cite{Klebanov:1991qa,Maldacena:2016hyu} and vector models \cite{Aharony:2020omh,Gubser:2002tv}.}. In particular, we generalize Gopakumar's derivation of Witten diagrams from free field theory to interacting theories\footnote{See \cite{Shailesh:2020} for generalization to higher spin amplitudes and our proceedings paper \cite{DomingoGallegos:2022ttp} where some of the earlier results are outlined.}. As a prototype, we take a real, massless N$\times$N matrix-valued scalar field $\Phi$ with $\Phi^h$ interaction, for integer $h>2$, in  d dimensions, and consider two and three-point functions of both canonical fields $\Phi(x)$ and composite operators  $\tr\, \Phi(x)^J $. The two-point function is given by Feynman diagrams summed over the number of independent quantum loops $\ell$ which, in the large N limit, can further be classified in terms of 2D Euclidean Riemann surfaces embedded in d dimensions. We show that each term in the sum over $\ell$ can be mapped onto a boundary-to-boundary propagator of a scalar field with mass $m$ related to $\ell$, in $(d+1)$-dimensional AdS space. We find a similar structure for the three-point function which we express as a sum over products of three bulk-to-boundary AdS propagators. This provides a dual ``closed string'' picture of the two and three-point functions in terms of a {\em generalized Witten diagram} given by sum over AdS Witten diagrams. Even though we perform our calculations in a simple scalar ungauged theory, we will be assuming that our findings generalize to theories like ${\cal N}=4$ super-Yang-Mills without conceptual difficulties. 

We study this interacting star-triangle duality from different angles. After reviewing Gopakumar's construction in free field theory  in the next section, in section \ref{sec::2pf} we first review Schwinger's proper time formulation for interacting field theories and then generalize Gopakumar's computation to finite coupling. Our main findings, for two-point function of canonical fields in the massless field theory are given by Equations (\ref{fullO2}) and (\ref{fullO3}). We reproduce the latter here: 
%%%%%
\be\lab{fullO3int}
\Omega(x,y) = N^2\sum_{g=0}^\infty N^{-2g} \sum_{\ell=0}^\infty \lambda_h^{\frac{2\ell}{h-2}}  v_\Delta\, \beta_{\Delta+2}(x,y)\, ,
\ee
%%%%%
where $g$ is the genus of the graphs, $\beta_{\Delta+2}(x,y)$ is the boundary-to-boundary propagator of a scalar field in AdS, $\Delta$ is a ``scale dimension of the graph'' given in terms of $\ell$ as $\Delta = \left(\frac{d}{2}-\frac{h}{h-2}\right)\ell +\frac{d}{2}-1$ and $v_\Delta$ are coefficients that involve integrals over Schwinger parameters of the Symanzik polynomials of contributing Feynman diagrams. We derive similar expressions for the two-point function of composite operators in (\ref{fullOJ2}) and (\ref{fullOJ3}). 

These are expressed as an infinite sum over $\ell$ of boundary-to-boundary propagators in AdS$_{d+1}$. We observe from the aforementioned relation between $\Delta$ and $\ell$ that the space-time (or momentum) dependence of the two-point function  in (\ref{fullO3int}) becomes independent of $\ell$, hence factorizes from the $\ell$ sum when the field theory coupling is marginal i.e. $h=2d/(d-2)$. Further assuming that reguralization of divergences does not generate a new scale, then the entire dependence of the field theory correlator on $N$ and $\lambda_h$ is given by the sum over $\ell$ and $g$ of $v_\Delta$ which becomes an overall coefficient. 

In section \ref{sec::2pfm} we consider massive field theories and again rewrite the two-point function in terms of $d+1$-dimensional bulk propagators, albeit, in a non-AdS space-time. Feynman diagrams involve several UV and IR divergences. An important role is played by regulating and renormalizing these divergences. This is discussed in detail in sections (\ref{sec::div}),  (\ref{sec::reg}) and (\ref{sec::ren}). We find that regularization in Schwinger's representation renders the coefficients $v_\Delta$ in (\ref{fullO3int}) finite but does not alter the general structure. On the other hand imposing renormalization conditions at a finite RG scale $\mu$ makes $n$-point functions depend on a new parameter $k/\mu$ hence breaks the AdS structure seen in (\ref{fullO3int}). 

In section \ref{sec::3pf} we consider the three-point function (only for the massless case and without renormalization) and show that it can also be rewritten in terms of bulk-to-boundary AdS propagators. Our final expression for the three-point function is given by Equation (\ref{3pfF}). This concludes our analysis of the map between field theory and target space of the putative string dual. 

Another way to look at the QFT/string map is to consider directly the correspondence between field theory amplitudes and the world-sheet computation of the dual string amplitudes. This is discussed in section  \ref{sec::FTST}. This more direct link is suggested by an intriguing similarity between the Schwinger representation of $n$-point amplitudes in terms of Symanzik polynomials and the Polyakov path integral for the corresponding string amplitude. In short, the first and second Symanzik polynomials become related to the determinant of the Laplacian and the Green's function on the world-sheet\footnote{We only consider spherical world-sheets expected to be dual to planar field theory amplitudes.}. Furthermore, we argue that the string field that embeds the holographic dimension in the string target space, $R(\sigma)$, is dual to a continuum limit of the set of all Schwinger parameters on Feynman diagrams. This provides an interesting alternative description of emergence of the holographic dimension directly from the field-theory amplitudes. We also study the continuum limit of Feynman diagrams that is expected to arise in the limit $\ell\to\infty$. We show that this limit generically corresponds to tuning 't Hooft coupling to a critical value $\lambda_h\to \lambda_c$. We then argue that value of this critical coupling, $\lambda_c$, corresponds to curvature radius in the target space in string units. For a 4D CFT this becomes the AdS radius in string units. In the same section, we then focus on CFTs and show that the dual bulk space in the continuum limit is indeed AdS using the aforementioned relation between Schwinger parameters and the holographic direction. We finally discuss how to generalize this construction of the bulk geometry from field theory amplitudes to non-CFTs. 

While the continuum limit focuses on large $\ell$ part of the sum in (\ref{fullO3int}), we argue that finite $\ell$ contributions generalize the geometric, string limit of the duality to include non-geometric contributions. In particular these non-geometric contributions give rise to a new parameter $\lambda_h/\lambda_c-1$ which measures deviation from the perturbative string limit. Many open issues such as how to read off string amplitudes from field theory in the non-perturbative string limit, S-duality, generalization to higher point functions and so on  are discussed in section \ref{sec::discuss} where we also provide an overlook.  

Appendices \ref{2ptDetails} to \ref{SymanzikExamples} provide details of our calculations but they also introduce new material. In particular in Appendix we show how to express the two-point functions in terms of boundary-to-boundary AdS propagators,  Appendix \ref{FeynmanCounting} introduces a novel method --- based on ``creation/annihilation operators'' that create vertices --- to compute the number of Feynman diagrams at a given loop $\ell$ and Appendix \ref{Full2pf} uses the same idea to provide a compact expression for the full two-point function. Finally in Appendix \ref{SymanzikExamples} we devise a (as far as we know) new method to compute the zeros of Symanzik polynomials.

%%%%%%%%%%%%%%%%%%%%%%%%%%%%%%%%%%%%%%%%%%%%%%%%%%%%%%%%%%%%%%%
%%%%%%%%%%%%%%%%%%%%%%%%%%%%%%%%%%%%%%%%%%%%%%%%%%%%%%%%%%%%%%%
%%%%%%%%%%%%%%%%%%%%%%%%%%%%%%%%%%%%%%%%%%%%%%%%%%%%%%%%%%%%%%%
%%%%%%%%%%%%%%%%%%%%%%%%%%%%%%%%%%%%%%%%%%%%%%%%%%%%%%%%%%%%%%%
\section{Open-closed and star-triangle dualities}
\begin{figure}
\begin{center}
\includegraphics[width=15cm]{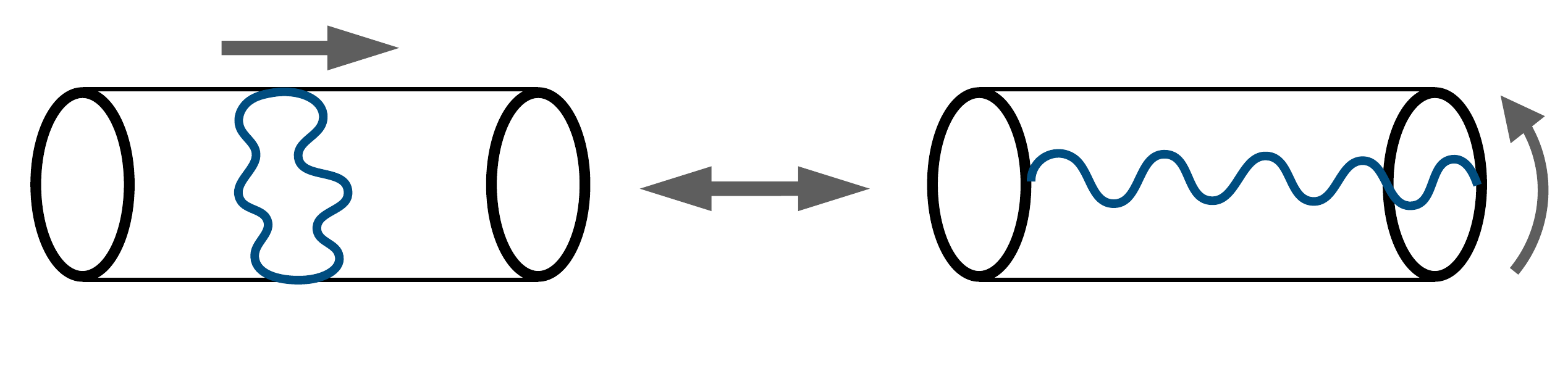}
\end{center}
\caption{\label{fig1} Equivalence between one-loop diagram of an open string and tree level propagation of a closed string.}
\end{figure}

\lab{sec::startriangle}
%%%%%%%%%%%%%%%%%%%%%%%%%%%%%%%%%%%%%%%%%%%%%%%%%%%%%%%%%%%%%%%
%%%%%%%%%%%%%%%%%%%%%%%%%%%%%%%%%%%%%%%%%%%%%%%%%%%%%%%%%%%%%%%
%%%%%%%%%%%%%%%%%%%%%%%%%%%%%%%%%%%%%%%%%%%%%%%%%%%%%%%%%%%%%%%
%%%%%%%%%%%%%%%%%%%%%%%%%%%%%%%%%%%%%%%%%%%%%%%%%%%%%%%%%%%%%%%
\noindent\\
The AdS/CFT correspondence originates from an equivalence between open and closed string  descriptions of a set of D3 branes in IIB string theory \cite{Maldacena:1997re}. Loosely speaking, and in the simplest case, this can be understood geometrically as in fig. \ref{fig1} which depicts an equivalence between one-loop partition function of open strings in d dimensions and propagation of a closed string in $d+1$ dimensions \cite{Polyakov:1987ez}.  
In the low energy limit where the massive string states decouple, open strings on N coincident D3 branes are effectively described by 4D $U(N)$ Yang-Mills gauge theory. On the other hand the closed string in fig. \ref{fig1} turns out to propagate in the AdS$_5\times S^5$ geometry which is generated by the backreaction of the brane system. More precisely, 
the $n$-point function of gauge invariant operators in the Yang-Mills theory is given 
in terms of the closed string world-sheet path integral  
%%%%
\begin{equation}
\label{eq:oc}
\langle {\cal O}_1 (k_1) \cdots  {\cal O}_n (k_n) \rangle_g = \int_{M_{g,n}} \langle {\cal V}_1(k_1,z_1) \cdots  {\cal V}_n(k_n,z_n)\rangle_{w.s.}\, ,
\end{equation}
%%%%
where the subscript $g$ on the RHS denotes the genus-g contribution to the Feynman diagrams and ${\cal V}$s are the closed string vertex operators which correspond to gauge theory operators on the LHS. The integral is over the moduli of Riemann surfaces with genus g and n punctures. 
%%%%%
%%%%%
%%%%%
%%%%%
To demonstrate this equivalence at the level of Feynman diagrams, one must show how the holes on the open string (gauge theory) side are glued together and generate closed string world-sheets with n punctures. This mechanism was first proposed by 't Hooft \cite{tHooft:1973alw}
in the double scaling limit, 
%%%
\begin{equation}\label{thooft}
 g_{YM} \to 0\, , \qquad N\to\infty\, , \qquad g_{YM}^2 N = \lambda \, ,
\end{equation}
%%%
where $g_{YM}$ is the Yang-Mills coupling constant. Emergence of a dual description in this limit can be made explicit in 2D string theory, where the quantum mechanics of $N\times N$ hermitean matrices become dual to 2D non-critical string theory, see for example \cite{Klebanov:1991qa}. 

A strong indication that the same ``gluing'' happens in higher dimensional {\em free} field theories was noted in \cite{Gopakumar:2003ns} utilising the proper time formulation of 
$n$-point functions, which we review below. 

Schwinger's proper time formulation makes the point-like feature 
of QFT manifest\footnote{See e.g. \cite{Edwards:2021} and references therein for the current developments on the worldline formulation of QFT}. In particular, correlators of a quantum field are
represented by propagation of a quantum mechanical particle in proper 
time $\tau$ embedded in space-time as world-line $x^\mu(\tau)$. 
To see this one exponentiates the denominator in the two-point function 
%%%%
\begin{equation}\label{sch1}
  \langle \Phi(x_1) \Phi(x_2) \rangle = i \int d^dk \frac{e^{i k(x_1-x_2)}}{k^2 + m^2 - i\epsilon} = \int_0^\infty d\tau \langle x_1 | e^{-i \tau (-\partial^2 + m^2)} | x_2 \rangle   \, .
\end{equation}
%%%%
The RHS is nothing else but the path integral of a particle propagating in $\tau$ with hamiltonian $H_{pp} = k^2 + m^2$. The integral over $\tau$ is moduli --- a consequence of the reparametization invariance of the worldsheet \footnote{Which can be removed by introducing an auxiliary worldsheet einbein $g^{\tau\tau}$ in the path integral.}. A generalization of this representation to $n$-point functions in a free field theory involves introduction of vertex operators inside the path integral
%%%%
\begin{equation}
\label{eq:npf}
    \langle \phi(x_1) \cdots \phi(x_n) \rangle = \int_0^\infty \frac{d\tau}{\tau} \prod_{i=1}^n d\tau_i \langle e^{i k_1 \hat X(\tau_1)}\dots e^{i k_n \hat X(\tau_n)} \rangle_{q.m.} 
\end{equation}
%%%%
where the RHS is the path integral with the point particle hamiltonian $H_{pp} = k^2 + m^2$. 
The integral is over the moduli of the Feynman diagram given by the total proper time for the process and proper times at insertions of the vertex operators. Note the structural similarity between (\ref{eq:oc}) and (\ref{eq:npf}) which already implies the utility of the 
Schwinger's formulation to explore the basic mechanism behind the gauge-string duality. 

\begin{figure}
\begin{center}
\includegraphics[width=15cm]{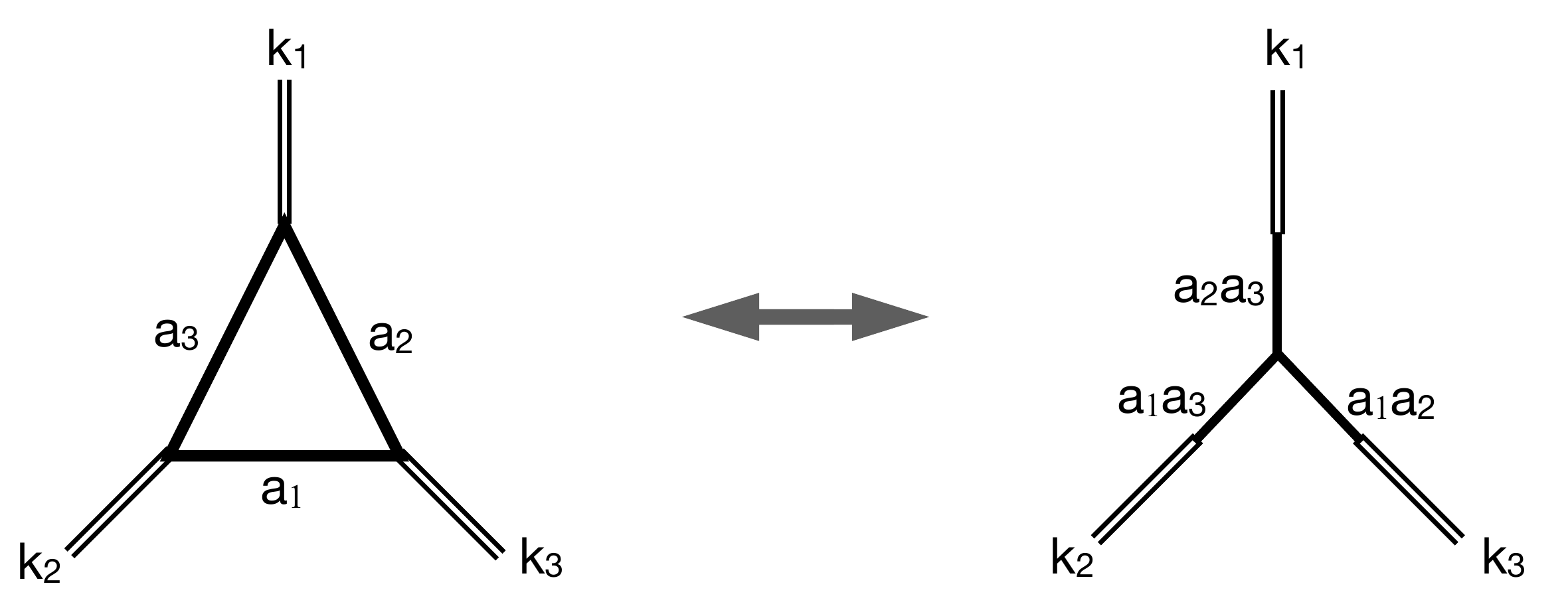}
\end{center}
\caption{\label{fig2} Star-triangle duality in free field theory. LHS shows the only Feynman diagram that contributes to the three-point function of $\tr \Phi^2$ where $k_i$ are external momenta and $\alpha_i$ are the Schwinger parameters. RHS shows its equivalent under the duality.}
\end{figure}

As the path integral in (\ref{eq:npf}) is Gaussian for free field theory, one can compute it explicitly \cite{Strassler:1992zr} and express the result solely in terms of moduli integrals. More interestingly, one can find a judicious change of variables of moduli to reformulate the result in terms of propagators of scalar fields in AdS$_{d+1}$ \cite{Gopakumar:2003ns,Gopakumar:2004ys}. Consider ${\cal N}=4$ super-Yang-Mills at large N and in the free limit $\lambda=0$, see (\ref{thooft}). For the purpose of demonstration let us consider the simplest non-trivial case of the three-point function and the operator $\tr \Phi^2$ where $\Phi$ is one of the 6 scalars in the theory. There is a single diagram that contributes to the connected three-point function $\langle \tr \Phi^2(k_1) \Phi^2(k_2) \Phi^2(k_3)\rangle $ that is shown on the left figure in fig. \ref{fig2}. 

Introducing a change of variables \cite{Gopakumar:2003ns} $\alpha_i = \epsilon_{ijk} |\tau_j - \tau_k|/\tau$ from the moduli $\tau_i$ to Schwinger parameters  one can rewrite the connected three-point function as follows 
%%%%
\begin{equation}
    \Omega(k_1,k_2,k_3) \propto \delta^d(\sum k_i) \int_0^\infty d\tau \int_0^1 \prod_{i=1}^3 d\alpha_i \, \delta(\sum \alpha_i -1)\,  e^{-\tau(k_1^2 \alpha_2\alpha_3 + k_2^2 \alpha_3\alpha_1 + k_3^2 \alpha_1\alpha_2)}
\end{equation}
%%%%
This is precisely in the form given by product of three propagators with 
dual Schwinger parameters $\alpha_1\alpha_2$, etc. as shown on the RHS of fig. \ref{fig2}. This procedure explicitly achieves the ``gluing'' mentioned above in the sense that the hole on the ``open string side'' i.e. the LHS of fig. \ref{fig2} is closed up on the ``closed string side'' i.e. the RHS of fig. \ref{fig2}. The RHS also resembles the Witten diagram for the three-point function in AdS and this resemblance can be made precise by another change of variables $\alpha_i = \rho_i/\sum_{j=1}^3 \rho_j$ \cite{Gopakumar:2003ns} and defining the radial coordinate of the AdS space $z_0$ in terms of these Schwinger moduli as 
%%%%
\begin{equation}
    z_0^2 = 4\tau \left(\sum_{i=1}^3 \rho_i \right)\prod_{i=1}^3 \alpha_i\, .
\end{equation}
%%%%
This results in the final expression after Fourier transforming to space-time as 
%%%%
\begin{equation}
    \Omega(x_1,x_2,x_3) \propto  \int_0^\infty \frac{dz_0}{z_0^{d+1}} \int d^dz \prod_{i=1}^3 K_{\Delta_i}(x_i;z,z_0)
\end{equation}
%%%%
where $K_{\Delta}(x_i,y;z)$ are the boundary-to-bulk propagators for a scalar with mass $m^2 = \Delta(d-\Delta)$, with $\Delta = 2$, corresponding to $\tr \Phi^2$ operator in AdS$_{d+1}$ on the Poincar\'e patch
%%%%
\begin{equation}
   ds^2 = \frac{1}{z_0^2} \left(dz_0^2 + \eta_{ab}\, dz^a dz^b \right)\, . 
\end{equation}
%%%%
 This computation can be generalized to an arbitrary string of $\Phi$ fields \cite{Gopakumar:2004qb}, presumably to other ${\cal N}=4$ super-Yang-Mills operators and higher point functions \cite{Gopakumar:2004ys}. 
%%%%
%%%%
%%%%
%%%%

%%%%%%%%%%%%%%%%%%%%%%%%%%%%%%%%%%%%%%%%%%%%%%%%%%%%%%%%%%%%%%%
%%%%%%%%%%%%%%%%%%%%%%%%%%%%%%%%%%%%%%%%%%%%%%%%%%%%%%%%%%%%%%%
%%%%%%%%%%%%%%%%%%%%%%%%%%%%%%%%%%%%%%%%%%%%%%%%%%%%%%%%%%%%%%%
%%%%%%%%%%%%%%%%%%%%%%%%%%%%%%%%%%%%%%%%%%%%%%%%%%%%%%%%%%%%%%%
\section{Generalization to interacting theories}
\lab{sec::2pf}
%%%%%%%%%%%%%%%%%%%%%%%%%%%%%%%%%%%%%%%%%%%%%%%%%%%%%%%%%%%%%%%
%%%%%%%%%%%%%%%%%%%%%%%%%%%%%%%%%%%%%%%%%%%%%%%%%%%%%%%%%%%%%%%
%%%%%%%%%%%%%%%%%%%%%%%%%%%%%%%%%%%%%%%%%%%%%%%%%%%%%%%%%%%%%%%
%%%%%%%%%%%%%%%%%%%%%%%%%%%%%%%%%%%%%%%%%%%%%%%%%%%%%%%%%%%%%%%
\noindent\\

In this section we demonstrate that the derivation of AdS propagators from Feynman diagrams carries over to interacting QFTs. For simplicity, we consider a matrix scalar field $\Phi$ in d dimensions with an interaction potential $\Phi^h$. We consider a generic QFT action in Euclidean signature 
%%%
\be
\mathcal{S} = \int d^dx\, \text{Tr}\, \left(-\frac{1}{2}(\partial\Phi)^2 -\frac{M^2}{2} \Phi^2 +  \sum_h\frac{\mathfrak{g}_h}{h!}\Phi^h\right)\, ,
\ee 
%%%
where $h>2$ is the coordination number of the vertex associated with the interaction term $\Phi^h$, $\mathfrak{g}_h$ is the associated coupling constant and $\Phi$ is a $ N\times  N$ real matrix with mass $M$. We consider the general case $d\geq 2$ in this paper. Now we rescale $\Phi \rightarrow \sqrt{N} \Phi$:
%%%
\be \label{action0}
\mathcal{S} = N\int d^dx\, \text{Tr}\, \left(-\frac{1}{2}(\partial\Phi)^2 -\frac{M^2}{2} \Phi^2 +\sum_h\frac{\lambda_h}{h!}\Phi^h\right)\, ,
\ee
%%%
where $\lambda_h \equiv  N^{(h-2)/2}\mathfrak{g}_h$. The theory is renormralizable for $h\leq 2d/(d-2)$.

We are interested in computing correlation functions of the scalar fields in \eqref{action0}. We consider a Feynman diagram $F$ of genus $g$, with $I$ internal lines, $V$ vertices and $\ell=I-V+1$ loops. Using Euler's formula, $V-I+f = 2-2g$, we can also relate the number of loops $\ell$ to the number of faces $f$ via 
%%%%
\be \label{loopface}
\ell = f-1+2g\,, 
\ee
%%%%
which will be useful in the following sections.

If we remove $\ell$ internal lines from $F$ such that there is no loop left, the remaining graph can be shown to be a simply-connected subgraph of $F$, which we call \textit{tree}, $T_1$. Its complement, i.e. the set of removed lines, is called a \textit{co-tree}. If $\ell+1$ lines are removed from $G$ such that we are left with two disconnected components (trees) with no loops, we call it a \textit{2-tree} or \textit{2-forest}, $T_2$. Its complement is called a \textit{co-2-tree}.

Given an amplitude with a Feynman diagram F with $n_e$ external momenta, such that $\vec k =(k_1,\dots, k_{n_e})$, $V$ vertices, $I$ internal legs and $\ell$ independent loop momenta, one can express the amplitude in terms of the Schwinger parameters $a_r$ associated to each internal leg of the diagram, see e.g. \cite{Itzykson:1980rh}, \cite{Lam:1969xk}: 
%%%%
\be \label{npt}
\Omega(\vec k) =\delta^{(d)}\left(k_1+\dots+k_{n_e}\right)\int_{0}^{\infty}\left( \prod_{r=1}^{I}da_r\right)\, \mathcal{U}(a_r)^{-d/2}\, e^{-\left(\sum_{r=1}^{I}a_r m_r^2+P\left(a_r;\vec k\right)\right)} {S}(a_r; \vec k) \, ,
\ee
%%%%
where $\mathcal{U}$ and $\mathcal{A}$  are called \textit{Symanzik polynomials} and they are non-negative homogeneous functions of the Schwinger parameters $a_r$'s of degree $\ell$ and $\ell+1$ respectively. They are defined by the graph theoretic 
input  
%%%
\begin{align}
\begin{split}
%\lab{Udef}
\mathcal{U}(a) &\equiv\sum_{T_1\in \mathcal{T}_1}\prod^{\ell}_{r\not \in T_1} a_r \, ,\\
\lab{Pdef}
P(a_r;\vec k)& \equiv\frac{\mathcal{A}(a_r;\vec k)}{\mathcal{U}(a_r)} = \frac{1}{\mathcal{U}(a_r)}\sum_{T_2\in  \mathcal{T}_2}\left(\prod^{\ell+1}_{r\not \in T_2}a_r\right)\left(\sum_{b\in \mathcal{J}}k_b \right)^2 \, ,
\end{split}
\end{align}
%%%
with $\mathcal{T}_1,\mathcal{T}_2$ being the sets of trees and 2-trees respectively,  and $\mathcal{J}$ is one of the two disconnected components of a 2-tree; $a_r$ is the Feynman parameter associated with branch $r$ in the graph  and  $m_r$ is the mass of the particle propagating along line $r$. Finally,  $S(a_r; \vec k)$ is in general a complicated function of  Schwinger parameters and external momenta, which is non-trivial when  the action contains vector and fermion fields. Here we consider a scalar theory without derivative interactions, so that this function is given simply by the numerator of the usual Feynman rules in momentum space i.e. $S(a_r; \vec k) \propto \lambda_h^V$ and the proportionality constant is a tensor depending on the indices of the matrix $\Phi$ and a power of $N$. We will suppress this dependence on the indices in what follows and we will not show the coupling constant dependence $\lambda_h^V$ in the amplitudes until we sum over all graphs. 

%%%%%%%%%%%%%%%%%%%%%%%%%%%%%%%%%%%
%%%%%%%%%%%%%%%%%%%%%%%%%%%%%%%%%%%
%%%%%%%%%%%%%%%%%%%%%%%%%%%%%%%%%%%
\subsection{Two-point function}
%%%%%%%%%%%%%%%%%%%%%%%%%%%%%%%%%%%
%%%%%%%%%%%%%%%%%%%%%%%%%%%%%%%%%%%
%%%%%%%%%%%%%%%%%%%%%%%%%%%%%%%%%%%

We first consider equation \eqref{npt} and compute it for a given diagram $F$ with two external legs and a fixed number $\ell$ of independent loop momenta. In particular, we will first consider a graph where the external legs are\footnote{We will generalize this to composite operators later in Section \ref{CompSect}.} $\Phi_{ij}$ and $\Phi_{kl}$. 

%%%%%%%%%%%%%%%%%%%%%%%%%%%%%%%%%%%
%%%%%%%%%%%%%%%%%%%%%%%%%%%%%%%%%%%
\subsubsection{Two-point function for the massless theory}
\lab{sec::2pfm0}
%%%%%%%%%%%%%%%%%%%%%%%%%%%%%%%%%%%
%%%%%%%%%%%%%%%%%%%%%%%%%%%%%%%%%%%

We demonstrate the computation first in the simpler case where the QFT fields are massless. We suppress the tensor structure of the correlator and the external propagators, but one  should remember to reinstate the function $ {S}$ when needed. Specializing eq. \eqref{npt} to a graph $F$ with $n_e=2$ and $m_r=0$, we have for the \textit{amputated} two-point amplitude:
%%%
\be\label{G20b}
\Omega_F(k_1,k_2)=\delta\left(k_1+k_2\right) \int_{0}^{\infty}\prod_{r=1}^{I}da_r\, \mathcal{U}_F^{-d/2}\, \exp\left(- \frac{\mathcal{A}_F}{\mathcal{U}_F} k_1^2\right) \, ,
\ee
%%%
where we defined with a slight abuse of notation 
%%%
\be \label{Adef}
\mathcal{A}_F \equiv  \mathcal{A}_F(a_r;k_1,k_2)/k_1^2 =  \sum_{T_2\in \mathcal{T}_2}\prod^{\ell+1}_{r \not \in T_2} a_r\, ,
\ee
%%%
where we have used, due to the delta function, that $k_1^2 = k_2^2$. First consider the tree-level contribution i.e $V=0$, $\ell =0$ and $I=1$. The two-point function is given by the free propagator\footnote{We drop inessential constants.}. 
In this case $\mathcal{A}_F(a_r;k_1,k_2)=a k_1^2$, $\mathcal{U}_F(a_r)=1$ 
and the two point function becomes
%%%
\be \label{G0k}
\Omega_0(k_1,k_2) =\delta(k_1+k_2) \int_0^{\infty} da\, e^{-a k_1^2}= \frac{\delta(k_1+k_2)}{k_1^2}\, .
\ee
%%%
In the position space it becomes 
%%%
\be \label{G0}
\Omega_0(x,y)  = 4^{\tfrac{d}{2}-1}\pi^{d/2}\frac{\Gamma(\tfrac{d}{2}-1)}{|x-y|^{d-2}} \, .
\ee
%%% 

Let's now move on to the non-trivial case with interactions $V>0$ and with multiple loop momenta $\ell\ge1$. We consider (\ref{G20b}) and use the following identity
%%%
\be \label{IdEll0}
\int_0^{\infty}d\tau\, \delta(\tau^\ell-\mathcal{U}_F)=\frac{1}{\ell\, \mathcal{U}_F^{\frac{\ell-1}{\ell}}}\, .
\ee
%%%
The quantity $\tau$ will play an important role throughout our analysis. Essentially, it will correspond to the holographic coordinate in the dual geometry. We continue by  the change of variable $a_r=\tau b_r$, under which
$\mathcal{A}_F(a_r)=\tau^{\ell+1}\mathcal{A}_F(b_r)$ and 
$\mathcal{U}_F(a_r)=\tau^{\ell}\mathcal{U}_F(b_r)$
%\, , \qquad \qquad \qquad \mathcal{M}(a_r)^2= \tau\mathcal{M}(b_r)^2 
and the two-point amplitude now becomes
%%%
\begin{align}
\begin{split}
\Omega_F (k_1,k_2)
%&=\delta\left(k_1+k_2\right)\int_0^{\infty}d\tau \int_{0}^{\infty}\left(\prod_{r=1}^{I} db_r\right) \mathcal{U}^{-d/2} \tau^{I-d\ell/2} \left[\frac{\delta\left(1-\mathcal{U}\right)}{\tau^\ell}\ell\,\mathcal{U}^{\frac{\ell-1}{\ell}} \tau^{\ell-1}\right]\exp \left(-\tau\frac{\mathcal{A}}{\mathcal{U}} k_1^2 \right)\\
	&=\ell \,\delta\left(k_1+k_2\right)   \int_{0}^{\infty}\left(\prod_{r=1}^{I} db_r\right) \delta\left(1-\mathcal{U}_F\right) \int_0^{\infty}d\tau\,\tau^{I-1-d\ell/2}\, e^{- \tau \mathcal{A}_Fk_1^2 } \, .
\end{split}
\end{align}
%%%
A final change of variable, $\tau\rightarrow \tau/\mathcal{A}_F$, results in the following compact expression
%%%
\be\label{2pftaub}
%\Omega_F(k_1,k_2)&=\ell \, \delta\left(k_1+k_2\right) \int_{0}^{\infty}\left(\prod_{r=1}^{I} db_r\right)\delta\left(1-\mathcal{U}_F \right) \int_0^{\infty}d \tau\, \tau^{I-1-d\ell/2}\mathcal{A}_F ^{-I+d\ell/2}\, e^{- \tau k_1^2 }\\
\Omega_F(k_1,k_2) = \mathcal{V}_F\, \delta\left(k_1+k_2\right)  \int_0^{\infty}\frac{d \tau}{\tau}\,\tau^{-2-\Delta +d/2}\, e^{- \tau k_1^2}\, , 
\ee
%%%
where all the dependence on the Schwinger parameters $b_r$ are hidden in the coefficient 
%%%
\be\lab{Vn}
\mathcal{V}_F\equiv\ell\int_0^\infty\left(\prod_{r=1}^{I} db_r\right) \, \delta\left(1-\mathcal{U}_F \right)\mathcal{A}_F^{\Delta+2-d/2}\, .
\ee 
%%%
We provide two alternative expressions for this coefficient in Appendix \ref{AltVF}. Apart from the overall coefficient, the only dependence on the Feynman diagram is given by the power $\Delta$,  
%%%
\be\label{Delta2pt}
\Delta\equiv \frac{d}{2}(1+\ell)-I-2  \, .
\ee
%%%
This power can be understood as a scaling dimension associated to the particular Feynman diagram. The integral in equation \eqref{2pftaub} can already be solved exactly without the need to make further manipulations. After solving the integral and going to real-space, we find 
%%%
\be \label{Omsol0}
\Omega_F(x, y)= 4^{\Delta+2} \pi^{d/2}\Gamma(\Delta+2) \frac{ \mathcal{V}_F }{|x-y|^{2\Delta+4}}\,. 
\ee
%%%
We observe that for the two-point function of the free theory the tree level propagator (with $I=1$ and $\ell=0$) is indeed of the right form with $\Omega_F(x, y) \propto |x-y|^{2-d}$ given by (\ref{G0}). 

Now, we can immediately express (\ref{Omsol0}) in terms of product of two bulk-to-boundary propagators in AdS. We will do this in two ways. First, we know from the standard AdS/CFT relation, see for example \cite{DHoker:2002nbb}, that 
%%%%
\bea\lab{adscft1} 
\frac{1}{|x-y|^{2\Delta}} & \propto& \lim_{\epsilon\to 0} \int d^d z \epsilon^{1-d} K_\Delta(\epsilon,z,x)\frac{\6}{\6z_0} K_\Delta(z_0,z,y)\bigg|_{z_0=\epsilon} \, \\
{}&=&- \int \frac{d^dz dz_0}{z_0^{d+1}} \le(z_0^2\6_\m K_{\Delta}(z_0,z,x) \6_\m K_{\Delta}(z_0,z,y) + m^2 K_{\Delta}(z_0,z,x)K_{\Delta}(z_0,z,y)\ri)\nn\, ,
\eea
%%%%
where the mass of the bulk field is related to the scale dimension $\Delta$ as 
%%%%
\be\lab{mass1}
m^2 = \Delta(\Delta-d)\, ,
\ee
%%%% 
and we assumed standard quantization where $\Delta$ is the power of the leading term of the bulk field near the boundary $z_0=0$. Here $K_{\Delta}$ is the standard bulk-to-boundary propagator for a scalar field 
where we introduced the AdS bulk-to-boundary propagator\footnote{See Appendix \ref{AdSProperties} for properties and our conventions of AdS propagators.} 
%%%
\be \label{AdSProp0}
%K_{\Delta}(z_0, x, z)=\frac{ z_0^{\Delta}}{\pi^{d/2}\Gamma(\Delta-d/2)}\int_0^{\infty}d\rho\, \rho^{\Delta-1}\,e^{-\rho\left( z_0^2+(x-z)^2\right)}=\frac{\Gamma(\Delta)}{\pi^{d/2}\Gamma(\Delta-d/2)}\frac{z_0^{\Delta}}{\left(z_0^2+(x-z)^2\right)^{\Delta}}\, .
K_{\Delta}(z_0, x, z) = \frac{\Gamma(\Delta)}{\pi^{d/2}\Gamma(\Delta-d/2)}\frac{z_0^{\Delta}}{\left(z_0^2+(x-z)^2\right)^{\Delta}}\, .
\ee
%%%
Putting all the proportionality factors together we see that the contribution to the two-point function from Feynman diagram $F$ can be expressed as 
%%%
\bea \label{2pf1}
\Omega_F(x, y) &=& -\mathcal{V}'_F \int \frac{d^dz dz_0}{z_0^{d+1}} \big(z_0^2\6_\m K_{\Delta+2}(z_0,z,x) \6_\m K_{\Delta+2}(z_0,z,y) \nn\\
{}&&  \qquad\qquad\qquad\qquad  +m^2 K_{\Delta+2}(z_0,z,x)K_{\Delta+2}(z_0,z,y)\big)\, ,
\eea
%%%
where $z_0$ is the radial direction in $AdS_{d+1}$ with boundary located at $z_0=0$. The modified coefficient is given by   
%%%
\be\lab{vfp} 
\mathcal{V}'_F \equiv \mathcal{V}_F \pi^d 4^{\Delta+2} \frac{\Gamma(\Delta+2-d/2)}{\Delta+2}\, .
\ee
%%%

We will assume a single type of vertex $\lambda_h$ with coordination number $h>2$ from now on. Then, for $\ell>0$ the power $\Delta$ can be expressed solely in terms of $\ell$ 
as
%%%
\be\label{Delta2pt1}
\Delta = \left(\frac{d}{2}-\frac{h}{h-2}\right)\ell +\frac{d}{2}-1 \, .
\ee
%%%
This expression is derived in Appendix \ref{AppEul}. We observe that, the coefficient of $\ell$ in (\ref{Delta2pt1}) is in general negative for a super-renormalizable theory for which $h\leq 2d/(d-1)$ and $\lambda_h$ has positive mass dimension. Then the AdS mass in (\ref{mass1}) can become arbitrarily negative for large $\ell$ and the particles in AdS become unstable. This is, perhaps, an indication that the two-point function in super-renormalizable theories cannot be expressed in terms of gravitational propagators in AdS. From a formal point of view, this is not worrisome as the expression (\ref{Omsol0}) is well-defined. In any case,  we will mostly be interested in the marginal case $h=2d/(d-2)$ for which $\Delta$ does not depend on $\ell$. 

Our result (\ref{Omsol0}) for a particular contribution from diagram $F$ to the two-point function, can easily be generalized to the full perturbative answer by summing over all Feynman diagrams as follows. A generic graph  $F$ can be completely characterized  
by the number of genii $g$, the number of independent loop momenta $\ell$ and the set of trees and two-trees $\mathcal{T}_1,\mathcal{T}_2$. To see this first note that the number of vertices in the diagram  $F$ can also be expressed in terms of only $\ell$ as $V = 2\ell/(h-2)$ for a generic genus $g$, see appendix \ref{AppEul}. Second, the power of $N$ multiplying a generic contribution follows directly from (\ref{action0}) as $V-I+f$ where $f$ is the number of faces. Euler theorem relates this power to $2-2g$. Then the full answer is given by 
%%%%%
\be\lab{fullO1}
\Omega(x,y) = N^2\sum_{g=0}^\infty N^{-2g} \sum_{\ell=0}^\infty \lambda_h^{\frac{2\ell}{h-2}}  \sum_{F\in \mathcal{F}_\ell} \frac{1}{\sigma_F}\Omega_{F}(x,y)\, ,
\ee
%%%%%
where the sum over $\ell$ runs over all graphs with genus $g$ and the final sum is over all Feynman diagrams with $\ell$ independent loop momenta and $\sigma_F$ is the associated symmetry factor. This sum can be absorbed into the overall coefficient and one finds 
%%%%%
\be\lab{fullO2}
\Omega(x,y) = N^2 \pi^{\frac{d}{2}} \sum_{g=0}^\infty N^{-2g} \sum_{\ell=0}^\infty \lambda_h^{\frac{2\ell}{h-2}} \, \mathcal{V}_\Delta\,  |x-y|^{-2\Delta-4} \, ,
\ee
%%%%%
where we defined 
%%%%
\be\lab{vf00}
\mathcal{V}_\Delta \equiv 4^{\Delta+2}\, \Gamma(\Delta+2) \sum_{F\in \mathcal{F}_\ell} \frac{\mathcal{V}_F}{\sigma_F}\, .
\ee
%%%%
In these expressions we use $\Delta_0 = d/2-3$ and $\mathcal{V}_F^0 = 1$ for the $\ell=0$ contribution. Note that the coefficient $\mathcal{V}_\Delta$ depends both on the number of independent momenta $\ell$ and genus $g$ because the sum over Feynman diagrams  in (\ref{vf00}) involve all possible genii of graphs with $\ell$ loops. The maximum genus for a given $\ell$ is finite. Recalling the identity $\ell = f-1 +2g$, with $f$ being the number of faces, the maximum number of genus $g^*_\ell$ is determined by the minimum number of faces in a graph which is 2. Therefore  $g^*_\ell = \lfloor \frac{\ell-1}{2} \rfloor$. Therefore, performing the genus sum in (\ref{fullO2}) one can write
%%%%%
\be\lab{fullO7}
\Omega(x,y) = N^2 \pi^{\frac{d}{2}}  \sum_{\ell=0}^\infty \lambda_h^{\frac{2\ell}{h-2}} \, \bar{\mathcal{V}}_\Delta\,  |x-y|^{-2\Delta-4} \, ,
\ee
%%%%% 
where
%%%
\be\lab{vbard} 
\bar{\mathcal{V}}_\Delta = \sum_{g=0}^{ \lfloor \frac{\ell-1}{2} \rfloor} N^{-2g}  \sum_{F\in \mathcal{F}_{\ell,g}} \frac{\mathcal{V}_F}{\sigma_F}\, ,
\ee
%%%
with $ \mathcal{F}_{\ell,g}$ denoting the set of all genus-g diagrams with $\ell$ independent internal momenta. 

Following (\ref{adscft1}) then, the full perturbative two-point function can be expressed in a form suggesting a holographic interpretation as 
%%%%%
\be\lab{adscft3} 
\Omega(x,y) = \frac{\delta^2}{\delta \varphi_0(x)\delta \varphi_0(y)} S_{gr}\, ,\quad S_{gr} =\int \frac{d^dz dz_0}{z_0^{d+1}}\sum_\Delta \lambda_h^{\frac{2\ell}{h-2}}\, \mathcal{V}'_\Delta \le(z_0^2\6_\m \varphi_{\Delta+2} \6_\m \varphi_{\Delta+2}  + m_{\Delta+2}^2 \varphi_{\Delta+2}^2\ri)\,
\ee
%%%%%
where $\mathcal{V}'_\Delta =  \sum_{F\in \mathcal{F}_\ell} \frac{\mathcal{V}'_F}{\sigma_F}$, $\ell$ is given in terms of $\Delta$ as in (\ref{Delta2pt1}) and the bulk fields are evaluated on-shell with the same boundary condition $\varphi_\Delta (z_0,z)\to \varphi_0(x)$ as $z_0\to 0$, that is, 
%%%%
\be\lab{bc}
\varphi_\Delta(z_0,z) = \int d^dx K_\Delta(z_0,z,x) \varphi_0(x)\, .
\ee
%%%%
Note that  \eqref{2pftaub} generically involve divergences. In particular, this happens when $\tau\rightarrow 0$ and when $b\rightarrow 0$ or $b\rightarrow \infty$. Divergence structure will depend on the value of $\Delta$ and the particular Feynman diagram $F$. These divergences can be regularized and the standard renormalization procedure can be carried out in different ways as we discuss this in detail below. Here we assume for simplicity that these divergences are taken care of in the appropriate way. 

We demonstrated that the standard AdS/CFT prescription for the two-point function generalizes to interacting (massless) quantum field theories where one has to sum over infinitely many Witten diagrams where mass of the bulk fields are determined by the conformal dimension $\Delta$ of the field theory graphs. While this result is similar to the standard AdS/CFT prescription, it is somewhat different than generic expectation for an $n$-point function with $n>2$ which involves only a product of $n$ bulk-to-boundary AdS propagators, as reviewed for the free  three-point function in section \ref{sec::startriangle}. Our result above is different because it involves derivatives of AdS propagators. We can indeed put the two-point function above also in this more generic form and rewrite it as sum over products of two AdS bulk-to-boundary propagators. This alternative expression is derived in Appendix \ref{2ptDetails} and it will be useful to re-express the final result for the two-point function in terms AdS boundary-to-boundary propagators. The techniques introduced in this appendix will be generalized to higher-point functions below. 

In Appendix \ref{2ptDetails} we show that \eqref{2pftaub} can equivalently be rewritten as
%%%
\be \label{2pfell}
\Omega_{F}(x, y)=\lim_{\epsilon\rightarrow 0} v _{F,\epsilon}   \int\frac{dz_0 d^dz}{z_0^{1+d}}\, z_0^{2\epsilon}\,K_{\Delta+2+\epsilon}(z_0, z,x) K_{\Delta+2+\epsilon}(z_0, z, y) \, .
\ee
%%%
where $K_{\Delta+2+\epsilon}(z_0, z,x)$ is the AdS bulk-to-boundary propagator and 
%%%%
%\be \label{AdSProp0}
%K_{\Delta}(x; z_0, z)=\frac{ z_0^{\Delta}}{\pi^{d/2}\Gamma(\nu)}\int_0^{\infty}d\rho\, \rho^{\Delta-1}\,e^{-\rho\left( z_0^2+(x-z)^2\right)}=\frac{\Gamma(\Delta)}{\pi^{d/2}\Gamma(\nu)}\frac{z_0^{\Delta}}{\left(z_0^2+(x-z)^2\right)^{\Delta}}\, ,
%\ee
%%%%  
 we defined 
%%%
\begin{align}
\begin{split} \label{velleps}
v_{F,\epsilon}
%&\equiv2\pi^d \Gamma(\Delta-d/2+\epsilon)^2 4^{\Delta+d/2}\bar v_\ell \\
&= 2\pi^d4^{\Delta+2}\frac{\Gamma(\nu+2+ \epsilon)^2 \Gamma(2\epsilon)}{\Gamma(\nu+2+ 2\epsilon) \Gamma(\epsilon)^2}\, \ell\, \int_0^\infty \left( \prod_{r=1}^I db_r\right)\, \delta\left(1-\mathcal{U}_F\right)\mathcal{A}_F^{\nu+2}\, .
\end{split}
\end{align}
Here $\nu$ is defined as $\nu \equiv \Delta- d/2$. 
%%%
The subscript $F$ reminds us that this quantity depends on the specific Feynman diagram  $F$, not only on the number of loops. This dependence only enters in the Symanzik polynomials, see Appendix \ref{SymanzikExamples} for some examples. It is straightforward to show that (\ref{2pfell}) is finite in the limit $\epsilon\to 0$. 

Taking our derivation one step further, we use identity \eqref{2BBIdentityFin} to rewrite the integral of two bulk-to-boundary propagators as the boundary limit of one bulk-to-bulk propagator. This results in 
%%%
\begin{align}
\begin{split}\label{GG2}
\Omega_F(x,y) &= 2 (\nu+2)\,v_F \lim_{(x_0,y_0)\rightarrow (0,0)} (x_0y_0)^{-\Delta-2}  \mathcal{G}_{\Delta+2 }(x_0,x;y_0,y) \\
&=  \frac{v_F\Gamma(\Delta+2)}{\pi^{d/2} \Gamma(\nu+2)} \frac{1}{|x-y|^{2\Delta+4}} \\
& =  4^{\Delta+2}\pi^{d/2}\Gamma(\Delta+2)\frac{\mathcal{V}_F }{|x-y|^{2\Delta+4}} \, ,
\end{split}
\end{align}
%%%
where $\mathcal{G}$ is the AdS bulk-to-bulk propagator (see Appendix \ref{AdSProperties}) and we defined
%%%
\be
v_F \equiv   4^{\Delta+2}\pi^d \Gamma(\nu+2) \ell\, \int_0^\infty\left(\prod_{r=1}^I db_r\right)\, \delta\left(1-\mathcal{U}_F\right)\mathcal{A}_F^{\nu+2} =   4^{\Delta+2}\pi^d \Gamma(\nu+2)\mathcal{V}_F\, .
\ee
%%%
It is easy to see that \eqref{GG2} is equal to  \eqref{Omsol0}. 
Note that this correlator can compactly be rewritten as
%%%
\be
\Omega_F(x,y) = v_F\beta_{\Delta+2}(x,y)\,, \\
\ee
%%%
where we introduced the AdS boundary-to-boundary propagator $\beta_{\Delta}(x,y)$ via
%%%
\be \label{bulkLimit}
\lim_{(x_0,y_0)\rightarrow (0,0)}(x_0y_0)^{-\Delta}\mathcal{G}_{\Delta}(x_0,x;y_0,y) 
= \frac{1}{2\nu} \beta_{\Delta}(x,y) 
= \frac{\Gamma(\Delta)}{2\nu\,  \pi^{d/2}\Gamma(\nu)} \frac{1}{(x-y)^{2\Delta}}\, .
\ee
%%%
The final expression for the two-point function is then given by summing over all Feynman diagrams as in (\ref{fullO1}). It is expressed completely in terms of boundary-to-boundary AdS propagators as
%%%%%
\be\lab{fullO3}
\Omega(x,y) = N^2\sum_{g=0}^\infty N^{-2g} \sum_{\ell=0}^\infty \lambda_h^{\frac{2\ell}{h-2}}  v_\Delta\, \beta_{\Delta+2}(x,y)\, ,
\ee
%%%%%
where we defined
%%%
\be\lab{vf} 
v_\Delta = \sum_{F\in \mathcal{F}_\ell} \frac{1}{\sigma_F}\, v_F\, .
\ee
%%%
For the tree level contribution we use the convention above with $\Delta_0 = d/2-3$ and $\mathcal{V}_F=1$. 
Our result (\ref{fullO3}) means that each contribution to the full two point function of the massless field theory is proportional to a boundary-to-boundary AdS propagator, with the proportionality constant given by $v_F$.

We can also rewrite this result as an integral over $\Delta$ using \eqref{Delta2pt1}, which provides an interesting Kallen-Lehmann representation for the two-point function in the massless theory: 
%%%%
\begin{align}
\begin{split}\label{GG2loops}
&\Omega(x,y) =N\Omega_0(x,y) + N\lambda^{-2\frac{d-2}{d(h-2)-2h}}\int d\Delta\, n(\Delta)v_{\Delta}\, \lambda^{\frac{4\Delta}{d(h-2)-2h}} \beta_{\Delta+2}(x,y)\, ,
\end{split}
\end{align}
%%%%
where $\Omega_0$ is the free propagator, we defined the density  
%%%%
\be
n(\Delta)\equiv  \sum_{\ell=1}^{\infty} \delta\left(\Delta-\frac{d}{2}+1-\ell\left (\frac{d}{2}-\frac{h}{h-2}\right) \right)\, ,
\ee
%%%%
and we only considered planar graphs with $g=0$ for simplicity. 

Finally, one can also invert \eqref{fullO3} to express the AdS boundary-to-boundary propagators in terms of the field theory two-point function as follows. 
We can write \eqref{fullO3} (for $g=0$) as
%%%
\be
\Omega(\lambda) = \sum_{\ell=0}^{\infty}\, \lambda^{2\ell/(h-2)} f_\ell= \sum_{\ell=0}^{\infty}\,  z^{-\ell} f_\ell\, ,
\ee 
%%%
where we defined $z \equiv \lambda^{-2/(h-2)}$ and
%%%
\be\lab{fl1}
f_\ell   \equiv v_\Delta  \beta_{\Delta+2}(x,y)\,.
\ee
%%%
The full propagator is then precisely  in the form of a unilateral $\mathcal{Z}$-transform (which is usually regarded as the discrete version of a Laplace transform). Its inverse reads
%%%
\be\lab{fl2} 
f_\ell =  \frac{1}{2\pi i} \oint_{\mathcal{C}}dz\, \Omega(z) z^{\ell-1}\, ,
\ee
%%%
where the countour $\mathcal{C}$ is chosen such that that it encircles the origin and is entirely in the region of convergence of the integrand. Therefore we found that the boundary-to-boundary AdS propagator can be written as
%%%%
\be\lab{fl3} 
\beta_{\Delta+2}(x,y)= \frac{1}{v_\ell\, 2 \pi i } \oint_{\mathcal{C}}dz\, \Omega(z) z^{\ell-1}\, .
\ee
%%%%
The integrand on the right-hand side is essentially a weighted average of field theory propagators with measure $ \lambda^{-2\ell/(h-2)-1}d\lambda$.

%%%%%%%%%%%%%%%%%%%%%%%%%%%%%%%%%%%
%%%%%%%%%%%%%%%%%%%%%%%%%%%%%%%%%%%
\paragraph{Special interactions: $\tfrac{d}{2}=\tfrac{h}{h-2}$.}%\footnote{Fun fact: this condition is invariant under the exchange $d\leftrightarrow h$, i.e. if you solve for $h$ instead of $d$ you find $h/2= d/(d-2)$.}
%%%%%%%%%%%%%%%%%%%%%%%%%%%%%%%%%%%
%%%%%%%%%%%%%%%%%%%%%%%%%%%%%%%%%%%

In this case we have $h=2d/(d-2)$ and  $\Delta = \frac{d}{2}  - 1$ for all $\ell\ge1$, such that equation \eqref{fullO2} reduces to
%%%
\begin{align}
\begin{split}  \label{Gconf}
\Omega(x,y)& = 4^{\tfrac{d}{2}+1}\pi^{d/2}\frac{\Gamma(\tfrac{d}{2}+1)}{|x-y|^{d+2}}\left[1+\sum_{\ell=1}^{\infty}\lambda^{\ell \tfrac{d-2}{2}} \,\ell\sum_{F\in \mathcal{F}_{\ell}}\frac{1}{\sigma_F}\int_0^\infty\left(\prod_{r=1}^{\tfrac{d}{2}\ell-1}db_r\right) \, \delta\left(1-\mathcal{U}_F\right)\mathcal{A}_F\right] \, ,
\end{split}
\end{align}
%%% 
where we used that $\ell = (h-2)(I+1)/h = 2(I+1)/d$, equation \eqref{Iell}, as well as the expression for$\Omega_0$ \eqref{G0}. Using \eqref{VF3} we can also rewrite this as
%%%
\be\label{Gconf2}
\Omega(x,y)  = \frac{\pi^{d/2}}{|x-y|^{d+2}}\sum_{\ell=0}^{\infty}\lambda^{\ell \tfrac{d-2}{2}}\sum_{F\in \mathcal{F}_{\ell}}\frac{1}{\sigma_F}\int_0^\infty\left(\prod_{r=1}^{\tfrac{d}{2}\ell-1}da_r\right) \mathcal{A}_F^{-d/2}\exp\left(-\frac{\mathcal{U}_F}{4\mathcal{A}_F}\right)\,.
\ee
%%%
In this case the propagator manifestly exhibits conformal symmetry (i.e. it's invariant under Poincar\'e and covariant under rescaling). Note that, for integer $d$ and $h$, the condition $d=2h/(h-2)$ can only satisfied by $(d,h) = (3,6), (4,4), (6,3)$ for $d>2$. 
%and possibly for $d=2, h\rightarrow \infty$ (but that diagram would have $\ell-1$ internal lines which doesn't make sense (actually $\ell\rightarrow \infty$ in this case so maybe it makes sense)), or for $d\rightarrow \infty, h=2$ (which corresponds to a free massive scalar, kind of). {\color{red} Note that these last two cases probably don't make sense because the relation between $I$ and $\ell$ is only valid when $h\neq 2, \infty$.}

%{\color{purple} NZ: We have not introduced composite operators yet. Should this paragraph be here?}\DG{I agree that this should not be here. I think it can be removed completely as it is mentioned on the next section that the same results follow by changing the form of $\Delta$}
%For composite operators we would have instead (using \eqref{IellJ2})
%%%%
%\be
%\Delta_{J} =   \left(\frac{d}{2}-\sout{\mathcal{H}} {\color{purple} \frac{h}{h-2}}\right)\ell +\frac{2J-h}{h-2} + \frac{d}{2} =\frac{2J}{h-2} = \frac{J}{2}(d-2)  \, ,
%\ee
%%%%
%which is the correct dimension for a free scalar to the power of $J$. 

%%%%%%%%%%%%%%%%%%%%%%%%%%%%%%%%
\subsubsection{Two-point function of composite operators }\label{CompSect}
%%%%%%%%%%%%%%%%%%%%%%%%%%%%%%%%

Let us now consider the two-point function of composite operators $\langle \tr\Phi^J(x) \tr\Phi^J(y)\rangle$. The computation of the $\ell$-loop contribution to this correlator follows the same steps above, the main difference is that the minimum number of independent  loop momenta is $\ell=J-1$ for the free diagram. This is because only the total momentum of the operator $ \tr\Phi^J(k)$ is known, which leaves $J-1$ undetermined. Representation of the two-point amplitude in terms of Schwinger parameters are 
all the same and for the $\ell$-loop contribution one arrives at the expression (\ref{2pftaub}) with (\ref{Vn}) and (\ref{Delta2pt}). $\Delta$ is now expressed in terms of $\ell$ using (\ref{IVJ2}) and (\ref{ellVJ2}): 
%%%
\be\lab{DeltaJ}
\Delta_J = \frac{d}{2}(1+\ell)-I =   \left(\frac{d}{2}-\frac{h}{h-2}\right)\ell +\frac{2J-h}{h-2} + \frac{d}{2}-2\, .
\ee
%%%
In the massless case one finds
%%%
\be\lab{GnkJ}
\Omega_F^J(x,y)  =4^{\Delta_J+2}\pi^{d/2} \Gamma(\Delta+2) \frac{\mathcal{V}_F}{|x-y|^{2\Delta_J+4}}\, .
\ee
%%%
In the free case ($\ell=J-1, h=0$), this yields $\Delta +2 = \frac{d}{2}(J-1)-J+\frac{d}{2} = \frac{J(d-2)}{2}$ as expected. One can again sum over the loops as in (\ref{fullO2}) with the replacement $\Delta\to \Delta_J$ and the power of $\lambda_h$ is 
now given by $2(\ell+1-J)/(h-2)$. Then the analog of (\ref{fullO2}) for the composite operators are given by 
%%%%%
\be\lab{fullOJ2}
\Omega_J(x,y) = N^2 \pi^{\frac{d}{2}} \sum_{g=0}^\infty N^{-2g} \sum_{\ell=0}^\infty \lambda_h^{\frac{2(\ell+1-J)}{h-2}} \, \mathcal{V}_{\Delta_J}\,  |x-y|^{-2\Delta_J-4} \, .
\ee
%%%%%
Now, as in the end of the previous section, consider a theory with marginal interactions for which $h/(h-2) = d/2$ as in the case of $\mathcal{N} = 4$ super Yang-Mills. Then $\Delta_J$ in (\ref{DeltaJ}) becomes independent of $\ell$: $\Delta_J = 2J/(h-2) - 2$. 
From (\ref{GnkJ}) we observe that this agrees with the fact that, in a conformal field theory with $h=4$ the two point function $\Omega_J$ should scale as $|x-y|^{2J}$. Interactions can only change the proportionality factor. In fact we find the 
full answer 
%%%%%
\be\lab{fullOJ3}
\Omega_J(x,y) = \frac{C_J}{|x-y|^{2J}}\, , \qquad C_J =  N^2 \pi^{\frac{d}{2}} \sum_{g=0}^\infty N^{-2g} \sum_{\ell=0}^\infty \lambda_h^{\frac{2(\ell+1-J)}{h-2}} 4^J\Gamma(J) \sum_{F\in F_\ell} \frac{\mathcal{V}_F}{\sigma_F}  \, .
\ee
%%%%%
  with 
%%%
\be\lab{VnJ}
\mathcal{V}_F\equiv\ell\int_0^\infty\left(\prod_{r=1}^{I} db_r\right) \, \delta\left(1-\mathcal{U}_F \right)\mathcal{A}_F^{J-d/2}\, .
\ee 
%%%
Rest of the computations in section \ref{sec::2pfm0}, in particular expressions of the two point function as a sum over AdS propagators, remain valid with the aforementioned replacements. Having demonstrated that 
our results for $n$-point functions of $\Phi$ apply to $n$-point functions of composite operators with little change, we will not consider composites further below.

%%%%%%%%%%%%%%%%%%%%%%%%%%%%%%%%%%%
%%%%%%%%%%%%%%%%%%%%%%%%%%%%%%%%%%%
\subsubsection{Two-point function in massive theory}
\lab{sec::2pfm}
%%%%%%%%%%%%%%%%%%%%%%%%%%%%%%%%%%%
%%%%%%%%%%%%%%%%%%%%%%%%%%%%%%%%%%%
 
Following the same steps in Appendix \ref{2ptDetails}, it is not hard to derive (for $\ell\neq 0$) the following expression for the two-point amplitude that arises from a particular graph $F$.  
%%%
\begin{align} 
\begin{split} 
\Omega_F(k_1,k_2)&=\ell \, \delta\left(k_1+k_2\right) \int_{0}^{\infty}\left(\prod_{r=1}^{I} db_r\right)\delta\left(1-\mathcal{U}_F \right) \mathcal{A}_F ^{\nu+2}\int_0^{\infty}d \tau\, \tau^{-3-\nu}\, e^{- \tau(k_1^2 +M ^2_F)}\, , \label{2pftaubM}
\end{split}
\end{align}
where we defined an effective mass parameter
%%%
\be
M^2_F\equiv   \mathcal{A}_F(b_r)^{-1} \sum_{r=1}^I b_r m^2_r\,.
\ee
%%% 
Now we see that it is not possible to decouple the $\tau$ integral from the $b_r$ integrals. Therefore the integral over the Schwinger parameters does not decouple and become an overall factor. A natural question is whether we can still 
write the two point function as a sum of products of two bulk propagators in a putative higher dimensional background. The answer is in the affirmative and is instructive. We decompose the $\tau$ integral into two using equation (\ref{a1a2id}) and arrive at the following result using the same manipulations explained in the appendix and Fourier transforming to position space:  
%%%
\bea\lab{m2pf1}
\Omega_F(x_1,x_2) &=&  \frac{2^{-d}\ell\,\Gamma(2+2c)}{\Gamma(\nu+4+2c)\Gamma(1+c)^2} \int d^dz dt  \int_{0}^{\infty}\left(\prod_{r=1}^{I} db_r\right)\delta\left(1-\mathcal{U}_F \right) \mathcal{A}_F ^{\nu+2} \, t^{\nu + 3+2c}\nn\\
{}&& \qquad\qquad\qquad\qquad\qquad\qquad \times \, \prod_{i=1}^2 \int \frac{d\a_i}{\a_i^{d/2-c}} e^{-\a_i(t + M_F^2) - \frac{|x_i-z|^2}{4\a_i}} \, ,
\eea
%%%
where $c$ is an arbitrary real number as before. The second line above is almost in the same form as the product of two AdS propagators except that the mass term depends on the Schwinger parameters hence they do not completely decompose in two parts. They can be formally decomposed however by defining 
a generalized propagator that depends on the Schwinger parameters. Let us first carry out one of the $b$ integrals, say $b_I$, taking into account the delta function in the first line of (\ref{m2pf1}) which then gives a Jacobian $|\6 U_F^*/\6 b_I|^{-1}$ where quantities with a star denote its value under the substitution $b_I = b_I^*$ which solves $U_F = 1$. Now we define the measure 
%%%%
\be\lab{mmes} 
d\mu(b) \equiv \prod_{r=1}^{I-1} db_r \left|\frac{\6 U_F^*}{\6 b_I}\right|^{-\frac12} \left(A_F^*\ri)^{1+\frac{\nu}{2}}\, ,
\ee 
%%%%
and introduce a product of delta functions over the Schwinger parameters $\prod_{r=1}^{I-1} \delta(b_r-b_r') = \prod_{r=1}^{I-1}  \int_{-\infty}^{\infty} dw_r \exp(i w_r (b_r - b'_r))$ to glue two propagators with different Schwinger moduli at the junction 
$z,t$: 
%%%%% 
\be\lab{m2pf2}
\Omega_F(x_1,x_2) =  \frac{2^{-d}\ell \,\Gamma(2+2c)}{\Gamma(\nu+4+2c)\Gamma(1+c)^2} \int d^dz dt   \prod_{r=1}^{I-1}  \int_{-\infty}^{\infty} dw_r \, \tilde K(x_1,z,t;w) \tilde K(x_2,z,t;-w) \, ,
\ee
%%%%%
where we defined the bulk propagator 
%%%
\be\lab{stprop}
\tilde K(x,z,t;w) = \int d\mu(b) \, e^{i w\cdot b}\,  t^{\frac{\nu + 3+2c}{2}}  \int_0^\infty \frac{d\a}{\a^{d/2-c}}\,  e^{-\a(t + M_F^2) - \frac{|x-z|^2}{4\a}}  \, .
\ee
%%%
We note that there is no obvious choice for the arbitrary coefficient $c$ in this case --- it was fixed to obtain the standard AdS propagator form in the massless case --- however, the final answer should not depend on it. Dependence of the propagator on the Schwinger parameters in the massive case suggests non-criticality of the dual string. Indeed the mass term in the propagator in equation (\ref{npt}) is reminiscent of worldsheet cosmological constant: 
%%%%%
\be\lab{wscosmo} 
m^2 \sum_r a_r \leftrightarrow \int \mu \sqrt{g}\, .
\ee
%%%%%
It is natural to believe that the sum over $r$ --- which runs through all internal lines of the graph ---  becomes, in a putative continuum limit $I\to \infty$, an integral over the worldsheet and the mass term $m$ becomes related to a cosmological constant. 
Whether such a continuum limit exists is of course non-trivial and this connection can be made rigorous in special cases, see for example \cite{Klebanov:1991qa} in the case of 2D string theory. We will return to this question in section \ref{sec::FTST}. We conjecture that massive field theories can at most be dual to non-critical string theories where Weyl invariance is broken, hence the worldsheet metric can be gauge-fixed up to a conformal factor $g_{ab} = \exp(2w(\sigma)) \delta_{ab}$ which should then be integrated. It is then tempting to conjecture that the integral over the parameters $w_r$ in (\ref{m2pf2}) are related to this conformal mode. Put differently, the propagator $\tilde K(x_1,z,t;w)$ of a particular string mode above --- which corresponds to the particular field theory graph $F$ --- is specified not only by the embedding coordinates $z^m(\sigma)$ and $t(\sigma)$ but, in the case of non-critical string, also by the Liouville mode $w(\sigma)$.

In passing, let us also provide a compact expression for the two-point amplitude of the massive theory. We can Fourier transform to position space directly in (\ref{2pftaubM}) and perform the $\tau$ integral to obtain
%%%
\be \label{amp2pfM}
\Omega_F(x,y;m_r) = \frac{2^{ \Delta+3}\pi^{d/2}}{|x-y|^{\Delta+2}} \ell \, \int_{0}^{\infty}\left(\prod_{r=1}^{I} db_r\right)\delta\left(1-\mathcal{U}_F \right) \mathcal{A}_F ^{\nu+2}M_F^{\Delta+2}\mathcal{K}_{\Delta+2}\left(M_F|x-y|\right)\, ,
\ee
%%%
where we introduced the modified Bessel function of the second kind $\mathcal{K}_n$ via its integral representation
%%%
\be
\int_0^\infty d\rho\, \rho^{n-1}e^{-\rho R}e^{-T/\rho} = 2\left(\frac{T}{R}\right)^{n/2}\mathcal{K}_n\left(2\sqrt{RT}\right)\,,
\ee
%%%
valid for arbitrary $R>0 $ and $T>0$.
Note that
%%%
\be
\mathcal{K}_n(w) \sim w^{-n}2^{n-1} \Gamma(n)+w^n 2^{-n-1} \Gamma(-n)\,,
\ee
%%%
when $w\sim 0$, so that we obtain the massless two-point function \eqref{Omsol0} by expanding \eqref{amp2pfM} around $m_r\sim 0$.

%%%%%%%%%%%%%%%%%%%%%%%%%%%%%%%%%%%
%%%%%%%%%%%%%%%%%%%%%%%%%%%%%%%%%%%
\subsection{Regularization of loop integrals}
\lab{sec::loopdiv}
%%%%%%%%%%%%%%%%%%%%%%%%%%%%%%%%%%%
%%%%%%%%%%%%%%%%%%%%%%%%%%%%%%%%%%%

%%%%%%%%%%%%%%%%%%%%%%%%%%%%%%%%%%%
%%%%%%%%%%%%%%%%%%%%%%%%%%%%%%%%%%%
\subsubsection{Divergences}
\lab{sec::div}
%%%%%%%%%%%%%%%%%%%%%%%%%%%%%%%%%%%
%%%%%%%%%%%%%%%%%%%%%%%%%%%%%%%%%%%

So far we have ignored potential divergences contained in the two-point amplitudes. Strictly speaking the series of change of variables  that we performed to express the two-point function in terms of bulk propagators are not justified unless one can regularize these potential divergences. In this section we will list the potential divergences, focusing only on the two-point function for simplicity, show how to regularize them and discuss the subsequent step of renormalization. We assume that the field theory is renormalizable. 

Let us consider a particular graph with $I$ internal lines and $\ell$ independent loop momenta with $\ell\geq 1$. The superficial degree of divergence of the graph is given by the power of loop momenta in the integrand, that is 
$$ \Delta_{div} = d\ell - 2I = \left(d-\frac{2h}{h-2}\right)\ell + 1 \, ,$$
where we used (\ref{Iell}). The first term is non-positive for renormalizable theories i.e. for $h\leq 2d/(d-2)$, hence the diagram is naively expected to be less divergent for higher $\ell$, therefore one would in general only worry about a finite number of graphs with small $\ell$. Even though this naive expectation fails when the graph has sub-divergences, it still means that once a finite number of such sub-divergences are regularized, then the expression will be finite as there will be no new type of divergences arising for higher $\ell$. 

Classification of divergences can most easily be done by considering the Chisholm representation of the amplitude --- see e.g. \cite{Bogner:2010kv} and \cite{Arkani-Hamed:2022cqe} for a recent modern perspective on the nature of divergences in amplitudes with multiple loops: 
%%%%
\be\lab{Chrisholm} 
\Omega_F(k^2) \propto \Gamma(\frac{d}{2}-\Delta-2) \int \left(\prod_{r=1}^I a_r\right) \delta(1-\sum_{j=1}^I a_j)\frac{\mathcal{U}_F^{-\Delta-2}}{\left(\mathcal{A}_F k^2 + \mathcal{U}_F m^2  \right)^{\frac{d}{2}-\Delta-2}}\, ,
\ee
%%%%
where $\Delta$ is given by (\ref{Delta2pt1}). It is easier to classify the divergences in this representation because the range of $a_r$ integrals are bounded from above thanks to the condition  $\sum_{r=1}^I a_r = 1$. 
First of all, there is an overall UV divergence coming from the factor of $\Gamma(\frac{d}{2}-\Delta-2)$, which, in dimensional regularization $d\to 2-2\epsilon$ could give rise to a single pole as $1/\epsilon$ when
$\frac{d}{2}-\Delta-2 = -n$ for a non-negative integer $n$. In a renormalizable but non-super-renormalizable theory i.e. when \footnote{For example a $\Phi^4$ theory in 4D or $\Phi^3$ theory in 6D.} $h= 2d/(d-2)$, we have $n=1$ for all $\ell$ and the overall coefficient is expanded as $1/\ell \epsilon$. For a super-renormalizable theory, i.e. for $h< 2d/(d-2)$, this divergence arises only for first few $\ell$. For example for the $\Phi^3$ theory in 4D, $n=-\ell +1$ hence there is a divergence only for $\ell=1$. 
From (\ref{Chrisholm}) one finds the overall divergence, for $\frac{d}{2}-\Delta-2 = \ell \epsilon - n$ with $n\geq 0$ an integer. Denoting $\mathcal{K} = \mathcal{A}_F k^2 + \mathcal{U}_F m^2$ the integrand becomes 
%%%
\begin{align}
\begin{split}
\Gamma\left(-\nu-2\right)\frac{\mathcal{U}^{-\Delta-2}}{\mathcal{K}^{-\nu-2}} &= \Gamma\left(\ell \epsilon-n\right)\frac{\mathcal{U}^{\epsilon(1+\ell)-n-d/2}}{\mathcal{K}^{n-\ell \epsilon}} \\
& = \frac{(-)^n}{\ell  n!} \mathcal{U}^{-d/2}\left(\frac{\mathcal{K}}{\mathcal{U}}\right)^n \left[\frac{1}{\epsilon} +\log \frac{\mathcal{U}^{1+\ell}}{\mathcal{K}^\ell} + \ell \frac{\Gamma'(1+n)}{\Gamma(1+n)} \right] + O(\epsilon)\,.
\end{split}
\end{align}
%%%
This is fixed e.g. by adding a counterterm determined by the coefficient multiplying $1/\epsilon$ above.

Apart from this overall UV divergence, there may be UV sub-divergencies that arise from the poles of $ \mathcal{U}_F$. Recalling that all the coefficients in the polynomial $ \mathcal{U}_F$ are +1, this can only arise at the lower boundary of integrations when some $a_r =0$. 
For  $h= 2d/(d-2)$  this can happen for all $\ell >0$ but this divergence generically does not show up for a super-renormalizable theory, e.g. for $\Phi^3$ in 4D for which $-\Delta-2 = \ell -1$. 

Finally, there are potential IR divergences that arise from the poles of the polynomial $\mathcal{A}_F k^2 + \mathcal{U}_F m^2$. To see them one typically considers the Euclidean region where $k^2$ is non-negative and $m^2>0$. The physical region can be obtained by analytic continuation. In the Euclidean region,  divergences can arise again only on the boundary of the integrals i.e. when $ \mathcal{U}_F= \mathcal{A}_F=0$ for $m^2 \neq 0$ and for $\mathcal{A}_F=0$ for $m^2=0$. For $m^2=0$ this may happen for all $\ell$ in a renormalizable theory but only for the first few $\ell$ in a super-renormalizable theory. On the other hand such IR divergences are known to be absent in special theories e.g. $\Phi^4$ in 4D. For $m^2 \neq 0$ IR divergences will generically be absent in the Euclidean region.  

%%%%%%%%%%%%%%%%%%%%%%%%%%%%%%%%%%%
%%%%%%%%%%%%%%%%%%%%%%%%%%%%%%%%%%%
\subsubsection{Regularization}
\lab{sec::reg}
%%%%%%%%%%%%%%%%%%%%%%%%%%%%%%%%%%%
%%%%%%%%%%%%%%%%%%%%%%%%%%%%%%%%%%%

One can regularize all the UV divergences following a powerful generic method in the Schwinger parametric representation of amplitudes as discussed in \cite{Itzykson:1980rh}. The idea is to subtract all possible divergences contained in different possible 
parts of a graph  $F$ by subtracting a series of derivatives of the integrand with respect to rescaled Schwinger parameters. 
The fully regularized graph is given by the following expression
%%%% 
\be\label{regF}
\Omega_F(k_i) =  \int_0^{\infty} \prod_r \left( da_r e^{-m_r^2 a_r}\right) \mathcal{R}(a_r; k_i)\,,
\ee
%%%%
where 
%%%%
\be\label{Rdef}
\mathcal{R}(a_r;k_i) = \lim_{\gamma\rightarrow 1}\prod_{\sigma} \left(1- \mathcal{T}_{\gamma_\sigma}^{-2I_{\sigma}}\right) \left[\mathcal{U}( a_r)^{-d/2}\, e^{-P( a_r,k_i)}\right] \, .
\ee
%%%%
Here the product is over all $(2^I -1)$ non-empty subsets of $\{a_1,a_2,\cdots a_I\}$ and the differential operator $\mathcal{T}^{k}$ is defined as
%%%%
\be
\mathcal{T}^k f(\rho) = \gamma^{-p_1} \sum_{s=0}^{k+p_1} \frac{\gamma^s}{s!} \frac{d^s}{d\gamma^s} \left[\gamma^{p_1}f(\gamma)\right]_{\gamma=0}\,,
\ee
%%%% 
for all $p_1\ge p$ and $p$ is an integer such that $\gamma^{p} f(\gamma)$ is differentiable at $\gamma=0$.  In \eqref{Rdef} $I\sigma$ is the length of subset $\sigma$ of $\{a_1,a_2,\cdots a_I\}$ and the subscript $\gamma_\sigma$ indicates that only the elements of $\sigma$ are scaled by $\gamma$ while the other elements of $\{a_1,a_2,\cdots a_I\}$  are left untouched. One can show that (\ref{regF}) is independent of the choice of $p_1$. The regularized expression (\ref{regF}) is finite at all $a_r=0$, hence contains 
no UV divergences. 

One can explicitly check in the case of two-point function, that the regularized integrand $\mathcal{R}(a_r; k_i)$ is of the following generic form: 
%%%%
\be
\mathcal{R}(a_r;k) = \mathcal{U}( a_r)^{-d/2}\, e^{-P(a_r,k)} - \sum_{i=1}^{i_F} (k^2)^{n_i} F_i(a) \, ,
\ee
%%%%
where $i_F\geq 0$ is some integer that depends on the particular graph $F$ and $n_i$ is a set of integers which again depend on the graph $F$. The non-trivial fact is that the functions $F_i$ satisfy the following scaling relation
%%%%
\be
(k^2)^{n_i} F_i(a/k^2) = (k^2)^{d\ell/2} F_i(a) \, .
\ee
%%%%
As a result the regularized two-point amplitude can be written as 
%%%% 
\be\label{regF2}
\Omega_F(k_1) = \delta(k_1 + k_2) \int_0^{\infty} \prod_{r=1}^I \left( da_r e^{-\frac{m_r^2}{k_1^2} a_r}\right) (k_1^2)^{\frac{d\ell}{2}-I} \mathcal{R}(a;1)\, .
\ee
%%%%
We first consider the simpler case of massless theory. Using integral representation of the Gamma  function one has, 
%%%% 
\be\label{regF3}
\Omega_F(k_1) = \delta(k_1 + k_2) \, \mathcal{V}_F^R\, \int_0^\infty \frac{d\tau}{\tau} \tau^{-2-\Delta + \frac{d}{2}} e^{-\tau k_1^2}\, ,
\ee
%%%%
which is of the same form as equation (\ref{2pftaub}), now with the regularized coefficient 
%%%% 
\be\label{regVF}
\mathcal{V}_F^R =\frac{1}{\Gamma(\frac{d}{2}-2-\Delta)} \int_0^{\infty} \prod_{r=1}^I da_r \mathcal{R}(a;1)\, .
\ee
%%%%
The overall Gamma function should still be regularized for renormalizable theories with dimensionless coupling which can easily be done by dimensional regularization as explained above. As equation (\ref{regF3}) is of the same form as 
 (\ref{2pftaub}), the same manipulations in Appendix \ref{2ptDetails} go through and one obtains the same representations of the two-point function in terms of AdS bulk-to-boundary and boundary-to-boundary propagators as above where only the coefficients are now regularized and finite. 
 
 One may be surprised that in the marginal case $h=2d/(d-2)$ the two-point function is still in the AdS form $\Omega(x,y) \sim |x-y|^{-2\Delta-4}$ with $\Delta$ independent of $\ell$. This means that only the overall coefficient is modified by regularization and the full sum over $\ell$ is still proportional to $\Omega(k) \sim (k^2)^{\Delta-d/2-2}$. This will not be the case after proper renormalization, see below, where one determines the two-point function by imposing renormalization conditions at a given RG scale $\mu$. As we exemplified in Appendix \ref{renorm4}, renormalization renders the two-point function dependent on the ratio $k^2/\mu^2$, hence it will generically have the form   $\Omega(k) \sim (k^2)^{\Delta-d/2-2} R(k^2/\mu^2)$ where $R$ is some complicated function. 
 It is however true that if one takes the renormalization scale to UV then the additional dependence on $k^2/\mu^2$ disappears and one obtains the AdS form. Therefore the regularization above, in general, yields the UV limit of the theory which is expected to be conformal indeed. 
 
%%%%%%%%%%%%%%%%%%%%%%%%%%%%%%%%%%%
%%%%%%%%%%%%%%%%%%%%%%%%%%%%%%%%%%%
\subsubsection{Renormalization}
\lab{sec::ren} 
%%%%%%%%%%%%%%%%%%%%%%%%%%%%%%%%%%%
%%%%%%%%%%%%%%%%%%%%%%%%%%%%%%%%%%%

So far we only discussed how to subtract the divergences. The method above in fact corresponds to a particular renormalization scheme where the renormalization conditions are defined at $k^2=0$, see e.g. \cite{Itzykson:1980rh}. 
 One can also renormalize at a fixed scale $k^2 = \mu^2$ in the usual fashion. To see the general structure it is easiest to use dimensional regularization, introduce counterterms in the Lagrangian and carry out the computation 
 in the Schwinger representation. Consider $\Phi^4$ theory at the two-loop level. For simplicity of presentation we consider the massless theory. The Lagrangian with counterterms is
%%%%
\be
\mathcal{L} = -\frac{1}{2}(1+\delta_Z) \left(\partial \Phi\right)^2+ \frac{1}{2} (m^2+\delta_m)\Phi^2+ \frac{1}{4!} (\lambda+\delta_\lambda)\Phi^4\,.
\ee
%%%%
where the counterterms are of the form 
 %%%%
\be
\delta_\lambda\propto \frac{\lambda^2}{\epsilon} + O(\lambda^3)\,, \qquad  \delta_Z \propto \frac{\lambda^2}{\epsilon} + O(\lambda^3)\,, \qquad \delta m \propto \frac{\lambda}{\epsilon} + O(\lambda^2) 
\ee
%%%%
and they generally are expanded in powers of $\lambda$. This means that we can renormalize the $n$-point functions in the Schwinger representation (\ref{npt}) by redefining the field strength, mass and coupling constant as
%%%
\be
k^2\rightarrow (1+\delta_Z)k^2\,, \qquad \qquad m^2 \rightarrow m^2+\delta_m\,, \qquad \qquad \lambda\rightarrow \lambda+\delta_\lambda\,.
\ee
%%%
Therefore the generic form of our expressions for the amplitudes {\em at fixed order in} $\lambda$, e.g. (\ref{2pf1}) and (\ref{2pfell}) remain unchanged and only the coefficients will be modified. Note however that the full $n$-point functions 
with a sum over $\ell$ will look different as each amplitude for a given graph $F$ now contains an infinite expansion in $\lambda$, as do $\delta_Z,\delta_m$ and $\delta_\lambda$. One, in general also needs to evaluate the $h$-point --- where $h$ is the degree of interaction --- function to be able to implement the renormalization conditions. We provide an example for how to carry out renormalization in the Schwinger representation in appendix \ref{renorm4}. 

%%%%%%%%%%%%%%%%%%%%%%%%%%%%%%%%%%%
%%%%%%%%%%%%%%%%%%%%%%%%%%%%%%%%%%%
%%%%%%%%%%%%%%%%%%%%%%%%%%%%%%%%%%%
%%%%%%%%%%%%%%%%%%%%%%%%%%%%%%%%%%%
\subsection{Three-point function}
\lab{sec::3pf}
%%%%%%%%%%%%%%%%%%%%%%%%%%%%%%%%%%%
%%%%%%%%%%%%%%%%%%%%%%%%%%%%%%%%%%%
%%%%%%%%%%%%%%%%%%%%%%%%%%%%%%%%%%%
%%%%%%%%%%%%%%%%%%%%%%%%%%%%%%%%%%%

The formula (\ref{npt}) with $\mathcal{U}$ and $P$ defined below this equation apply directly also to higher point functions. The main difference from the two-point function is that now there are different families of two-trees with each side coupling to a different combination of momenta. In the simplest case of the three-point function these momenta will be $k_1$, $k_2$ and $k_3$ as the two-trees by definition divide the graph into two parts and in the case of the three-point function one part will always be connected to a single external leg. In the case of the four-point function these momenta will be $k_1$, $k_2$, $k_3$, $k_4$, $k_1+k_2$, $k_1+k_3$ and $k_1+k_4$. For simplicity we consider only the three-point function and set the mass of the field to zero.  
Then we write the function $P$ as 
%%%%
\be
 P(a,K)=\frac{1}{\mathcal{U}(a)} \sum_{i=1}^{3}\mathcal{A}^{(i)}k_i^2\, ,
 \ee
 %%%%
where $\mathcal{A}^{(i)}$ are sum over two-trees with $\ell+1$ Schwinger parameters $a_r$ that isolate external momentum $k_i$ on one part of the two-three. 
One can easily Fourier transform to the position space as the integrals are Gaussian: 
%%%%%
\begin{align}
\begin{split} \label{3pfgoal}
\Omega_F(x_i)&=\pi^{3 d/2} \int_{0}^{\infty}\prod_{r=1}^{I}da_r\, \mathcal{U}^{-d/2} \, \int d^d z\, \prod_{i=1}^3 \left[\left(\frac{\mathcal A^{(i)}}{\mathcal{U}}\right)^{-d/2} e^{ - \mathcal{U} \frac{(x_i-z)^2}{4 \mathcal{A}^{(i)} }}\right]
\end{split}
\end{align} 
%%%%% 
Defining $P_i=\mathcal{U}/\mathcal{A}^{(i)}$ we can rewrite this as 
%%%%%
\begin{align}\label{3pfgood}
\begin{split}
\Omega_F(x_i) &=\pi^{3 d/2} \int_{0}^{\infty}\prod_{r=1}^{I}da_r\,\mathcal{U}^{-d/2} \, \int d^d z\, \prod_{i=1}^3 \left[\int_0^{\infty}d\rho_i\, \rho_i^{d/2}\delta(\rho_i-P_i) e^{ - \rho_i (x_i-z)^2/4} \right]\\
&=\pi^{3d/2} \int\prod_{r=1}^{I}da_r\,\mathcal{U}^{-d/2}\prod_i P_i^{-c} \,  d^d z\, \prod_{i=1}^3 \left[\frac{d\rho_i d\alpha_i}{2\pi}  \rho_i^{d/2+c}e^{ - \frac{\rho_i}{4} (x_i-z)^2 + i\alpha_i(\rho_i-P_i)}  \right]
\end{split}
\end{align}
%%%%%
where we introduced delta functions $\delta(\rho_i-P_i)$ which are then represented as integrals over $\alpha_i$ from $-\infty$ to $\infty$ and also set $P_i^{d/2} = \rho_i^{d/2+c}P_i^{-c}$ with $c$ arbitrary. This is allowed because of the delta function that sets $\rho_i = P_i$. We also combined all the integrals under a single integral sign to reduce cluttering. We hope that range of these integrals are clear to the reader. Now change variable $a_r = (\sum_i \rho_i)^{-1} b_r$ so that $P_i\rightarrow ( \sum_j \rho_j) P_i$\footnote{Note that to do this change of variable we must assume that we can interchange the $\rho$ and $a$ integrals. This is allowed as long as the divergences are regulated as discussed in the previous section. We simply assume that these 
operations commute with the regularization procedure. Since we work in arbitrary dimension $d$, one can at least do dimensional regularization.}:
%%%%%
\be\label{3pfbad}
\Omega_F(x_i) =\pi^{3d/2} \int \prod_{r,i}\frac{db_r d^d z d\rho_id\alpha_i}{2\pi}\, \mathcal{U}^{-d/2}P_i^{-c} \,\rho_i^{d/2+c}e^{ - \frac{\rho_i}{4} (x_i-z)^2 + i \xi}  (\sum_i \rho_i)^{-I+d\ell/2-3c }
\ee
%%%%%
where 
%%%
\be \label{3ptxi}
\xi=\sum_{i=1}^{3}\alpha_i\left( \rho_i- (\sum_j \rho_j) P_i\right) = \sum_{i=1}^{3} \rho_i\left( \alpha_i(1-P_i)-\sum_{j\neq i}\alpha_j P_j\right)\equiv \sum_{i=1}^3\rho_i Q_i\, .
\ee
%%%
Now shift $\rho_i \rightarrow 4\rho_i$, and use the standard integral representation of the Gamma function to move the last term in (\ref{3pfbad}) into the exponential by introducing a new variable $\tau$:
%%%%
\be 
\Omega_F(x_i) =\frac{\pi^{3 d/2}4^{3 d/2+3-I+d\ell/2}}{\Gamma(I+3 c-d\ell/2)} \int \mathcal{U}^{-d/2} \tau^{I-d\ell/2+3 c-1}\prod_i P_i^{-c}  \rho_i^{d/2+c}e^{ - \rho_i\left[\tau+ (x_i-z)^2\right]} e^{i \xi}  \,.
\ee
%%%%
We now expand the exponential of $\xi$ as 
%%%
\be
e^{i\xi} =\prod_{i=1}^{3} \sum_{j_i=0}^{\infty} \frac{(i \rho_iQ_i)^{j_i}}{j_i!} \,,
\ee
%%%
change  variable $\tau = z_0^2$, and use the integral representations of the AdS bulk-to-boundary propagators, see (\ref{AdSbtb}), to write this expression succinctly as 
%%%%
\be\lab{3pfF}
\Omega_F(x_1,x_2,x_3) = \sum_{j_1,j_2,j_3=0}^\infty v^F_{j_1,j_2,j_3} \int \frac{d^{1+d}z}{z_0^{1+d}}  \prod_{i=1}^3 K_{j_i+\Delta+2}\left(z_0,z,x_i\right) \,,
\ee
%%%%
where, using the relation (\ref{Iell}) between then number of internal legs $I$ and the number independent loop momenta $\ell$ we defined the ``scale dimension of the graph"
%%%
\be
  \Delta = \frac{d}{3}\left(\ell+2\right) -\frac{2}{3}I = \frac23\left(\frac{d}{2} - \frac{h}{h-2}\right)\ell +\frac23 \left(d + \frac{h-3}{h-2}\right) \, .
\ee
%%%
We also used the freedom in choosing the constant $d$ and set $c = \Delta-d/2 + 1$. Equation (\ref{3pfF}) looks deceptively simple because all the complication is absorbed in the definition of the structure constants $v_{j_1,j_2,j_3}$: 
%%%%
\be\lab{vJ} 
v^F_{j_1,j_2,j_3} \equiv \frac{2\pi^{3 d}4^{ (3\Delta+d )/2} }{\Gamma( 3 (\frac{\Delta}{2}+ 1)-d/2)} \prod_{r,i} \int \frac{d b_r da_i}{j_i!}  \mathcal{U}^{-d/2} \Gamma(j_i+\Delta - d/2+2)P_i^{d/2-\Delta-1}(i  Q_i)^{j_i}\, ,
\ee
%%%%
where $Q_i$ is given in terms of $P_i$ and $\alpha_i$ as in (\ref{3ptxi}) and $P_i = \mathcal{U}/\mathcal{A}_i$. Equation (\ref{3pfF}) is of course one of the infinitely many contributions to the three-point function that are characterized by the choice of the Feynman diagram $F$. The full three-point function is obtained by summing (\ref{3pfF}) over $\ell$ and over all Feynman diagrams $F_\ell$ with $\ell$ loops including the symmetry factors.

%%%%%%%%%%%%%%%%%%%%%%%%%%%%%%%%%%%
%%%%%%%%%%%%%%%%%%%%%%%%%%%%%%%%%%%
%%%%%%%%%%%%%%%%%%%%%%%%%%%%%%%%%%%
%%%%%%%%%%%%%%%%%%%%%%%%%%%%%%%%%%%
\section{Relation to string amplitudes}
\lab{sec::FTST} 
%%%%%%%%%%%%%%%%%%%%%%%%%%%%%%%%%%%
%%%%%%%%%%%%%%%%%%%%%%%%%%%%%%%%%%%
%%%%%%%%%%%%%%%%%%%%%%%%%%%%%%%%%%%
%%%%%%%%%%%%%%%%%%%%%%%%%%%%%%%%%%%

Our final goal is to express $n$-point functions we consider in interacting holographic theories in d dimensions in terms of string amplitudes in a putative $d+1$-dimensional string theory. More precisely, we would like to read off properties of the 
$d+1$-dimensional bulk geometry directly from these $n$-point functions. This will be beyond the scope of this paper but, in this section, we would like to draw similarities between the two quantities and make some conjectures for a more detailed 
comparison between the two theories. For simplicity we consider planar contributions to the $n$-point functions in bosonic string that are expected to correspond to string amplitudes on the two-sphere $S_2$ with n punctures. 

%%%%%%%%%%%%%%%%%%%%%%%%%%%%%%%%%%%
%%%%%%%%%%%%%%%%%%%%%%%%%%%%%%%%%%%
%%%%%%%%%%%%%%%%%%%%%%%%%%%%%%%%%%%
\subsection{Flat target space as warm-up}
\lab{sec::FlatFTST} 
%%%%%%%%%%%%%%%%%%%%%%%%%%%%%%%%%%%
%%%%%%%%%%%%%%%%%%%%%%%%%%%%%%%%%%%
%%%%%%%%%%%%%%%%%%%%%%%%%%%%%%%%%%%

To set the stage we start with string theory on a flat $(d+1)$-dimensional background\footnote{Strictly speaking $d$ should be 25 for the bosonic string to cancel the Weyl anomaly but we will keep it arbitrary for later convenience.}, which we will generalize to a curved background later. 
Consider a string $n$-point function on the sphere embedded in flat $(d+1)$-dimensional space-time
%%%
\be\lab{snpf1} 
A_{S_2}^n(k_i,\sigma_i) = \la \left[ e^{i k_1 \cdot X(\sigma_1)}\right]_r  \cdots \left[ e^{i k_n \cdot X(\sigma_n)}\right]_r  \ra\, ,
\ee
%%%
where the subscript $r$ denotes renormalized vertex operators and the quantum average is given by the path integral over the world-sheet metric $g_{\a\b}(\sigma)$ and the embedding functions $X^M(\sigma)$. We take the momenta $k_i$ to lie in the 
$d$-dimensional subspace of the full bulk geometry. The result can be found for example in \cite{Polchinski:1998rq}, ch. 6, which we follow closely below. For the critical string one can fix the world-sheet metric completely to the conformal gauge 
%%%
\be\lab{wsm}
g_{\a\b} = e^{2w(\sigma)} \delta_{\a\b}\, .
\ee 
%%%
Using Poincar\'e invariance of the $(d+1)$-dimensional space, one can also expand the string embedding functions $X^M$ in string modes as 
%%%
\be\lab{expX}
X^M(\sigma) = \sum_I x_I^M X_I(\sigma)\, , 
\ee
%%%
where $x_I^M$ are constants to be integrated over and the string modes $X_I$ satisfy the world-sheet equation of motion and the orthogonality relation 
%%%
\be\lab{propX} 
\nabla^2 X_I = - \omega^2_I X_I \, , \qquad \int g^{1/2} X_I X_J = \delta_{IJ}\, . 
\ee
%%%
Decomposition (\ref{expX}) with (\ref{propX}) reduces the path integral over $M$ to ordinary integrals over $x_I^M$ which are all Gaussian. One should recall, however, that there is a constant zero-mode $X_0 = \left(\int d^2\sigma g^{1/2}\right)^{-1/2}$, hence 
the corresponding $x_0^M$ integral produces a $d$-dimensional delta function $\delta^d(\sum_i k_i)$ ($d$ instead of $d+1$ because we take $k_i$ along the $d$ directions). The final result of the Gaussian integrals is     
%%%
\bea\lab{snpf2} 
A_{S_2}^n(k_i,\sigma_i) &=& i X_0^{-d} \left(\det' -\frac{\nabla^2}{4\pi^2\alpha'} \right)^{-\frac{d}{2}}_{S_2}(2\pi)^d \delta(\sum_i k_i) \nn\\
{}&& \exp\left[ -\sum_{i<j}^n k_i \cdot k_j G'(\sigma_i,\sigma_j) - \frac12 \sum_{i=1}^n k_i^2 G'_r(\sigma_i,\sigma_j) \right]\, ,
\eea
%%%
where the determinant excludes the zero-mode 
%%%
\be\lab{detX} 
\det'\left( -\nabla^2\right) = \prod_{I\neq 0} \omega_I^2\, .
\ee
%%%
World-sheet Green's functions are given by 
%%%
\be\lab{wsG} 
G'(\sigma_1,\sigma_2) = \sum_{I\neq 0} \frac{2\pi \alpha'}{\omega_I^2} X_I(\sigma_1) X_I(\sigma_2)\, ,
\ee 
%%% 
and they satisfy 
%%%
\be\lab{wsG2} 
-\frac{1}{2\pi\alpha'} \nabla^2 G'(\sigma_1,\sigma_2) = g^{-1/2} \delta^2(\sigma_1-\sigma_2) - X_0^2\, .
\ee 
%%% 
The renormalized Green's function $G_r$ includes a subtraction of geodesic distance between the points $\sigma_1$ and $\sigma_2$: 
%%%
\be\lab{wsG3} 
G_r(\sigma_1,\sigma_2) = G'(\sigma_1,\sigma_2) + \frac{\alpha'}{2} \log d^2(\sigma_1,\sigma_2)\, ,\qquad d^2(\sigma_1,\sigma_2) \approx (\sigma_1-\sigma_2)^2 e^{2w(\sigma_1)} \, .
\ee 
%%% 
We will not need the full detailed expression for the geodesic distance as it cancels in the final expression. 

Let us now focus on the two-point amplitude for simplicity. Equation (\ref{snpf2}) reduces in this case to 
%%%
\be\lab{s2pf} 
A_{S_2}^n(k_i,\sigma_i) = i X_0^{-d} \left(\det' -\frac{\nabla^2}{4\pi^2\alpha'} \right)^{-\frac{d}{2}}_{S_2}(2\pi)^d \delta^d(k_1+k_2)\exp\left[ k_1^2 \bar G (\sigma_1,\sigma_2) \right]\, ,
\ee
%%%
where we defined a new Green's function 
%%%
\be\lab{wsG4} 
\bar G(\sigma_1,\sigma_2) = G'(\sigma_1,\sigma_2) - \frac12 G_r(\sigma_1,\sigma_1) - \frac12 G_r(\sigma_2,\sigma_2)  \, .
\ee 
%%% 
This is the expression that we want to compare with the field theory two-point function. We consider the massless theory for simplicity. The field theory two-point function is then given by 
%%%
\be\lab{FT2pf1} 
\Omega(k_1,k_2)  = \delta^d(k_1+k_2) \sum_{\ell} \sum_{F\in F_\ell} \frac{1}{\sigma_F} \lambda_h^{\frac{\ell}{h-2}} \prod_{r=1}^I \int_0^{\infty} da_r\, \mathcal{U}_F(a)^{-d/2} e^{-\frac{\mathcal{A}_F(a)}{\mathcal{U}_F(a)} k_1^2} \, ,
\ee 
%%% 
which can be rewritten in terms of graph structures $M$, $Q$ and $J$ defined in Appendix \ref{sec::Symanzik}, see equation (\ref{MQJ}), as follows 
%%%
\be\lab{FT2pf2} 
\Omega(k_1,k_2)  = \delta^d(k_1+k_2) \sum_{\ell} \sum_{F\in F_\ell} \frac{1}{\sigma_F} \prod_{r=1}^I \int_0^{\infty} da_r \, \det M(a)^{-d/2} e^{Q(a) M(a)^{-1} Q(a) + J(a)} \, ,
\ee 
%%% 
using equation (\ref{Sym2}). This can further be simplified, using (\ref{qmqj}) as 
%%%%
\be\lab{FT2pf3} 
\Omega(k_1,k_2)  = \delta^d(k_1+k_2) \sum_{\ell} \sum_{F\in F_\ell} \frac{1}{\sigma_F} \prod_{r=1}^I \int_0^{\infty} da_r \, \det M(a)^{-d/2} e^{k_1^2 \left(\mathcal{G}(1,2) - \frac12 \mathcal{G}(1,1) - \frac12 \mathcal{G}(2,2) -\Delta(1,1) - \Delta(2,2) \right)} \, ,
\ee 
%%%% 
where we defined 
%%%%
\be\lab{defsd}
\mathcal{G}(i,j) = -\bar \sigma_i \cdot M^{-1} \cdot \bar \sigma_j\, , \qquad (\bar \sigma_i)_m = \sum_r \lambda_{rm} \sigma_{ri} a_r\, ,\qquad \Delta(i,j) = \sum_r \sigma_{ri}\sigma_{rj} a_r\, .
\ee 
%%%%
We note that, as in Appendix \ref{sec::Symanzik}, one can define the total proper time by inserting $\delta(\tau - \sum_r a_r)$ in this expression and the rescaling Schwinger parameters as $a_r = \tau b_r$ which are then constrained as $\sum_r b_r =1$. Ignoring for the moment the sums over $\ell$ and $F_\ell$ in (\ref{FT2pf1}) we observe the following similarities 
between (\ref{s2pf}) and (\ref{FT2pf3}):  
%%%%
\be\lab{comp1} 
\nabla^2 \leftrightarrow M_{mn} \, , \qquad G'(\sigma_i,\sigma_j) \leftrightarrow \mathcal{G}(i,j)\, , \qquad \frac{\alpha'}{4} \log d^2(\sigma_i,\sigma_j) \leftrightarrow \Delta(i,j)\, .
\ee
%%%%  
These identifications are further supported by the fact that, from (\ref{wsG2}), one has $G'(\sigma_i,\sigma_j) \sim \frac{1}{\nabla^2}$ and indeed $M^{-1}$, appears in  $\mathcal{G}(i,j)$. Finally we note that the constraint on $G'$ and $\det' \nabla^2$ 
which we denote by a prime, arise from excluding the string zero mode in these quantities, and there is a similar constraint in the field theory computation that arises from the condition $\sum_r b_r = 1$. Thus the string zero mode is expected to be related to $\tau$. 

Our comparison is incomplete for several reasons. First, we compared string amplitude in {\em flat} $d+1$-dimensional space with the field theory amplitude which is, in general, expected to be dual to a {\em curved} $d+1$-dimensional space.  
Second, we ignored the sum over $\ell$ and $F_\ell$ in (\ref{FT2pf1}). Finally, it is unclear what the integrals over $b_r$ in (\ref{FT2pf1}) correspond\footnote{In the large N limit, a dictionary between the Schwinger parameters $b_r$ and the Strebel parametrization of moduli space was proposed in \cite{Gopakumar:2005fx}, giving a definite prescription  in which the open-closed string duality is realized. This prescription was refined in \cite{Razamat:2008} and an explicit realization was shown in \cite{Gaberdiel:2021}.}.  In the discussion below --- which will be unavoidably speculative --- we will improve on our comparison by taking these points into account one by one. 

%%%%%%%%%%%%%%%%%%%%%%%%%%%%%%%%%%%
%%%%%%%%%%%%%%%%%%%%%%%%%%%%%%%%%%%
%%%%%%%%%%%%%%%%%%%%%%%%%%%%%%%%%%%
\subsection{Curved target space}
\lab{sec::curvedFTST} 
%%%%%%%%%%%%%%%%%%%%%%%%%%%%%%%%%%%
%%%%%%%%%%%%%%%%%%%%%%%%%%%%%%%%%%%
%%%%%%%%%%%%%%%%%%%%%%%%%%%%%%%%%%%

The putative $d+1$-dimensional bulk spacetime should involve Poincar\'e invariance in its $d$-dimensional subspace. Hence we consider the following bulk metric: 
%%%
\be\lab{bulkmetric} 
ds^2 = e^{2A(r)} \left(dr^2 + \eta_{\mu\nu} dx^\m dx^\n\right)\, ,
\ee 
%%%
where $\eta_{\m\n}$ is the $d$-dimensional Minkowski metric. Using diffeomorphism invariance to fix the world-sheet metric in the conformal form (\ref{wsm}) one can write the corresponding Polyakov path integral in (\ref{snpf1}) as 
\be\lab{sr1} 
A_{S_2}^n(k_i,\sigma_i) = \int \mathcal{D} X^\mu(\sigma)  \mathcal{D} R(\sigma) \mathcal{D}w(\sigma) \, e^{-S_P} \, e^{i \int d^2 g^{1/2} \sigma J(\sigma)\cdot X(\sigma)} \, 
\ee
where $X^\mu$ and $R$ denote the embedding coordinates in the Minkowski and the ``holographic'' $r$ directions, $w$ denotes the conformal factor and the source is given by $J(\sigma) = \sum_{i=1}^n k_i \delta^2(\sigma-\sigma_i)$. We can expand the embedding coordinates in string modes as in the flat case as 
%%%
\be\lab{expXR}
X^\mu(\sigma) = \sum_I x_I^M X_I(\sigma)\, , \qquad R(\sigma) = \sum_I r_I  Y_I(\sigma)\, .
\ee
%%%
Therefore the source term in (\ref{sr1}) can be written as  $\int d^2\sigma J(\sigma)\cdot X(\sigma) = \sum_{I} x_{I,\mu} J_I^\mu$ where $J_I^\mu = \sum_i k_i^\mu X_I(\sigma_i)$. It is clear that the integral over the zero-mode 
$x_0^\mu$ will still yield a $d$-dimensional delta function $\delta^d(\sum_i k_i)$. On the other hand, the Polyakov action reads 
%%%%
\be\lab{Polyakov} 
S_P = \int d^2\sigma \delta^{\a\b} \left( \6_\a X^\mu \6_\b X^\nu \eta_{\mu\nu} e^{2A(R)} +\6_\a R \6_\b R e^{2A(R)}  \right)\, . 
\ee
%%%%
The string modes $X_I$  in (\ref{expXR}) satisfy $\nabla^2 X_I = \omega_I X_I$ for the metric (\ref{wsm}) with the conformal factor $w$. Then the $X^\mu$ kinetic term in the Polyakov action (\ref{Polyakov})   can be written as 
%%%%
\be\lab{Poly1} 
\int d^2\sigma \left( - x_I^\m x_{J,\m} X_I e^{2(w(\sigma)+ A(\sigma))} \nabla^2 X_J - 2 \6^\b X_J X_I  x_I^\m x_{J,\m} \6_\b R A'(R) e^{2A}  \right) \, .
\ee
%%%%
The second term above is a cubic interaction term and complicates the path integral. We will ignore it for the sake of the general discussion here by assuming, for example, that an orthogonality condition\footnote{One can also include it in the Gaussian integrals over $x_I^\m$, then it would modify the determinant and the Green's function in (\ref{snpf2}).} $X \6_\a X \6^\a R =0$ can be imposed. This condition will hold, for our discussion around equation (\ref{wstsi}) below, which concerns the saddle point of the $R$ path integral. 
More generally it can be achieved for example by gauge fixing the residual conformal symmetry on the worldsheet. We will not discuss gauge fixing of the conformal symmetry here in detail. 

We then make a change of variables in the $w(\sigma)$ path integral $w(\sigma) \to w(\sigma) - A(R(\sigma))$ for  every curve $R(\sigma)$ in the path integral over $R$. This would normally introduce a kinetic term for $w$ with the proportionality constant given by the conformal anomaly. Here we simply assume that the Weyl anomaly cancels\footnote{This can be assured by including ghosts that arise from gauge-fixing but, again, we will not worry about them for the heuristic discussion here.}.
%%%%%
%\be\lab{wR} 
%w(\sigma) = -A(R(\sigma)) \, , \qquad \forall R(\sigma)\, .
%\ee  
%%%%%
Then the first term reduces to $-\int d^2\sigma \sum_I x_I \cdot x_I \omega^2_I$. On the other hand the source term in (\ref{sr1}) can also be written as  $\sum_I x_I^\mu \cdot J_{I,\mu} = \sum_I x_I \cdot k $. Therefore the path integral over $X_I$ again becomes 
Gaussian\footnote{This is of course not true in general, for example when the second term in (\ref{Poly1}) is not ignored. In that case the $x_I^\m$ integrals are still Gaussian but one is still left with path integrals over $X_I$.} and the calculation for the flat bulk metric above almost goes through with the replacement   
$$ \nabla^2 \to e^{2A(\sigma)} \nabla^2\, .$$

We can still identify $M \leftrightarrow e^{2A(\sigma)} \nabla^2$ as in (\ref{comp1}) but this will now relate the Schwinger parameters with the function $A(\sigma)$. Indeed, from (\ref{MQJ2}) one finds that the non-trivial eigenvalues of $M$ are given by\footnote{Here we denote the eigenvalues of the matrix $M$ by $a_m$ with a slight abuse of notation.  Rank of $M$ is $\ell$ and its elements comprise of linear combinations of the Schwinger parameters $a_r$, with $r= 1,\cdots I$ where $I$ is the number of internal lines.} for $m=1,\cdots \ell$. Then it becomes tempting to identify
%%%%
\be\lab{bRid} 
a_m \leftrightarrow e^{2A(\sigma)} \, .
\ee
%%%%
In passing we note that, in principle, the worldsheet cosmological constant in (\ref{wscosmo}) can also be included in our analysis. This would correspond to the QFT mass term in (\ref{npt}) which is hidden in the $J$-term in (\ref{FT2pf2}), see (\ref{MQJ2}).  Indeed the mass term in (\ref{npt}) is proportional to $m^2\sum_r b_r = m^2 \sum_r 1/I \to \int \sqrt{g} d^2\sigma $  on the worldsheet in the continuum limit as we expect all $b_r$ approach the same value in this continuum limit\footnote{We expect this because of an emergent permutation symmetry of the dominant graphs in the large $\ell$ limit which permutes all $b_r$'s and leads to an emergent reparametrization symmetry on the worldsheet, see below.} and  because the sum $\sum_r b_r$ is restricted to be 1. 

The identification (\ref{bRid}) requires the existence of a continuum limit $\ell\to\infty$ upon which a continuous worldsheet arises from the dominant Feynman diagrams. We argue below for the existence of such a continuum limit in holographic QFTs. Furthermore, it is natural to assume that the large N limit of the QFT corresponds to a classical saddle $A(R)$ in the target space (\ref{bulkmetric}). Therefore the function $A(\sigma)$ amounts to the knowledge of $R(\sigma)$\footnote{We assume that $R(\sigma)$ can be determined uniquely from $A(\sigma)$ and $A(R)$. Our discussion easily generalizes to the non-unique case.}, and, through the identification (\ref{bRid}), integrals over $b_r$'s in (\ref{FT2pf1}) in the continuum limit is expected to approach to the path integral over the holographic coordinate $R(\sigma)$! That is 
%%%%
\be\lab{bRid2} 
\prod_r \int_0^\infty db_r  \leftrightarrow \mathcal{D} R(\sigma) \, .
\ee
%%%%
Here we note that the precise identification will involve non-trivial factors in the $R$ path integral, e.g. the second term in (\ref{Polyakov}). We leave a detailed study of the correspondence between the Schwinger parameters and the holographic coordinate to future work and discuss below only some generic, interesting observations that follow from this identification. 

The identification (\ref{bRid}) gives an interesting meaning to the Schwinger parameters, namely that they correspond to $A(\sigma)$. Suppose that in the continuum limit a particular solution $\{a_m^*\}$ extremizes these integrals. Using (\ref{bRid2}) this would correspond to the classical solution $R(\sigma) = R_{cl}(\sigma)$ which solves the $R$ equation of motion in (\ref{Polyakov}). Reversing the logic, then, one would first obtain the function $A(\sigma)$ from $\{a_m^*\}$ in the continuum limit. Second, as one also knows $R=R_{cl}$ in this limit, one would be able to construct the function target space $A(R)$ in the large-N limit. This would then determine the dual background function from the spherical QFT two-point function! 

%%%%%%%%%%%%%%%%%%%%%%%%%%%%%%%%%%%
%%%%%%%%%%%%%%%%%%%%%%%%%%%%%%%%%%%
%%%%%%%%%%%%%%%%%%%%%%%%%%%%%%%%%%%
\subsection{Continuum limit}
\lab{sec::contFTST} 
%%%%%%%%%%%%%%%%%%%%%%%%%%%%%%%%%%%
%%%%%%%%%%%%%%%%%%%%%%%%%%%%%%%%%%%
%%%%%%%%%%%%%%%%%%%%%%%%%%%%%%%%%%%

How can we make this proposed duality between the sum over QFT Feynman diagrams and string amplitudes more precise? We consider planar (genus-0) graphs for simplicity. One idea, motivated by the old correspondence between matrix quantum mechanics and 2D string theory \cite{Ginsparg:1993is}, is taking a continuum limit of Feynman diagrams where the number of vertices is sent to infinity, such that the graphs approximate the continuous world-sheets of a dual string theory. For example, in case of $\Phi^3$ theory, Feynman diagrams can be dualized and a graph with a large number of vertices can be viewed as the triangulation of a 2D surface which then becomes the world-sheet. This idea can be generalized to a $\Phi^h$ theory and to higher dimensions \cite{tHooft:1973alw}. In this limit one takes the number of loops to infinity and the area of triangles $a_{tr}\to 0$. Existence of this limit typically also requires tuning the coupling to a critical value $\lambda \to \lambda_c$. In our case as well, we would like to explore the existence of such a continuum limit. 

The two-point function is given by 
%%%
\be\lab{FT2pf5} 
\Omega(k_1,k_2)  =  \delta^d(k_1+k_2) \sum_{\ell} (k_1^2)^{\Delta+2-\frac{d}{2}} \lambda_h^{\frac{2\ell}{h-2}} v_\ell \equiv  \delta^d(k_1+k_2), \bar\Omega(z,k_1) 
%{}&& \equiv  \delta^d(k_1+k_2) \sum_{\ell} (k_1^2)^{\Delta+2-\frac{d}{2}} \lambda_h^{\frac{\ell}{h-2}} v_\ell\,
\ee
%%% 
where $z = \lambda_h^{-2/(h-2)}$ and 
%%%
\be\lab{vg}
v_\ell = \sum_{F\in F_\ell} \frac{1}{\sigma_F} \prod_{r=1}^I \int_0^{\infty} da_r\, \frac{e^{-\frac{\mathcal{A}_F(a)}{\mathcal{U}_F(a)}}}{ \mathcal{U}_F(a)^{d/2}} \, .
\ee 
%%%
Here $\Omega(z,k_1)$ has precisely the form of a unilateral $\mathcal{Z}$ transform, which we discussed above equation (\ref{fl2}). Therefore one can invert this relation and write
%%%
\be\lab{ft6} 
(k_1^2)^{\Delta+2-\frac{d}{2}} = \frac{1}{v_\ell\, 2 \pi i } \oint_{\mathcal{C}}dz\, \bar \Omega(z,k_1) z^{\ell-1}\, .
\ee
%%%
We want to know how the coefficient $v_\ell$ scales with $\ell$. This is hard to figure out for an arbitrary theory. However, recall that 
we expect the string dual to be a {\em critical} string only when there is no explicit breaking of scale invariance in the theory, which means $\lambda_h$ is dimensionless, that is $h = 2d/(d-2)$. Then, using (\ref{Delta2pt1}) we see that $\Delta = d/2-1$ --- or $\Delta = 2J/(h-2) -2$ for composite operators, see equation (\ref{DeltaJ}) --- and the LHS of (\ref{ft6}) becomes independent of $\ell$! For the RHS also to be independent, we need to have
%%%
\be\lab{vlim} 
v_\ell = \lim_{z\to 0} \frac{\bar\Omega(z,k_1)}{(k_1^2)^{\Delta+2-\frac{d}{2}}} \, z^\ell, \qquad \textrm{or}\qquad v_\ell =  z_c^\ell  \lim_{z\to z_c} \frac{\bar\Omega(z,k_1)}{(k_1^2)^{\Delta+2-\frac{d}{2}}} \frac{z-z_c}{z_c}\, 
\ee
%%%
when the integrand has a single pole at the origin $z=0$ or  when it has a single pole\footnote{We assume a single pole for simplicity.} at $z_c$, respectively. In either case we find that $v_\ell \propto \lambda_c^{-2\ell/(h-2)}$, where in the first case $\lambda_c = \infty$. This means that the expression (\ref{FT2pf5}) could indeed allow for a continuum limit. To see this,  consider dividing the sum over loops in (\ref{FT2pf5}) into two parts $\ell< \ell_{cut}$ and $\ell> \ell_{cut}$ for $\ell_{cut} \gg 1$: 
%%%
\be\lab{FT2pf4} 
\bar\Omega(z, k_1)  = \Omega_{ng} + \Omega_g \equiv   (k_1^2)^{\Delta+2-\frac{d}{2}} \left( \sum_{\ell<\ell_{cut}} + \sum_{\ell\geq \ell_{cut}}\right)  \lambda_h^{\frac{2\ell}{h-2}}\, v_\ell\, ,
\ee 
%%%
where the subscripts ``g'' and ``ng'' refer to geometric and non-geometric and the reason for this nomenclature will become clear below. It is clear that if there exists a continuum limit which will then be identified with a string world sheet, this would then emerge from $\Omega_g$. We therefore focus on $\Omega_{g}$ and take the limit $\ell_{cut}\to \infty$.  Then $\Omega_g$ will have a finite limit $\ell_{cut}\to\infty$ for $z\to z_c$ (or $\lambda_h \to \lambda_c$) precisely because of the scaling in (\ref{vlim}).  Obviously the converse is also true and we conclude that the continuum limit $\ell_{cut}\to\infty$ implies criticality $\lambda\to\lambda_c$. 

Let us analyze this limit in more detail. In general $v_\ell$ in (\ref{vlim}) will be given by $v_\ell = v_0(z_c) z_c^\ell$ where $v_0$ has no dependence on $\ell$. Then $\Omega_g$ in the continuum limit will become $\lim_{\ell_{cut}\to\infty} \Omega_g = (k_1^2)^{\Delta+2-\frac{d}{2}} \lim_{\ell\to\infty} (\lambda_h/\lambda_c)^{2\ell/(h-2)}$. For a generic $\lambda_h\leq \lambda_c$ this is non-vanishing only at $\lambda_h\to\lambda_c$ and existence of the continuum limit of $\Omega_g$ necessitates this critical coupling. For example when $z_c=0$ one obtains the result $\Omega_g \to (k_1^2)^{\Delta+2-\frac{d}{2}} \lim_{z_c\to0} v_0(z_c)$. One can, however, sum the full series over $\ell$ and obtain
%%%
\be\lab{lsum}
\Omega = (k_1^2)^{\Delta+2-\frac{d}{2}} \frac{v_0(z_c)}{1- \left(\frac{\lambda_h}{\lambda_c}\right)^{\frac{2}{h-2}}}\, .
\ee
%%%
In the most interesting case of $h=4, d=4$ we find a single pole\footnote{This is not surprising, of course, as we assumed this  above equation (\ref{vlim}). What is surprising is that we do not find the full Laurent series, i.e. there are no terms that go like $(\lambda_h-\lambda_c)^n$ for some positive integer $n$.} at $\lambda_h = \lambda_c$, $ \Omega = (k_1^2)^{\Delta+2-\frac{d}{2}} v_0(z_c) \lambda_c (\lambda_c- \lambda_h)^{-1}$. Here is an interesting observation. When $\lambda_c=\infty$ this full series become identical to the critical limit of $\Omega_g$. Therefore, in this case the full two-point function is given by its $\ell_{cut}\to\infty$ limit and dependence on $\lambda_h$ disappears. Below, we will see that the parameter $\lambda_c$ is related to AdS radius in string units. In the more general case however, while $\ell_{cut}\to\infty$ limit of $\Omega_g$ is independent of $\lambda_h$, the full series over $\ell$ generates this dependence, as in (\ref{lsum}). We conclude that the two-point function generically has a pole, its residue at this pole yields the continuum ``geometric'' limit, and the non-geometric contributions in (\ref{FT2pf4}) generates dependence on an additional parameter $\lambda_h$ with $\lambda_h-\lambda_c$ measuring proximity to the geometric limit. 

%%%%%%%%%%%%%%%%%%%%%%%%%%%%%%%%%%%
%%%%%%%%%%%%%%%%%%%%%%%%%%%%%%%%%%%
%%%%%%%%%%%%%%%%%%%%%%%%%%%%%%%%%%%
\subsection{Extrema of Schwinger parameters and criticality}
\lab{sec::extFTST} 
%%%%%%%%%%%%%%%%%%%%%%%%%%%%%%%%%%%
%%%%%%%%%%%%%%%%%%%%%%%%%%%%%%%%%%%
%%%%%%%%%%%%%%%%%%%%%%%%%%%%%%%%%%%

\begin{figure}
\begin{center}
\includegraphics[width=15cm]{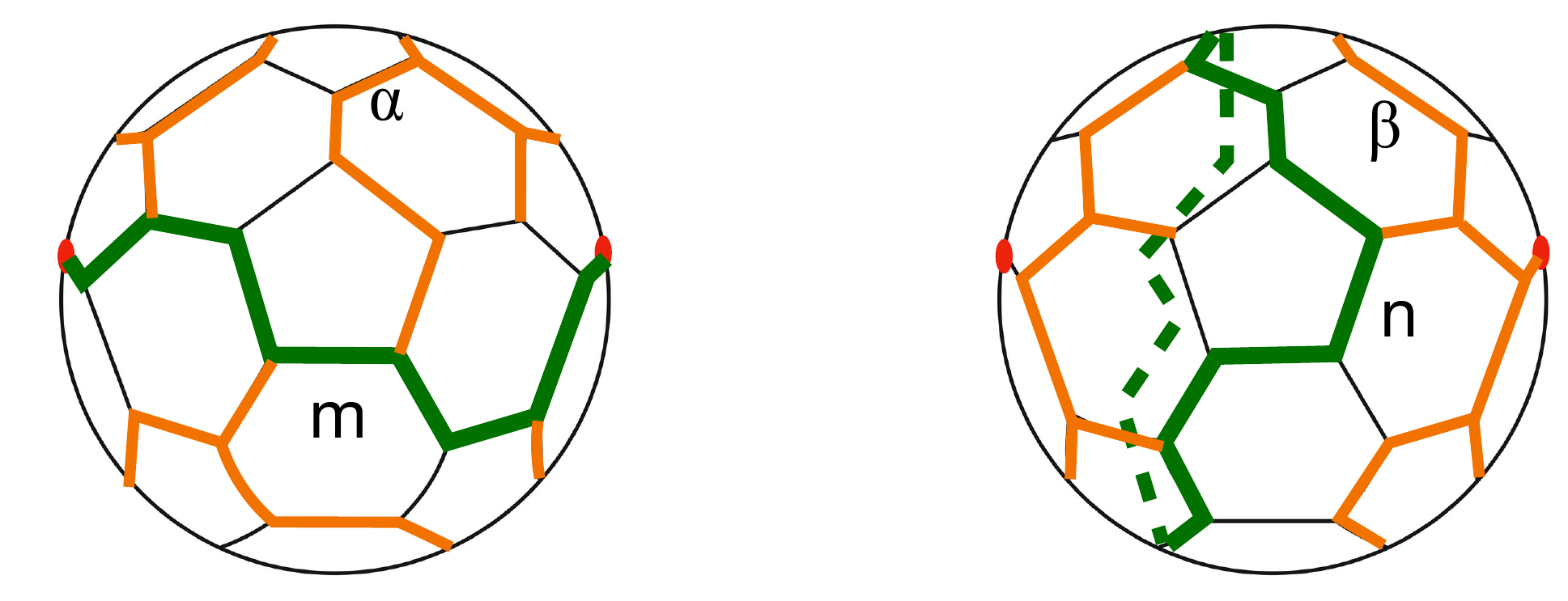}
\end{center}
\caption{\label{fig4} Example of a ``symmetric'' Feynman diagram with a fixed number of loops $\ell$ for the two-point function in $\Phi^3$ theory. (Left) Characterization of one-tree contributions to the first Symanzik polynomial $\mathcal{U}$ in terms of curves (green) that connect the two punctures i.e. external vertices (shown by red dots) --- we label different choices of this left-right curve by integer $m$ --- and the leftover uncut lines (orange) transverse to this curve --- we label different choices of a collection of orange lines by $\alpha$. A one-tree is given by all un-highlighted (thin/black) lines.  (Right) Characterization of two-tree contributions to the second Symanzik polynomial $\mathcal{A}$ in terms of non-contractible loops (green solid/dashed for front/back) that divide the graph into two parts (labelled by $n$) each containing one puncture (red dot) and cuts transverse to this loop (orange), labelled by $\beta$. }
\end{figure}

We will now argue that the critical limit we found above is related to extrema of the Schwinger parameters $a_r$. One expects from (\ref{bRid}) that a classical string $R=R_{cl}$ contribution to the string path integral will correspond to extremization of the field theory two-point function over the Schwinger parameters $a_r$. Let's try to get an idea about the dependence of  $\Omega_g$ on $\ell_{cut}$ by assuming that all Schwinger parameters are the same, i.e. $a_r=a$, on the dominant saddle. Even though we will not attempt to prove it, this is not unreasonable to expect in the continuum limit because of the following argument.  The saddle configuration $\{a_r^*\}$ is obtained by extremizing the exponent in (\ref{vg}) $\mathcal{A}_F(a)/\mathcal{U}_F(a) -d/2 \log  \mathcal{U}_F(a)$ with respect to all $a_r$. In the continuum limit $\ell \to \infty$ we expect the path integral to be dominated by ``symmetric graphs'' which we define as the graphs that are {\em not} n-particle reducible where $n\ll I$ (recall that $I$ is the total number of internal lines) when one cuts the graph at an arbitrary place\footnote{By arbitrary we mean not next to the punctures where the external momenta are inserted. These will necessarily be $(h-1)$-particle reducible where $h$ is the coordination number of the vertex.}, see figure \ref{fig4} as an example. Even though such ``finitely-reducible" graphs contribute to the amplitude, their contribution will be negligible in the continuum limit. Now, in the continuum limit --- which we expect to be dominated by such symmetric graphs --- one also expects an emergent permutation symmetry under all permutations of $\{a_r^* \}$: consider a symmetric graph, Symanzik polynomials of which are fixed in terms of the characteristic matrices $\lambda_{rm}$ and $\sigma_{ri}$, see Appendix \ref{sec::Symanzik}. 
For a (large) fixed value of $\ell$, one should sum over many of such symmetric graphs with the same $\ell$. Given any symmetric graph, it is natural to assume in the limit $\ell\to\infty$ that all permutations of $a_r$s will yield another symmetric graph that should be contained in the sum, which then gives rise to a large permutation symmetry that permutes the labels $r$\footnote{In passing we note that it is tempting to identify the continuum limit of this large permutation symmetry with {\em reparametrization} invariance of the emerging world-sheet because the permutation symmetry of Schwinger parameters will be inherited by elements of $M$ which are made of $a_m$ where $m$ labels loops. In the continuum limit $\ell\to\infty$, $a_m \to a(\sigma)$ and we expect the symmetry that permutes $m$ to turn into general coordinate transformations $\sigma^a \to \tilde \sigma^a(\sigma)$ on the world-sheet. See \cite{Klebanov:1991qa} for the same phenomenon in the old matrix models.}. Our assumption that $a_r=a$ on a saddle of Schwinger integrals in the continuum limit then follows from this argument. 

%%%%%
%%%%%
%%%%%
%%%%%

In this case the Symanzik polynomials simplify and are given by $\mathcal{U}_F(a) = N_1 a^\ell$ and $\mathcal{A}_F(a) = N_2 a^{\ell+1}$ where $N_1$ and $N_2$ are the number of one-trees and two-trees for a given graph $F$. It is in general hard to figure out the $\ell$ dependence of the exponent $\mathcal{A}_F/\mathcal{U}_F$ in (\ref{FT2pf5}) but we can make the following estimate in the continuum limit $\ell\to\infty$. Consider a planar graph which we can draw on a sphere. For the two point function we have two punctures on this sphere where vertex operators are inserted. Take them to be located at the left and right poles of the sphere, see figure \ref{fig4}. The number $N_1$ of one-trees is given by counting the number of curves that connect the left and the right poles (green curves on left figure \ref{fig4}) together with all possible cuts of the loops on the sphere that do not intersect this curve (orange curves in left figure \ref{fig4})\footnote{Note that in this picture we are classifying one-trees by their ``dual one-trees'', see \cite{Lam:1969xk},  i.e. the collection of lines in the Feynman diagram that are left over after removing a one-tree.}. The area of the sphere scales as the number of loops. Therefore the radius, hence a typical size of the left/right curve scales as $\sqrt{\ell}$. All other such curves are obtained by fluctuations of this curve embedded on the sphere. Suppose that the number of such fluctuations --- labelled by $m$ on figure \ref{fig4} --- is $N_f$. Call the total number of possible cuts of all the loops that are not intersecting the  curve $m$, which is counted by the orange curves in (left) figure \ref{fig4} and labelled by $\alpha$ --- as $N_c(m)$. Their number, i.e. the range of $\alpha$, will depend on the choice of curve $m$ in general. However in the large $\ell$ limit one expects this dependence to be negligible, hence $N_c(m) \sim N_c$ for all $m$ in this limit. All in all we have $N_1 \sim N_f N_c$ in the large $\ell$ limit.  
 
Now, apply the same counting to the number of two-trees shown on the right figure \ref{fig4}. Two-trees are given by all possible overall cuts of the lines on the sphere, such that the graph is divided in two parts which respectively contain the left and the right poles.  This can also be classified by the number of non-contractible loops inserted between the left and the right poles, see right figure \ref{fig4}. One can estimate their number by inserting rigid loops, multiplicity of which scales as $\sqrt\ell$\footnote{To see this imagine inserting a rigid (non-wiggling) and non-contractible circle on the two-punctured sphere. For simplicity think of inserting them vertically i.e. the line connecting the north pole and the south pole would be perpendicular to the line connecting the left and the right poles i.e. the two punctures. Their number is then proportional to the radius of the sphere hence $\sqrt\ell$. There are of course also tilted rigid circles that can be inserted, but they will be accounted for by the fluctuations of the vertical rigid ones.}  accounting for possible embedding of the mid-point of the one-loop on the sphere,  times all possible fluctuations times all possible cuts of the remaining quantum loops on the sphere.  In the continuum limit, the latter are expected to be proportional to respectively $N_f$ and $N_c$ ,which we defined above when counting one-trees. Hence we learn that $N_2/N_1 = \sqrt \ell \, \kappa$, where $\kappa$ is some exponent that characterizes the continuum limit of the Feynman diagrams and depends on  the properties of the underlying QFT, e.g.  the coordination number of the interaction $h$, number of dimensions $d$, etc. 

Therefore the exponential in (\ref{vg}) scales as $\sqrt \ell a \kappa - d/2 \ell \log a$ and all in all we found that 
%%%
\be\lab{vg1}
\lim_{\ell\to\infty}  v_\ell  \sim N_F(\ell) \,  a^{-d\ell/2}e^{-a \kappa \sqrt\ell}  \, .
\ee 
%%%
Here $N_F(\ell)$ is the number of all possible Feynman diagrams including their symmetry factors, which diverges in the limit but luckily it factors out. Comparison of this with (\ref{vlim}) yields the value of the critical coupling constant 
%%%
\be\lab{lambdac}
 z_c = \lambda_c^{-\frac{2}{h-2}} \sim a^{-d/2}\, .
 \ee
 %%%
%The critical value $\lambda_c =  \infty$ of the classical AdS/CFT corresponds to $a=\infty$ that is the IR limit of the Feynman diagrams. 
The value of $a$ should further be obtained by extremizing the full expression (\ref{vg}). Note that determining the precise relation between $a$ and $\lambda_c$ requires obtaining the dependence of $a$ on $\ell$ and of $\Omega_g$ on $z_c$ as proportionality constants depend on these relations. 
We will not do this here. On the other hand we argue below that in the case of a conformal field theory the value of $a$ is given in terms of the AdS radius in string units. 

%%%%%%%%%%%%%%%%%%%%%%%%%%%%%%%%%%%
%%%%%%%%%%%%%%%%%%%%%%%%%%%%%%%%%%%
%%%%%%%%%%%%%%%%%%%%%%%%%%%%%%%%%%%
\subsection{AdS/CFT and beyond}
\lab{sec::AdSCFT} 
%%%%%%%%%%%%%%%%%%%%%%%%%%%%%%%%%%%
%%%%%%%%%%%%%%%%%%%%%%%%%%%%%%%%%%%
%%%%%%%%%%%%%%%%%%%%%%%%%%%%%%%%%%%

A check of the identification (\ref{bRid}) and our arguments on the continuum limit above follows from considering a conformal field theory. If the field theory is conformal, one expects the target space scale factor in (\ref{bulkmetric}) to be given by $A(r) = -\log r$. Let us see how this arises from the field theoretic $n$-point function. Conformality of the two-point amplitude implies homogeneity under the transformation $k^\mu \to k^\mu \Lambda$, $x^\mu \to x^\mu/\Lambda$. From (\ref{FT2pf1})  this implies homogeneity under $a_r\to a_r \Lambda^2$. We argued that in the continuum limit the saddle point of the $a_m$ integrals should be given by $a_r = a$. Therefore conformality in the continuum limit implies symmetry of the two-point amplitude under $a\to a \Lambda^2$. On the other hand, this permutation symmetry of $\{a_r\}$ that arises in the continuum limit should be equivalent to the diffeomorphism symmetry of the worldsheet that we identify with the continuum limit of the Feynman diagrams. Using the diffeomorphism symmetry we can then write the worldsheet metric as $ds_{ws}^2 = e^{2w(\tau)}(d\tau^2 + \tau^2 d\theta^2)$. One can think of $\tau$ as defining a radius and $\theta$ as an angle on the stereographic projection of the Feynman diagrams drawn on a sphere. In terms of the original worldsheet coordinates one has $\sigma^1 = \tau \cos\theta$ and $\sigma^2= \tau \sin\theta$.  $\tau$ should be charged under scale transformation of the QFT as the line element on the world sheet is identified with the line element of the target space. Then dimensional analysis dictates $\tau\to \tau/\Lambda$ under the scalings above. Scale invariance of the world-sheet, which is inherited from the scale invariance of the QFT in the continuum limit, then requires $\exp(2w(\tau)) \propto \tau^{-2}$. Embedding this string in the bulk target space 
%%%%
\be\lab{wstsi} 
ds_{ws}^2 \propto \frac{1}{\tau^2} \left( d\tau^2 + \tau^2 d\theta^2\right) = e^{2A(R)} \left(dR^2 + \delta_{\mu\nu} dX^\mu dX^\nu\right)\, 
\ee
%%%%
and using the identification (\ref{bRid}) --- which implies  $\exp(2A) \to \Lambda^2 \exp(2A)$ under $a\to a \Lambda^2$  --- then scale invariance of the RHS fixes $\tau\propto R$ and $A(R) = -\log(R/l_{AdS})$ where $l_{AdS}$ is the AdS radius. Note that this argument is only valid when the field theory is conformal at the full quantum level. For example the scaling symmetry of (\ref{FT2pf1}) will be broken by the counterterms in (\ref{regVF}).  

For a CFT the critical value of $a$ in fact becomes a moduli that can be related to the AdS radius in string units in the dual string theory. As a specific example consider $d=4$, $h=4$. From (\ref{lambdac}) we find that $\lambda_c = (a\Lambda_{UV}^2/E^2)^2$ where we inserted an energy scale $E$ of the process in units of some UV cut-off energy $\Lambda_{UV}$. The energy scales as $E\to E \Lambda$ and accounts for invariance of $\lambda_c$ under $a\to a \Lambda^2$ rendering the 't Hooft coupling a moduli of the CFT. It is clear from the discussion above that this energy scale should be identified with the radial coordinate $R$ as\footnote{This follows from the scaling property $x\to x /\Lambda$ which leads to $R\to R/\Lambda$ by (\ref{wstsi}), and that energy should obey $E \to E \Lambda$ by dimensional analysis.} $R\propto 1/E$. Then using the identification in the previous paragraph we find $a\Lambda_{UV}^2/E^2 \propto (l_{AdS}/R)^2 (\Lambda_{UV}/E)^2 \propto (l_{AdS}\Lambda_{UV})^{2}$. Identification of the UV cut-off with the string length then yields 
%%%%
\be\lab{lambdac2}
\lambda_c \propto (l_{AdS}/l_s)^4\, .
\ee
%%%%
This is indeed the correct identification in the usual $AdS_5/CFT_4$ correspondence and the limit of Einstein gravity corresponds to $\lambda_c\to\infty$.

This argument can be generalized to QFTs that are classically scale invariant but possessing a non-trivial beta-function quantum mechanically. We expect the identification (\ref{bRid}) and the fact that the world-sheet metric arises from the continuum limit 
of the Feynman diagrams, i.e. $ds_{ws}^2 = e^{2w(\tau)}(d\tau^2 + \tau^2 d\theta^2)$, still to hold in this case. However the continuum saddle of Schwinger parameters $a$ will have a generic dependence $a=\tau^{-2} F[\tau\mu]$ where $\mu$ is the RG scale 
and $F$ is to be determined from the RG equations. For example one can choose $\mu$ as the dynamically generated energy scale $\Lambda_{QCD}$ in QCD. Assuming that we determine this function from Callan-Symanzik equations, then inverting the relation between $a$ and $\tau$ as $\tau = G[a/\mu^2]$, and using $a = \exp{2A}$ we find $\tau = H[A-\log \mu]$ where $G$ and $H$ are some functions. Then embedding of the world-sheet in bulk space would yield $A= w[H[A-\log\mu]]$, hence $w = H^{-1}$ from which one can determine $w(\tau)$ hence $A$ as a function of $R$ upon using the same identification $\tau \propto R$ as above.

%%%%%%%%%%%%%%%%%%%%%%%%%%%%%%%%%%%
%%%%%%%%%%%%%%%%%%%%%%%%%%%%%%%%%%%
%%%%%%%%%%%%%%%%%%%%%%%%%%%%%%%%%%%
%%%%%%%%%%%%%%%%%%%%%%%%%%%%%%%%%%%
\section{Discussion}
\lab{sec::discuss} 
%%%%%%%%%%%%%%%%%%%%%%%%%%%%%%%%%%%
%%%%%%%%%%%%%%%%%%%%%%%%%%%%%%%%%%%
%%%%%%%%%%%%%%%%%%%%%%%%%%%%%%%%%%%
%%%%%%%%%%%%%%%%%%%%%%%%%%%%%%%%%%%

Our main results for canonical fields Equations (\ref{fullO2}) and (\ref{fullO3}), and for composite fields,  (\ref{fullOJ2}) and (\ref{fullOJ3}), constitute novel forms of Kallen-Lehmann representations for the two-point function. These equations, generically, include the genus summation. In particular one can carry out the genus summation as in  (\ref{fullO7}). Therefore our expressions are valid for a large but finite $N$. Let us consider a CFT in 4D. Then the string coupling will be $g_s \propto g_{YM}^2 = \lambda_h/N$. Assuming S-duality of string theory, under $g_s \to 1/g_s$ this corresponds to 
$\lambda_h\to N^2/\lambda_h$. Therefore an S-duality in the 't Hooft coupling $\lambda_h$ only arises for $N$ finite, except for $\lambda_h=\infty$. In the latter case S-duality does not place any restrictions on the theory. When $N$ is finite however, one can then use the S-duality of the field theory under  $\lambda_h\to N^2/\lambda_h$ to obtain the non-perturbative limit $g_s\geq 1$ of string theory. Performing both the sum over $\ell$ and $g$ in the two-point function, one would arrive at an expression
$\Omega \propto F_1(1/N^2, \lambda_h) = F_2(g_s,\lambda_h)$. S-duality then implies $F_2(g_s,\lambda_h) = F_2(1/g_s,\lambda_h/g_s^2)$.  Therefore one finds $F_2(1/g_s,\lambda_h) = F_2(g_s,\lambda_h g_s^2)$. This means that the strong coupling limit of string theory amplitude is determined by both the weak string and weak 't Hooft coupling. Knowledge of function $F_2$ as a function of $\lambda_h$ allows one to find an expression for the string amplitude at inverse string coupling. Note that the string amplitude 
becomes self S-dual for $\lambda_h=0$ and $\lambda_h=\infty$. In both cases therefore one expects the dual background to be AdS with constant dilaton. This is consistent with both Gopakumar's finding in free field theory \cite{Gopakumar:2003ns} and the AdS/CFT duality in Einstein's gravity limit. On the other hand, for finite $\lambda_h$ S-duality of the string amplitude also modifies the AdS length scale $l_{AdS}/l_s$ to  $l_{AdS}/(l_s g_s^{\frac12})$. This is, of course, long known, see for example \cite{Argurio:1998cp} for a review.

In section  \ref{sec::FTST} we separated small and large $\ell$ contributions to field theory amplitudes, called them ``non-geometric" and ``geometric" respectively, and argued that the latter, in the weak string coupling limit, leads to perturbative string amplitudes where Feynman diagrams become smooth string worldsheets \`a la 't Hooft \cite{tHooft:1973alw}. We also argued that this continuum limit necessitates criticality of `t Hooft coupling $\lambda_h\to \lambda_c$. In this limit there are only two parameters in the dual theory, string coupling $1/N^2 \sim g_s \ll 1$ and $\lambda_c$ that is related to $l_{AdS}$ in string units. On the other hand, we found that inclusion of non-geometric contributions yields an additional parameter $\lambda_h$ which characterizes the non-geometric contributions and is different than $\lambda_c$ away from the geometric string limit. What does this parameter correspond to in string theory? It is tempting to conjecture that these non-geometric contributions generate a new scale which characterize collective excitations of strings that become the weakly coupled when $g_s\sim 1$. We can determine this ``Planck scale'' as follows. Consider computing free energy of the field theory from the path integral. This will be proportional to $N^2 \propto \lambda_h^2/ g_s^2$. The corresponding quantity is the 5D gravity action that is proportional to $(l_{AdS}/l_5)^3/g_s \propto (l_{AdS}/l_{10})^8/g_s $ where $l_5$ and $l_{10}$ are the 5 and 10 dimensional Planck scales respectively. Recalling Equation (\ref{lambdac2}) from previous section, we arrive at the following relation for the ratio, 
%%%%
\be\lab{lhlc2} 
\frac{\lambda_h}{\lambda_c} \propto g_s^\frac12 \le(\frac{l_s}{l_5}\ri)^4 \le(\frac{l_5}{l_{Ads}}\ri)^\frac52  =  g_s^\frac12 \le(\frac{l_s}{l_{10}}\ri)^4 \, .
\ee
%%%%   
Therefore we can understand the new parameter $\lambda_h/\lambda_c$ as the Planck scale in string units which is not necessarily 1. 

Our results can be extended in many new directions. Perhaps the most interesting will be deriving the dual gravitational background, in particular the scale factor $A(r)$ from field theory $n$-point amplitudes. We outlined one possible route in the previous section, by extremizing the integrals over the Schwinger parameters and extracting the function $A(r)$ from this extremum configuration in the continuum limit. Whether this can indeed be done reliably in examples beyond AdS remains to be seen. Another possible route would be reading off the dual geometry from the field theory two-point function, i.e. equation (\ref{fullO3}), given as a sum over boundary-to-boundary AdS propagators. Perhaps, one can extract the scale factor $A(r)$ from the coefficients in this sum, regarding AdS propagators with different masses as a complete set of solutions in which one can expand the solution to Green's function equation in the full geometry. This is likely to work at least near the asymptotically AdS boundary. 

%%%%%
%%%%%
\begin{figure}
\begin{center}
\includegraphics[width=15cm]{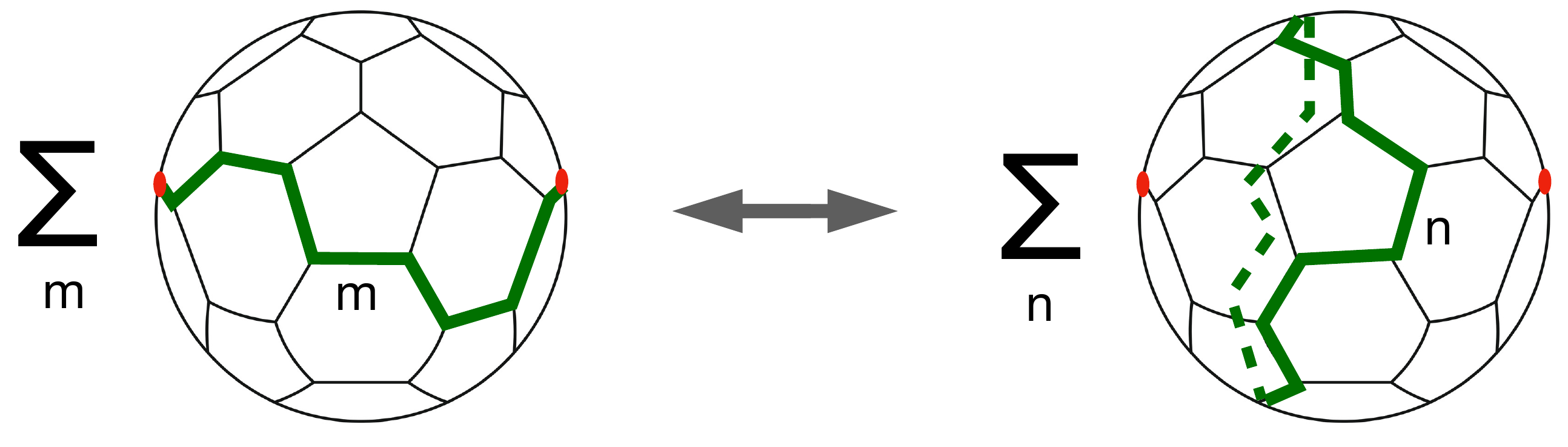}
\end{center}
\caption{\label{fig3} A discrete analog of the open-closed duality in spherical Feynman diagrams in the example of $\Phi^3$ theory. Red dots denote punctures i.e. insertions of external momenta. Left sum is over paths connected to punctures, right sum over non-contractible closed paths. Duality arises in the limit where the number of loops diverge, assuming that such a continuum limit exists.}
\end{figure}
%%%%%
%%%%%

Another possible extension involves 
our result in Appendix \ref{Full2pf} for the closed-form expression for the two-point function. As discussed this appendix, we believe our result can be extended to an arbitrary interaction $\Phi^h$, to arbitrary $n$-point function and arbitrarily large genus. Whether this can indeed be done, and if so, whether this new method could lead to new advances in perturbative quantum field theory remains to be seen. 

Finally, consider the symmetry of the two-point function under exchanging Symanzik polynomials $\mathcal{A}_F \leftrightarrow \mathcal{U}_F$ which we noted at the end of Appendix \ref{AltVF}. As we discussed in the previous section, we expect that different contributions to the two-point amplitude with a given (large) $\ell$ will approximate in the continuum limit to the symmetric type Feynman diagrams. Then the aforementioned symmerty under exchange of Symanzik polynomials acquires a geometric meaning: Consider a large $\ell$ graph contributing to the two-point function. For example this can be a graph dual to a possible triangulation with $\ell$ triangles of a world-sheet in the case of $\Phi^3$ theory, as in Fig. \ref{fig3}. As we discussed in the previous section, first Symanzik polynomial $\mathcal{U}_F$ can be formulated as a sum over all possible curves between the external end-points of the diagram (which correspond to the two punctures of the worldsheet in the continuum limit) modulo the extra transverse cuts (that are shown by the orange curves in figure \ref{fig4}). Note that in the continuum limit this sum resembles all possible {\em open string} configurations that can be drawn on a sphere extending between the two punctures. Similarly, the second Symanzik polynomial can also be formulated as a sum, but this time, over all possible cuts that divide the discretization of the sphere into two parts each containing one external point (again modulo the extra transverse cuts). This then resembles a sum over all possible non-contractible {\em closed string} configurations that can be drawn on a sphere with two punctures. Therefore, it is tempting to propose that this symmetry exchanging the first and the second Symanzik polynomials, see Fig. \ref{fig3}, is a primitive, discrete version of the open-closed duality in string theory, see Fig. \ref{fig1}. Examining this relation and making it more precise seems to be of fundamental importance to uncover origins of holographic duality in general.

%%%%%%%%%%%%%%%%%%%%%%%%%%%%%%%%%%%
%%%%%%%%%%%%%%%%%%%%%%%%%%%%%%%%%%%
%%%%%%%%%%%%%%%%%%%%%%%%%%%%%%%%%%%
%%%%%%%%%%%%%%%%%%%%%%%%%%%%%%%%%%%
\section*{Acknowledgments}

We are indebted to Rajesh Gopakumar for very helpful correspondence and insightful comments on our manuscript and we thank Thomas Grimm for inspiring discussions. UG's work is supported by the Netherlands Organisation for Scientific Research (NWO) under the VICI grant VI.C.202.104.

%%%%%%%%%%%%%%%%%%%%%%%%%%%%%%%%%%%
%%%%%%%%%%%%%%%%%%%%%%%%%%%%%%%%%%%
%%%%%%%%%%%%%%%%%%%%%%%%%%%%%%%%%%%
%%%%%%%%%%%%%%%%%%%%%%%%%%%%%%%%%%%

\appendix

%%%%%%%%%%%%%%%%%%%%%%%%%%%%%%%%%%%%%%
%%%%%%%%%%%%%%%%%%%%%%%%%%%%%%%%%%%%%%
%%%%%%%%%%%%%%%%%%%%%%%%%%%%%%%%%%%%%%
\section{Rewriting field theory propagator as AdS propagator}\label{2ptDetails}
%%%%%%%%%%%%%%%%%%%%%%%%%%%%%%%%%%%%%%
%%%%%%%%%%%%%%%%%%%%%%%%%%%%%%%%%%%%%%
%%%%%%%%%%%%%%%%%%%%%%%%%%%%%%%%%%%%%%
We start with equation \eqref{2pftaub}:
%%%
\begin{align} 
\begin{split}
\Omega_F(k_1,k_2)&= \mathcal{V}_F\, \delta\left(k_1+k_2\right)  \int_0^{\infty}d \tau\,\tau^{-3-\nu}\, e^{- \tau k_1^2}\, ,
\end{split}
\end{align}
%%% 
and recall the definition $\nu = \Delta-d/2$.
We then rewrite this by using the following identity:
%%%
\be \label{a1a2id}
\int_0^{\infty} d\tau\, \tau^A e^{-\tau k^2} = \frac{\Gamma(2+2c)}{\Gamma(1+c)^2}\int_0^{\infty} d \alpha_1 d\alpha_2 \, (\alpha_1 \alpha_2)^c (\alpha_1+\alpha_2)^{A-1-2c}\, e^{-(\alpha_1+\alpha_2)k^2}\, ,
\ee
%%%
with $A=-3-\nu$ and $c\ge-1$\footnote{   
For the special case $c=-1$ one needs to take the limit $c=\epsilon-1$:
%%%
$$
\int dt\, t^a e^{-t k^2} = \lim_{\epsilon\rightarrow 0}\frac{\epsilon}{2} \int d \alpha_1 d\alpha_2 (\alpha_1 \alpha_2)^{\epsilon-1}(\alpha_1+\alpha_2)^{A+1-2\epsilon}e^{-(\alpha_1+\alpha_2)k^2}\, ,
$$
%%%
as the decoupled integral will yield a factor of $ 2/\epsilon$.}.
To prove this identity one performs two changes of variables $\alpha_1\rightarrow \alpha_1 \alpha_2$ and $\alpha_2\rightarrow \alpha_2/(\alpha_1+1)$ respectively. 
This allows us rewrite \eqref{2pftaub} as  
%%%
\be\label{2pfa1a2}
\Omega_F(k_1,k_2)= \mathcal{\bar V}_F\, \delta\left(k_1+k_2\right)  \int_0^{\infty} d\alpha_1 d\alpha_2\,(\alpha_1\alpha_2)^c (\alpha_1+\alpha_2)^{-\nu-4-2c}\, e^{-\alpha_1k_1^2-\alpha_2 k_2^2}\, ,
\ee
%%%
since $k_1^2=k_2^2$ and we defined $ \mathcal{\bar V}_F\equiv  \frac{\Gamma(2+2c)}{\Gamma(1+c)^2}\mathcal{V}_F $. Using the definition of the $\Gamma$ function\footnote{  Note that this is only true if $\nu +4+2c>0$. When this condition is not satisfied we can use the definition of the \textit{incomplete} gamma function instead, by substituting the lower limit of integration with $\epsilon$, which is effectively a UV cutoff.}
%%%
\be \label{schw1}
\Gamma(\nu + 4 + 2c)=(\alpha_1+\alpha_2)^{\nu +4+2c}\int_0^{\infty}dt\, t^{\nu + 3+2c}\, e^{-(\alpha_1+\alpha_2)t}\, ,
\ee
%%% 
we find
%%%
\begin{align}
\begin{split}
\Omega_F(k_1,k_2)&=\frac{  \mathcal{\bar V_F} \  \delta\left(k_1+k_2\right)}{\Gamma(\nu+4+2c)} \int_0^{\infty}dt\, d\alpha_1 d\alpha_2\,(\alpha_1 \alpha_2)^c \, t^{\nu+3+2c}e^{-(\alpha_1+\alpha_2)t}\, e^{-\alpha_1k_1^2-\alpha_2 k_2^2}\, .
\end{split}
\end{align}
%%%
%This can be rewritten in a nicer way using the following changes of variable {\color{red} Note that this is only possible if $m_r\neq 0$}:
%\be
%\alpha_{1,2} \rightarrow M(b)^{-2} \alpha_{1,2}, \qquad \qquad t\rightarrow M(b)^{2} t,
%\ee
%which gives
%\be
%G(k_1,k_2)= \mathcal{V}M(b)^{-2(1+N+d\ell/2)} \frac{\delta\left(k_1+k_2\right)}{\Gamma(2-I+d\ell/2)}  \int_0^{\infty}dt d\alpha_1 d\alpha_2\,t^{1-I-d\ell/2}e^{-(\alpha_1+\alpha_2)t}\, e^{- \alpha_1\left(  k_1^2 M(b)^{-2} +1 \right)-\alpha_2\left(  k_2^2M(b)^{-2} +1 \right)},
%\ee

After Fourier transforming to real space%(using the convention $\int dk\, e^{-ik x}$)
, the correlator becomes 
%%%
\begin{align}
\begin{split} \label{propaa}
\Omega_F(x_1,x_2)&= \bar v_F   \int d^dz\,dt \frac{d\alpha_1 d\alpha_2}{\alpha_1^{d/2-c}\alpha_2^{d/2-c}}\,t^{\nu+3+2c}e^{-\alpha_1t-\frac{(x_1-z)^2}{4\alpha_1}}e^{-\alpha_2t-\frac{(x_2-z)^2}{4\alpha_2}}
\end{split}
\end{align}
%%%
where we introduced $z$ as a Lagrange multiplier for the delta function, i.e.
%%%
\be
\delta(k_1+k_2)=\frac{1}{(2\pi)^d}\int d^d z \,e^{iz(k_1+k_2)}\, ,
\ee
%%%
and then we performed the Gaussian integrals. We also defined
%%%
\be
\bar v_F =2^{-d} \frac{\mathcal{\bar V}_F}{\Gamma(\nu+4+2c)}\, .
\ee
%%%
Now we make the change of variables $\alpha_{1,2}=\frac{1}{4\rho_{1,2}}$,
%%%
\be
\Omega_F(x_1,x_2)=  \bar v_F\,  4^{d-2c-2} \int d^dz\,dt\, d\rho_1 \, d\rho_2\,(\rho_1\rho_2)^{d/2-c-2}\, t^{\nu+3+2c}\, e^{-t\left(\frac{1}{4\rho_1}+\frac{1}{4\rho_1}\right)}e^{-\rho_1(x_1-z)^2}e^{-\rho_2(x_2-z)^2}\, ,
\ee
%%%
followed by a final change of variables $t = 4 z_0^2 \rho_1\rho_2$, which yields
%%%
\begin{align} 
\begin{split} 
G_F(x_1,x_2)
%&= \bar v \, 4^{d-2c-2} \, 4^{2-I+d\ell/2+2c} \int d^dz\,d t\,  t^{1-I+d\ell/2+2c}\, \prod_{i=1}^2 d\rho_i\, \rho_i^{d/2 -I+d\ell/2+c}\,e^{-\rho_i\left(\bar t+(x_i-z)^2\right)} \\
&= 2\bar v \, 4^{\Delta+2 +d/2} \int \frac{d^dz\,d  z_0}{z_0^{1+d}} \, z_0^{2\Delta+8+4c}  \prod_{i=1}^2 d\rho_i \, \rho_i^{\Delta +2+c}\, e^{-\rho_i\left( z_0^2+(x_i-z)^2\right)}\,  .
\end{split}
\end{align}
%%%  
We can now introduce the AdS bulk-to-boundary propagator (see also Appendix \ref{AdSProperties}),
%%%
\bea  \lab{AdSbtb}
K_{\Delta+2}(x; z_0, z) &=&\frac{ z_0^{\Delta+2}}{\pi^{d/2}\Gamma(\nu+2 )}\int_0^{\infty}d\rho\, \rho^{(\Delta+2)-1}\,e^{-\rho\left( z_0^2+(x_2-z)^2\right)}\nn \\
{}&& =\frac{\Gamma(\Delta+2)}{\pi^{d/2}\Gamma(\nu+2 )}\frac{z_0^{\Delta+2}}{\left(z_0^2+(x-z)^2\right)^{\Delta+2}}\, ,
\eea
%%%  
and, after setting  $c=-1+\epsilon$, we finally arrive at
%%%
\be  
\Omega_{F}(x_1,x_2)=\lim_{\epsilon\rightarrow 0} v _{F,\epsilon}  \int\frac{dz_0 d^dz}{z_0^{1+d}}\, z_0^{2\epsilon}\,K_{\Delta+2+\epsilon}(x_1; z_0, z) K_{\Delta+2+\epsilon}(x_2; z_0, z) \, ,
\ee
%%% 
with
%%%
\begin{align}
\begin{split}
v_{F,\epsilon}  &  \equiv 2\pi^d \Gamma(\nu+2+\epsilon)^2 4^{\Delta+2+d/2}\bar v_F  \\
&= 2\pi^d4^{\Delta+2}\frac{\Gamma(\nu+2+ \epsilon)^2 \Gamma(2\epsilon)}{\Gamma(\nu+2+ 2\epsilon) \Gamma(\epsilon)^2}\, \ell\, \int_0^\infty \left( \prod_{r=1}^I db_r\right)\, \delta\left(1-\mathcal{U}_F\right)\mathcal{A}_F^{\Delta+2-d/2}\, .
\end{split}
\end{align}
This quantity is defined for a specific Feynman diagram $F$ with $\ell$ loops. If we had two diagrams $F_1$ and $F_2$ both with $\ell$ loops, we would need to compute two different quantities, $v_{F_1,\epsilon} $ and $v_{F_2,\epsilon} $, since two different diagrams with the same number of loops can have different Symanzik polynomials, see Appendix \ref{SymanzikExamples} for simple examples.

%%%%%%%%%%%%%%%%%%%%%%%%%%%%%%%%%%%%%%
%%%%%%%%%%%%%%%%%%%%%%%%%%%%%%%%%%%%%%
%%%%%%%%%%%%%%%%%%%%%%%%%%%%%%%%%%%%%%
\section{Alternatyive expressions for $\mathcal{V}_F$} \label{AltVF}
%%%%%%%%%%%%%%%%%%%%%%%%%%%%%%%%%%%%%%
%%%%%%%%%%%%%%%%%%%%%%%%%%%%%%%%%%%%%%
%%%%%%%%%%%%%%%%%%%%%%%%%%%%%%%%%%%%%%

In Equation (\ref{Vn}) we gave an expression for the coefficient $\mathcal{V}_F$. Here we provide two more alternative expressions for this coefficient in the massless case and observe an interesting symmetry between these expressions. Start with the two-point amplitude in real space %{\color{purple} I am adding spaces after each measure to avoid clutter}
%%%
\be\lab{VF1}
\Omega_F(x-y) = \int d^dk\,  e^{-i k\cdot (x-y)} \int \prod_r da_r\, \mathcal{U}_F^{-d/2} e^{-\frac{\mathcal{A}_F}{\mathcal{U}_F}k^2}\, .
\ee
%%%
Carrying out the Gaussian momentum integrals we arrive at 
%%%
\be\lab{VF2}
\Omega_F(x-y) = \pi^{d/2} \int \prod_r da_r\,  \mathcal{A}_F^{-d/2} e^{-\frac{\mathcal{U}_F}{4\mathcal{A}_F}(x-y)^2} \, .
\ee
%%%
Rescaling $a_r\to a_r(x-y)^2$ yields 
%%%
\be\lab{VF3}
\Omega_F(x-y) = \pi^{d/2} \left((x-y)^2\right)^{-\Delta-2} \int \prod_r da_r\,  \mathcal{A}_F^{-d/2} e^{-\frac{\mathcal{U}_F}{4\mathcal{A}_F}} \, ,
\ee
%%%
where we used Equation (\ref{Delta2pt}). 

Now again start with (\ref{VF1}) but instead of carrying out the momentum integrals rescale $a_r\to a_r/k^2$. Then one finds 
%%%
\be\lab{VF4}
\Omega_F(x-y) = \int d^dk\,  \left(k^2\right)^{d\ell/2-I} e^{-i k\cdot (x-y)} \int \prod_r da_r\,  \mathcal{U}_F^{-d/2} e^{-\frac{\mathcal{A}_F}{\mathcal{U}_F}}\, .
\ee
%%%
Now use the definition of Gamma function to write 
$$  \left(k^2\right)^{d\ell/2-I} = \frac{1}{\Gamma(I-d\ell/2)}\int_0^\infty dt\,  t^{I-d\ell/2-1} e^{-k^2 t}\, $$
and to rewrite (\ref{VF4}) as 
%%%
\be\lab{VF5}
\Omega_F(x-y) = \frac{1}{\Gamma(I-d\ell/2)}  \int d^dk\int_0^\infty dt\, t^{I-d\ell/2-1} e^{-k^2 t -i k\cdot (x-y)} \int \prod_r da_r\, \mathcal{U}_F^{-d/2} e^{-\frac{\mathcal{A}_F}{\mathcal{U}_F}}\, .
\ee
%%%
Now carry out the Gaussian $k$ integrals and rescale $t\to t (x-y)^2$ to obtain
%%%
\be\lab{VF6}
\Omega_F(x-y) = \frac{\pi^{d/2}\left((x-y)^2\right)^{-\Delta-2}}{\Gamma(I-d\ell/2)} \int_0^\infty dt\, t^{I-d(\ell+1)/2-1} e^{-\frac{1}{4t}} \int \prod_r da_r\, \mathcal{U}_F^{-d/2} e^{-\frac{\mathcal{A}_F}{\mathcal{U}_F}}\, .
\ee
%%%
The integral over $t$ is generically divergent for $h\leq 2d/(d-2)$. However in the marginal case, that is the main interest in this paper, it is convergent and one obtains the formal result  
%\sout{After performing the $t$ integral one arrives at the expression}
%%%
\be\lab{VF7}
\Omega_F(x-y) =  \frac{\pi^{d/2}2^{d+2}\Gamma(d/2+1)\left((x-y)^2\right)^{-\Delta-2}}{\Gamma(-1)} \int \prod_r da_r\, \mathcal{U}_F^{-d/2} e^{-\frac{\mathcal{A}_F}{\mathcal{U}_F}}\, .
\ee
%%%
where the divergence from $\Gamma(-1)$ can be regulated by $d\to d-2\epsilon$. Ignoring the numerical coefficients, comparison of (\ref{VF3}) and (\ref{VF7}) yields an interesting symmetry under $\mathcal{A}_F\leftrightarrow \mathcal{U}_F$.  
 
%Finally, we remark that these expressions are only valid for $\ell\ge 1$ because we used identity \eqref{IdEll0}. The $\ell=0$ case was given in \eqref{G0}. 
Let us finally derive an alternative expression for the coefficient $\mathcal{V}_F$. We can rewrite \eqref{G20b} in real-space as
%%%
\be
\Omega_F(x,y)=\int d^dk e^{-ik\cdot(x-y)} \int_{0}^{\infty}\prod_{r=1}^{I}da_r\, \mathcal{U}_F^{-d/2}\, \exp\left(- \frac{\mathcal{A }_F}{\mathcal{U }_F} k^2\right) = 4^{\Delta+2} \pi^{d/2}\Gamma(\Delta+2) \frac{ \mathcal{V}_F }{|x-y|^{2\Delta+4}}\, ,
\ee
%%%
where 
%%%%
\be\lab{altVf} 
\mathcal{V}_F = \frac{4^{-\Delta-2}}{\Gamma(\Delta+2)}  \int_{0}^{\infty}\left(\prod_{r=1}^{I}da_r\right)\, \mathcal{A}_F^{-d/2}\exp\left(-\frac{\mathcal{U}_F}{4\mathcal{A}_F}\right)\,.
\ee
%%%%
We note that this expression for the two-point amplitude is also valid when $\ell=0$.

%%%%%%%%%%%%%%%%%%%%%%%%%%%%%%%%%%%%%%
%%%%%%%%%%%%%%%%%%%%%%%%%%%%%%%%%%%%%%
%%%%%%%%%%%%%%%%%%%%%%%%%%%%%%%%%%%%%%
\section{Some properties of AdS propagator} \label{AdSProperties}
%%%%%%%%%%%%%%%%%%%%%%%%%%%%%%%%%%%%%%
%%%%%%%%%%%%%%%%%%%%%%%%%%%%%%%%%%%%%%
%%%%%%%%%%%%%%%%%%%%%%%%%%%%%%%%%%%%%%

We define the bulk-to-bulk propagator  in Euclidean AdS via (see for example \cite{DHoker:2002nbb})
%%%
\be \label{AdSPropEquation}
D_X^{(\Delta)} \mathcal{G}(X,Y) \equiv (-\nabla^2_X +m^2_\Delta)\mathcal{G}(X,Y)= \frac{1}{\sqrt{g}}\delta^{d+1}(X-Y)\, ,
\ee
%%%
where $X=(x_0,x)$ and $Y=(y_0,y)$ are points in AdS with radial components $x_0$ and $y_0$.  Note that
%%%
\be
\nabla^2_X =x_0^2\left(\partial_{x_0}^2 -\frac{d-1}{x_0} \partial_{x_0} +\partial_x^2\right)\,.
\ee
%%%
The solution to \eqref{AdSPropEquation} can be found as
%%%
\be
\mathcal{G}(X,Y)=\frac{\Gamma(\Delta)}{2\nu\,\pi^{d/2} \Gamma(\nu)} \left(\tfrac{\xi}{2}\right)^{\Delta} F\left(\tfrac{\Delta}{2},\tfrac{\Delta}{2}+\tfrac{1}{2};\nu+1,\xi^2\right) \, ,
%\equiv  \frac{2C_{\Delta}}{\nu } g_{\Delta}(\xi)
\ee
%%%
with
%%%
\be
\xi \equiv \frac{2x_0y_0}{x_0^2+y_0{}^2+(x-y)^2}, \qquad\quad \nu\equiv\sqrt{\tfrac{d^2}{4}+m^2_{\Delta} L^2} \equiv\Delta-\tfrac{d}{2}\, , \qquad\quad  C_{\Delta}  \equiv  \frac{\Gamma(\Delta)}{\pi^{d/2} \Gamma(\nu)} \, .
\ee
%%%
This is related to the bulk-to-boundary propagator via
%%%
\be \label{BbtoBB} 
K_{\Delta}(x_0,x;y) = \lim_{y_0\rightarrow 0} \frac{2\nu}{ y_0^{\Delta}} \mathcal{G}(x_0,x;y_0,y) =C_{\Delta} \frac{x_0^\Delta}{\left[x_0^2+(x-y)^2\right]^\Delta}\, .
\ee
%%%
Similarly we can define the boundary-to-boundary propagator as
\be \label{bbDef}
\beta_{\Delta}(x,y) \equiv\lim_{(x_0,y_0)\rightarrow (0,0)}\frac{2\nu}{(x_0y_0)^{\Delta}} \mathcal{G}(x_0,x;y_0,y)= \lim_{x_0\rightarrow 0} \frac{1}{x_0^{\Delta}} K(x_0,x;y) = C_{\Delta} (x-y)^{-2\Delta}\,. 
\ee

 Using \eqref{BbtoBB} we can then rewrite \eqref{2pfell} as 
%%%
\begin{align}
\begin{split} \label{GG}
\Omega_F(x,y)& = \lim_{\epsilon\rightarrow 0} v_{F,\epsilon}  \lim_{(x_0,y_0)\rightarrow (0,0)}\int_{AdS} z_0^{2\epsilon}  \frac{(2(\nu_{\epsilon}+2))^2}{(x_0 y_0)^{\Delta+2}} \mathcal{G}_{\Delta+2+\epsilon }(x_0,x;z_0,z)\mathcal{G}_{\Delta+2+\epsilon }(z_0,z;y_0,y)\\
&=  \lim_{\epsilon\rightarrow 0} 4(\nu_{\epsilon}+2)^2\,v_{F,\epsilon} \lim_{(x_0,y_0)\rightarrow (0,0)} (x_0y_0)^{-\Delta-2}\int_{AdS} z_0^{2\epsilon}\,   \mathcal{G}_{\Delta+2 +\epsilon}(X,Z)\mathcal{G}_{\Delta+2+\epsilon }(Z,Y)\, ,
\end{split}
\end{align}
%%%
where we set $\nu_{\epsilon} \equiv  \Delta+\epsilon-d/2$. 

At this point we would like to use an identity to rewrite the integral of two bulk-to-bulk propagators as a single bulk-to-bulk propagator.  However, we already know that the final result of this calculation is \eqref{GG2}, which means that the identity we are looking for is 
%%%
\be \label{2BBIdentityFin}
 \int_{Z\in \text{AdS}}z_0^{2\epsilon}\, \mathcal G_{\Delta+2+\epsilon} (X ,Z)\mathcal  G_{\Delta+2+\epsilon} ( Z,Y) \sim  \frac{\mathcal{G}_{\Delta+2}(X,Y)}{2(\nu+2) \epsilon}\, ,
\ee
%%%
where the $"\sim"$ means that this identity is only valid for small $\epsilon, x_0$, and $y_0$. Using this in \eqref{GG} we find
%%%
\begin{align}
\begin{split}
\Omega_F(x,y) &=  \lim_{\epsilon\rightarrow 0} 4(\nu_{\epsilon}+2)^2\,v_{F,\epsilon} \lim_{(x_0,y_0)\rightarrow (0,0)} (x_0y_0)^{-\Delta-2}\frac{\mathcal{G}_{\Delta+2}(X,Y)}{2(\nu_{\epsilon}+2) \epsilon} \, ,
\end{split}
\end{align}
%%%
which indeed reduces to the expression we showed in the main text, \eqref{GG2}, after substituting the definition \eqref{velleps}.

%%%%%%%%%%%%%%%%%%%%%%%%%%%%%%%%%%%%%%
%%%%%%%%%%%%%%%%%%%%%%%%%%%%%%%%%%%%%%
%%%%%%%%%%%%%%%%%%%%%%%%%%%%%%%%%%%%%%
\section{Identities relating number of loops, lines and vertices}\label{AppEul}
%%%%%%%%%%%%%%%%%%%%%%%%%%%%%%%%%%%%%%
%%%%%%%%%%%%%%%%%%%%%%%%%%%%%%%%%%%%%%
%%%%%%%%%%%%%%%%%%%%%%%%%%%%%%%%%%%%%%

The following relation holds in a connected graph with $\ell$ number of independent loop momenta, $V$ vertices and $I$ internal lines:
%%%
\be\lab{momcons} 
\ell = I - V + 1\, .
\ee 
%%%
This is easy to obtain, as there are $I$ total momentum running in the graph subject to momentum conservation at $V$ vertices and one such constraint is not independent due to overall momentum conservation. 

There is an additional equation that relates $V$ and $I$. To derive this consider a $n_e$-point function of the canonical field $\Phi$. Start from a fully connected  diagram with $n_e$ external lines. The number of vertices and internal lines for this diagram are $V_0=n_e$ and $I_0=n_e(h-1)/2$ where $h$ is the coordination number of the vertex\footnote{We assume a single type of vertex for simplicity. The analysis can be generalized.}. Each time one attaches a new internal vertex to this diagram one has $h-2$ free lines to connect somewhere. The minimum number of new vertices $\Delta V$ in general will be more than 1 if we want $n_e$ to stay the same. Assuming no tadpoles (which indeed can be removed by a renormalization condition), then after connecting ends of these lines to other parts of the diagram, one finds that the number of internal lines increase by $\Delta I = \Delta V + \Delta V(h-2)/2$. The first term here arises from the fact that an internal vertex divides the original line the vertex is introduced to, into two lines. The second arises from connecting the rest of the lines emanating from the new internal vertices either with each other or with the rest of the diagram. This also determines the increase in the number of independent loop momenta as $\Delta \ell = \Delta V (h-2)/2$. Note that this is consistent with (\ref{momcons}) as $\Delta \ell = \Delta I - \Delta V$. Writing $I = I_0 + \Delta I$ and $V = V_0 + \Delta V$ and rearranging, we find 
%%%
\be\lab{Nn} 
I = \frac{V h}{2} - \frac{n_e}{2} \, .
\ee 
%%%
Now  we use (\ref{momcons}) to express $\ell$ in terms of $V$ 
%%%
\be\lab{elln} 
\ell = V\frac{h-2}{2} + 1 -\frac{n_e}{2} \, .
\ee 
%%%
This shows that the Feynman diagrams contributing to a $n_e$-point function can be labelled by $V$ only. Also note that we can combine \eqref{Nn} and \eqref{elln} to find a relation between the number of loop momenta and number of internal lines:
%%%
\be \label{Iell}
\ell = I\frac{h-2}{h} + 1 -\frac{n_e}{h}\, . 
\ee
%%%
A word of caution when using these formulas for the free propagators. A free two-point function should be thought of as two external lines connected to the ends of an internal line. Hence the number of internal lines should be counted as $I=1$, the number of vertices $V=2$ and the coordination number is $h=2$ consistent with (\ref{Nn}) and (\ref{elln})\footnote{One might think that the derivation relies on the initial graph with $V_0=n_e$ hence the formulas above only apply to certain type of graphs but this is not so. One can derive the same formulas e.g. starting with an initial tree graph with  $\ell_0=0$, $V_0=(n_e-2)/(h-2)$ and $I_0 = V_0-1$.}. Note also that nowhere in these derivations we used the number of genii $g$. Addition of vertices in principle can increase the number of genii but the formulas still apply. What depends on the number of genii is the number of {\em faces} of a graph that is related to $\ell$ as $f = \ell +1 - 2g$. Indeed, depending on how one connects the internal lines one may change the genus of the graph, hence $\Delta f$ need not be the same as $\Delta \ell$.  

This calculation can easily be generalized to $n$-point functions of composite operators e.g. $\langle \tr\Phi^{J_1}(x_1)\cdots \tr\Phi^{J_n}(x_n)\rangle$ as follows. The basic diagram we start with, see for instance the figure below, has $V_0=n_e$ vertices and $I_0=\sum_i J_i /2$ internal lines. The number of independent loop momenta is determined by (\ref{momcons}) as  $\ell_0 = \sum_i J_i/2 - n_e +1$. When we add $V-n_e$ internal vertices (so that $V$ is the total number of vertices of the diagram, including the $n_e$ 'external' vertices), we also have to add $(V-n_e)h/2$ internal lines. Then
%%%%
\be \label{IVJ2}
I = \frac{(V-n_e)h}{2} +\frac{\sum_i J_i}{2}\, .
\ee
%%%%
Using (\ref{momcons}) we also determine 
%%%%
\be\label{ellVJ2}
\ell = \frac{V(h-2)}{2}+\frac12 \sum_i J_i - \frac{n_e h}{2} +1 \,.
\ee
%%%%

These identities reduce (formally) to the case of non-composite operators when $J=h-1$ and $b=1$. They also apply to the tree level contribution with only external vertices when $h$ is set to 0.

Let's check for example the case $n_e=3, J=4$ with the following diagram:
\begin{center}
\begin{tikzpicture}
  \begin{feynman}
    \vertex (a) ;
    \node [right=of a, crossed dot] (b);
    \vertex [right=of b] (c);
    \vertex [right=of b] (x);
    \node [right=of x, crossed dot] (c);
    \vertex [right=of c] (d);
    \node [above=of x, crossed dot] (e);
    \vertex [above=of e] (f);

    \diagram* {
     (a) --  [boson, edge label = $k_1$] (b) --  [quarter right](c) --  [boson, edge label = $k_2$](d) ,
      (b) -- (c) -- [quarter right](e) --  [boson, edge label = $k_3$](f),
      (b) -- [quarter left](e),
      (b) -- (e) -- (c)
          };
  \end{feynman}
\end{tikzpicture}
\end{center}
The number of internal lines and loops are correctly given by $I=6$ and $\ell= 4$ using the formula above with $h=0$. 

%%%%%%%%%%%%%%%%%%%%%%%%%%%%%%%%%%%%%%
%%%%%%%%%%%%%%%%%%%%%%%%%%%%%%%%%%%%%%
%%%%%%%%%%%%%%%%%%%%%%%%%%%%%%%%%%%%%%
\section{Renormalization of massless $\Phi^4$} \label{renorm4}
%%%%%%%%%%%%%%%%%%%%%%%%%%%%%%%%%%%%%%
%%%%%%%%%%%%%%%%%%%%%%%%%%%%%%%%%%%%%%
%%%%%%%%%%%%%%%%%%%%%%%%%%%%%%%%%%%%%%

We consider $\Phi^4$ theory, with vanishing mass for simplicity of discussion and demonstrate renormalization of the two-point function in the Schwinger representation up to two-loop level. At this order, there are only two diagrams contributing, i.e. the free propagator and the sunset diagram:
%%%%
\be
(1+\delta_Z)k^2 + \Omega_2(k) = (1+\delta_Z)k^2 + 2(\lambda+\delta_\lambda)^2 \int db_1db_2db_3 \delta(1-\mathcal{U}) \Gamma(\epsilon-1) \left[(1+\delta_Z)k^2\right]^{1-\epsilon}\mathcal{A}^{1-\epsilon}\,.
\ee  
%%%%
Generally we can write $\delta_Z = \sum_{i=1}^{\infty} \lambda^{i}\delta_Z^{(i)}$. Expanding around $\lambda\sim 0, \epsilon \sim 0$ and keeping only the terms of order $\lambda^2$ we find (notice that $\delta_\lambda$ drops out)
%%%%
\be
\lambda^2 k^2 \left[\delta_Z^{(2)} - \frac{\mathcal{I}_1}{\epsilon} +\mathcal{I}_1 \log k^2  +   \mathcal{I}_1(\gamma-1)+\mathcal{I}_2\right]
\ee
%%%%
with
%%%%
\be
\mathcal{I}_1 \equiv 2 \int db_1db_2db_3 \delta(1-\mathcal{U}) \mathcal{A}=1/2 \, , \qquad \qquad \mathcal{I}_2 \equiv 2\int db_1db_2db_3 \delta(1-\mathcal{U}) \mathcal{A}\log{\mathcal{A}} = - 9/8\,.
\ee
%%%%
Then it is easy to see that we can set 
%%%%
\be
\delta_Z = \mathcal{I}_1\left(\frac{1}{\epsilon} \lambda^2 - \log \mu^2\right) + O(\lambda^3)
\ee
%%%%
 to obtain a finite expression for the two-point function up to two loops:
%%%%
\be
\lambda^2 k^2 \left[  \frac{1}{2} \log\frac{ k^2}{\mu^2}  +   \frac{1}{2}(\gamma-1)-\frac{9}{8}\right]\,.
\ee
%%%%
This procedure can of course be generalized to higher orders, although it is definitely more cumbersome than the usual method of adding new vertices to the theory.

%%%%%%%%%%%%%%%%%%%%%%%%%%%%%%%%%%%%%%
%%%%%%%%%%%%%%%%%%%%%%%%%%%%%%%%%%%%%%
%%%%%%%%%%%%%%%%%%%%%%%%%%%%%%%%%%%%%%
\section{Counting the number of Feynman diagrams} \label{FeynmanCounting}
%%%%%%%%%%%%%%%%%%%%%%%%%%%%%%%%%%%%%%
%%%%%%%%%%%%%%%%%%%%%%%%%%%%%%%%%%%%%%
%%%%%%%%%%%%%%%%%%%%%%%%%%%%%%%%%%%%%%

%%%%%%%%%%%%%%%%
\begin{figure}[t]
\centering
\includegraphics[width=0.7\textwidth]{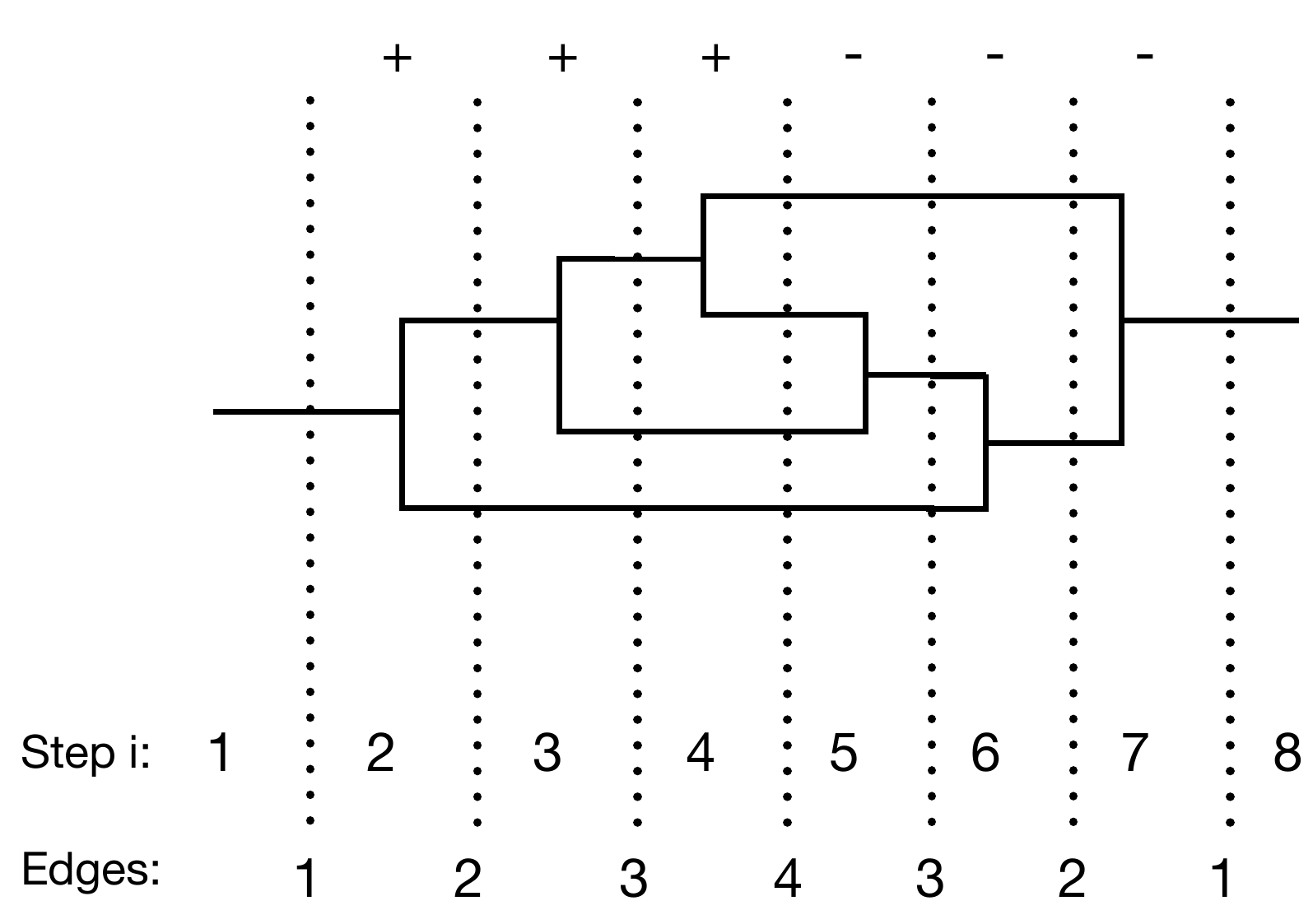}%
\caption{Example of a 3-loop planar graph with $\Phi^3$ interaction. Construction is from left to right with a single vertex appearing in each step. The signs above denote the type of vertices, bifurcation (+) and recombination (-). We also show the number of edges in the graph crossed by the dashed lines at the end of each step.}
\label{fig3loop}
\end{figure}
\begin{figure}[t]
\centering
\includegraphics[width=0.9\textwidth]{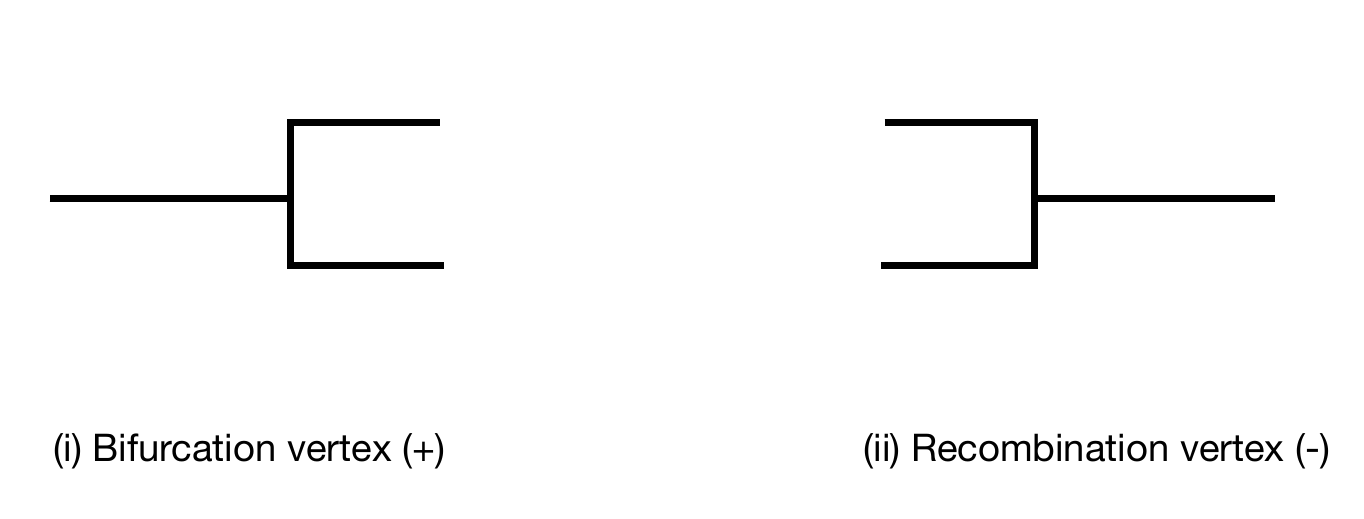}
\caption{The two type of vertices that appear in $\Phi^3$ theory in a left-to-right construction.}
\label{figver}
\end{figure}

%%%%%%%%%%%%%%%%
In this appendix we devise a method to count the number of planar $\Phi^3$ Feynman diagrams with $\ell$ loops that contribute to the scalar two-point function. For simplicity we focus only on the planar graphs but the arguments below can be generalized to higher genus. Our main idea is to construct the diagram with $V$ vertices sequentially with $V$ steps drawn from left to right such that at every step one and only one vertex appears. Number of vertices are fixed in terms of the number of loops, coordination number of the vertex and genus. For planar $\Phi^3$ graphs we have  $V = 2\ell$ See Fig. \ref{fig3loop} for an example of a 3-loop diagram. In this left-to-right sequential construction there exist two type of vertices, a bifurcation and a recombination which we denote by + and - type respectively, see Fig. \ref{figver}. 
%%%%%%%%%%%%%%%%
%%%%%%%%%%%%%%%%
Now, in order to create an $\ell$-loop diagram one needs to use $\ell$ recombination type vertices. Using the relation between the number of vertices and loops $V= 2\ell$ above, this means we need to construct arrays of $2\ell$  + and - signs with exactly $\ell$ + and $\ell$ -. There are a few rules. First, in this left-to-right construction, clearly, the first vertex should be of + type and the last one - type. Then, we need to insert $(\ell-1)$ + and $(\ell-1)$ - vertices in between. Also clearly, when read from left to right the number of - vertices should not exceed the number of + vertices. An iterative way of constructing the set of arrays would then be starting with the ``fat graph'' array $++\cdots + -- \cdots -$ with $\ell$ + and - each and move the +'s towards the right one by one. This problem can be mapped to the spectrum of free fermions where the ``fat graph'' array corresponding to the ground state with $\ell$ fermions. However we still need to respect the rule that the number of - vertices (empty fermion sites) should not exceed the + vertices (filled fermion sites) as read from bottom up in energy. This can be imposed on the free fermion spectrum as an external restriction and one can indeed compute the ``partition function of the Feynman diagrams'' in this manner. 

We will however, use a simpler trick, and describe the +/- arrays in terms of creation annihilation operators for the number of edges $n(i)$ in a given step $i$, see Fig. \ref{fig3loop} as an example. Introduce a state with $n$ edges by $|n\rangle$. A bifurcation vertex acts on this as a creation operator that we denote by $a_+$. Similarly a recombination vertex acts as an annihilation operator, $a_+ |n\rangle \propto  |n+1\rangle$,  $a_- |n\rangle \propto  |n-1\rangle$. We will fix the proportionality constants in these expressions by counting the number of all possible edges the vertices can be attached to. Clearly a bifurcation in step i can be placed on $n(i)$ edges and a recombination in a planar graph can be attached to $n(i)$ pairs of adjacent edges. Therefore we introduce the following operations 
%%%
\be\lab{a+a-}
a_+ |n \rangle = n  |n+1 \rangle\, \qquad a_- |n \rangle = (n-1)  |n-1 \rangle\, \qquad a_- |1\rangle = 0\, .
\ee
%%%
It is easy to check that these operators satisfy $[a_+,a_-]=-1$. This defines the ``algebra of vertices'' on this hypothetical Hilbert space of Feynman diagrams. 

We now arrive at our final result. Denote the set of all permutations of $(\ell-1)$ + and $(\ell-1)$ - signs by $S_\Pi$ and let $\Pi(n) \in {+, -} $ be the nth element in a particular permutation $\Pi\in S_\Pi$. Then the total number of Feynman diagrams with $\ell$ loops is given by 
%%%%
\be\lab{numberGraphs}
N(\ell) \approx \langle 1 | a_- \left[\sum_{\Pi \in S_\Pi} a_{\Pi(1)} \cdots a_{\Pi(2\ell-2)} \right]a_+ |1\rangle\, .
\ee
%%%%
Note that the algebra of annihilation/creation operators assures that there is never more - type vertices than the + types when a graph is read from left to right or when the expression above is read from right to left.
This counting is imprecise for two reasons: (1) it ignores special graphs where our counting should be corrected by accounting for the symmetry factors. For instance a sub array of the form $a_-a_+$ when bifurcating and recombining the same edge then there is an additional factor of 1/2 from symmetry. For large values of $\ell$ these symmetric graphs will be measure 0 and we can safely ignore associated symmetry factors in the generic counting. (2) There may be an overcounting due to the fact that one can obtain the same diagram in different ways in this construction by acting on with $a_+$ on different legs. For example the same graph shown in (\ref{fig3loop}) obtained by the construction there can also be obtained by acting with $a_+$ on the lower legs. But this happens also because there is a symmetry factor of 2 in this graph. Hence, also this miscounting can be safely ignored when we consider a contribution with large $\ell$. We can therefore consider (\ref{numberGraphs}) as exact for a generic graph with large enough $\ell$. 

Let's work out the example of $\ell=2$. In this case the only series of creation/annihilation operators allowed are $a_-a_-a_+a_+$ and $a_-a_+a_-a_+$. The first one gives a multiplicity factor 4 whereas the second one gives 1 with a total of 5 diagrams. The correct number of distinct diagrams is 4 on the other hand. This is because the same diagram is obtained in the first array by acting the middle $a_-a_+$ on two separate legs. 

%%%%%%%%%%%%%%%%%%%%%%%%%%%%%%%%%%%%%%
%%%%%%%%%%%%%%%%%%%%%%%%%%%%%%%%%%%%%%
%%%%%%%%%%%%%%%%%%%%%%%%%%%%%%%%%%%%%%
\section{Full expression for the two point function in $\Phi^3$ theory} \label{Full2pf}
%%%%%%%%%%%%%%%%%%%%%%%%%%%%%%%%%%%%%%
%%%%%%%%%%%%%%%%%%%%%%%%%%%%%%%%%%%%%%
%%%%%%%%%%%%%%%%%%%%%%%%%%%%%%%%%%%%%%

Interestingly, the same method that we devised above to count the number of diagrams can be generalized to obtain the full expression for the loop-$\ell$ contribution to the two-point function. For this we need to assign Schwinger parameters $\alpha_i$ and momentum $k_i$ on the edges and define the creation/annihilation operators to reproduce momentum conservation at the junctions. Consider step i as above with $n_i$ edges. This defines a state $| j_1, j_2\cdots, j_{n(i)}\rangle$ with the Schwinger parameters 
and leg momenta $\alpha_{j_m}$ and $k_{j_m}$ with $m=1,2\cdots, n(i)$. Denote by $J$ the maximum value of $j_m$ in the set $\{ j_1, j_2\cdots j_{n(i)}\}$. Then define the bifurcation and recombination operators acting on the state as 
%%%
\bea\lab{a+2}
a_+| j_1, j_2,\cdots j_n\rangle &=& \delta(k_{j_1} - k_{J+1}-k_{J+2})   |J+1, J+2, j_2,\cdots j_n  \rangle + \cdots  \\ \nn 
{}&& \cdots +\delta(k_{j_n} - k_{J+1}-k_{J+2})   |j_{1}, \cdots j_{n-1}, J, J+1 \rangle 
\eea
%%%%
\bea\lab{a-2}
a_-| j_1, j_2,\cdots j_n\rangle &=& \delta(k_{j_1} + k_{j+2}-k_{J+1})   |J+1, j_3, \cdots j_n  \rangle + \cdots \\ \nn
{}&&  \cdots +\delta(k_{j_{n-1}} + k_{j_{n}} - k_{J+1})   |j_{1}, \cdots j_{n-2}, J+1 \rangle \, .
\eea
%%%
The series of creation/annihilation operators acting on the state $|1\rangle$ in the end produces a state $|I+2\rangle$ where $I = 3\ell -1$. One should then impose the normalization $\langle I +2 | 1\rangle =1$. 
Then the loop $\ell$, genus-0 contribution to the two-point function is given by 
%%%%
\be\lab{ell2pf}
\Omega_\ell(k_1,k_{I+2}) \approx  \prod_{r=2}^{I+1} \int_{-\infty}^{\infty}d^dk_r  \int_{0}^\infty d\alpha_r e^{-\alpha_r (m^2 + k_r^2)}\langle 1 | a_- \left[\sum_{\Pi \in S_\Pi} a_{\Pi(1)} \cdots a_{\Pi(2\ell-2)} \right]a_+ |1\rangle\, ,
\ee
%%%%
 One can easily check that this is proportional to $\delta(k_1-k_{I+2})$. The full answer for the two-point function is then given by the sum over $\ell$. Note that the expectation value of the series of $a_\pm$ above is a sum over product of $2\ell$ delta functions of momenta $k_i$. Integral over the internal momenta can be carried out as described in the text and gives the same dependence on external momentum $k_1$ for all the terms in the expectation value with  $\alpha$ dependence is factored out completely. What the expression above provides is a systematic construction of all Feynman diagrams contributing at loop $\ell$. That is, each specific product of delta functions corresponds to and can be used to label a graph $F$ with independent loop momenta $\ell$. 
 
This construction can be generalized to the two-point function of an arbitrary theory $\Phi^{h}$ where the definition of $a_\pm$ should be modified such that $a_+$ produces $h-1$ adjacent legs from a single one and  $a_-$ joins $h-1$ adjacent legs into one. It is also possible to generalize to $n$-point functions by taking the difference of number of $a_+$ and $a_-$ to be $n-2$. Finally, the same construction can be applied to contributions with arbitrary genus by allowing $a_-$ act on non-adjacent legs.

 %%%%%%%%%%%%%%%%%%%%%%%%%%%%%%%%%%%%%%
%%%%%%%%%%%%%%%%%%%%%%%%%%%%%%%%%%%%%%
%%%%%%%%%%%%%%%%%%%%%%%%%%%%%%%%%%%%%%
\section{Basic properties of Symanzik polynomials}
\lab{sec::Symanzik} 
%%%%%%%%%%%%%%%%%%%%%%%%%%%%%%%%%%%%%%
%%%%%%%%%%%%%%%%%%%%%%%%%%%%%%%%%%%%%%
%%%%%%%%%%%%%%%%%%%%%%%%%%%%%%%%%%%%%%
We review some of the useful properties of the Symanzik  polynomials that appear in the $n$-point amplitudes. For a more extensive review, see for example \cite{Bogner:2010kv}. We have used two equivalent expressions for the $n$-point amplitudes in throughout the paper,  the Schwinger representation (\ref{npt}) and the Chrisholm (or Feynman) representation (\ref{Chrisholm}). Let us first show how the two can be obtained from one another. We
demonstrate this in the case of the two-point amplitude. Start from the Schwinger representation, which we reproduce here,  
%%%
\be\lab{Schw1} 
\Omega_F(k_1,k_2) = \delta^d(k_1+k_2) \int_0^\infty \prod_{r=1}^I da_r \, \mathcal{U}_F^{-d/2} e^{-\frac{\mathcal{A}_F}{\mathcal{U}_F}k_1^2 + m^2 a_r} \, ,
\ee
%%%
 and introduce the delta-function $ 1 = \int_0^\infty d\tau \delta(\tau - \sum_r a_r)$ inside the integral. Then rescale $a_r = \tau b_r$ and perform the $\tau$ integral. One immediately obtains 
%%%
\be\lab{Chrs1} 
\Omega_F(k_1,k_2) = \delta^d(k_1+k_2) \Gamma(-\Delta-2+d/2)\int_0^\infty \prod_{r=1}^I db_r \, \delta(1-\sum_r b_r) \frac{\mathcal{U}_F^{-\Delta-2}}{\left( \mathcal{A}_F k_1^2 + m^2 \mathcal{U}_F\right)^{\frac{d}{2}-\Delta-2} }\, ,
\ee
%%%
where $I-d\ell/2 = d/2 - \Delta - 2$.  

There are various different expressions for the Symanzik polynomials. First of all they are given in terms of trees and two-trees as follows 
%%%
\be
 \mathcal{U}(a)=\sum_{T \in \mathcal{T}_1}\prod^{\ell}_{r\not\in T_1} a_r,\qquad\qquad\qquad \mathcal{A}(a,k)= \sum_{T_2\in \mathcal{T}_2}\prod^{\ell+1}_{r \not \in T_2} a_r k^T
\ee
%%%
where $\mathcal{T}_1$ is the set of trees of a graph $G$, $\mathcal{T}_2$ is the set of 2-trees (or 2-forests) of $G$ and  $k^T$ is the sum of the external momenta flowing into one of the two disconnected components of $T_2\in \mathcal{T}_2$. For clarity, note that the set of trees of $G$  has the form $\mathcal{T}_1=\left(T_1,T_2,\dots\right)$, so that the dummy variable $T$ can be $T=T_1,T_2,\dots$; on the other hand the set of 2-forests of $G$ is $\mathcal{T}_2=\left((T^{(1)}_1,T^{(2)}_1), (T^{(1)}_2,T^{(2)}_2),\dots\right)$ so that the dummy variable $T_2$ is now a vector with two components. From these definitions one immediately observes the following useful properties:
\begin{itemize}
\item They are homogeneous functions of the Feynman parameters $a_r$, $\mathcal{U}$ has degree $\ell$ and $\mathcal{A}$ has degree $\ell+1$.
\item They are linear in each Feynman parameter.
\item Each monomial of $\mathcal{U}$ has coefficient $+1$.
\end{itemize}

One can represent the Symanzik polynomials also using the structure matrices of the Feynman diagram which we define now. Denote the momenta on each edge $r$ of the graph as $q_r$, loop momenta on loop $m$ as $p_m$ and the external momenta as $k_i$. $r$ runs from 1 to $I$, $m$ from 1 to $\ell$ and for an $n$-point function $i$ runs from 1 to $n$. Now the structure matrices $\lambda$ and $\sigma$ are defined as follows 
%%%
\be\lab{structure}
q_r = \sum_{m=1}^\ell \lambda_{rm}\, p_m + \sum_{i=1}^n \sigma_{ri} \,k_i\, \qquad \lambda_{rm} ,\, \sigma_{ri} \in \{-1,0,1\}\, .
\ee
%%%
Now define the following matrices 
%%%
\be\lab{MQJ} 
\sum_{r=1}^I a_r (q_r^2 +m^2) = \sum_{m,n=1}^\ell p_m M_{mn} p_n + \sum_{m=1}^\ell 2 p_m \cdot Q_m + J\, ,  
\ee
%%%
where each $Q_m$ are d-vectors given by combinations of the external momenta $k_i$. One can express $M(a)$, $Q(a)$ and $J(a)$ in terms of the structure matrices $\lambda$ and $\sigma$: 
%%%%
\be\lab{MQJ2} 
M_{mn} = \sum_{r=1}^I a_r \lambda_{rm}\lambda_{rn}\,, \qquad Q_m = \sum_{r=1}^I \sum_{i=1}^n a_r \lambda_{rm}\sigma_{ri} k_i\, ,  \qquad J = \sum_{r=1}^I \sum_{i,j=1}^n a_r\left( \sigma_{ri}\sigma_{rj} k_i\cdot k_j + m^2\right) \, .
\ee
%%%%
Now we can write down another representation of the Symanzik polynomials are now given in terms of $M$, $Q$ and $J$ as
%%%
\be\lab{Sym2} 
\mathcal{U}_F(a) = \det M\, \qquad  \mathcal{A}(a,k_i)+ m^2 \mathcal{U}_F(a) = \det M \left(J  + Q M^{-1} Q \right)\, ,
\ee 
%%%
where $\mathcal{A}$ is defined in (\ref{Pdef}). Specifying to the two-point function it is easy to see, using $k_2^2 = k_1^2$ and $k_1\cdot k_2 = -k_1^2$ that $QM^{-1}Q + J$ can be written in this case as
%%%%
\be\lab{qmqj} 
QM^{-1}Q + J = -k_1^2 \sum_{i,j=1}^2 (-1)^{\delta_{ij}} \left(\bar \sigma_i \cdot M^{-1} \cdot \bar\sigma_j + \sum_r \sigma_{ri}\sigma_{rj} a_r \right) + m^2\, ,
\ee
%%%%
where we defined 
%%%
\be\lab{sdef}
(\bar \sigma_i)_m \equiv \sum_r \lambda_{rm} \sigma_{ri} a_r\, .
\ee
%%%
We provide yet another representation of the $n$-point amplitude in terms of creation/annihilation operators in (\ref{ell2pf}) for $n=2$.

Finally we define the Kirchoff polynomial of a graph:
\be
\mathcal{K}(a_1,\dots a_N) = \sum_{T\in \mathcal{T}_1} \prod_{r\in T}^{\ell} a_r
\ee
This is very similar to the definition of the first Symanzik polynomial $\mathcal{U}$, but now the product is over the Feynman parameters \textit{contained} in the tree $T$. Also note that $\mathcal{K}$ is an homogeneous function of the $a$'s of order $(I-\ell)$, while $\mathcal{U}$ is of order $\ell$. This function is related to $\mathcal{U}$ via
\be
\mathcal{U}(a_1,\dots a_N) = a_1\dots a_N\, \mathcal{K}(a_1^{-1}, \dots a_N^{-1}), \qquad \mathcal{K}(a_1,\dots a_N) = a_1\dots a_N\, \mathcal{U}(a_1^{-1}, \dots a_N^{-1})
\ee
Now define the Laplacian of a graph:
\be
L_{ij} = \begin{cases}
\sum_k {a_k} & \text{if $i=j$ and edge $e_k$ is attached to $V_i$ and is not a self-loop}\\
-\sum_k a_k & \text{if $i\neq j$ and edge $e_k$ conects $V_i$ and $V_j$}
\end{cases}
\ee
If two vertices are connected by e.g. two edges $e_x$ and $e_y$ then the Laplacian only depends on the sum $a_x+ a_y$. If an edge is a self-loop (i.e. a tadpole) then it does not contribute to the Laplacian.
Using these definitions we can now state the \textit{matrix-tree theorem}:
\be
\mathcal{K} = \det L[i]
\ee
where $L[i]$ is the minor of the Laplacian obtained by removing row $i$ and column $i$, where $1\leq i \leq \ell$ is arbitrary. We can also count the number of trees of a given graph, by setting all the Feynman parameters to one ($a_1=\dots=a_N=1$): 
\be
|\mathcal{T}_1| = \mathcal{K}(1,\dots 1) = \mathcal{U}(1,\dots 1).
\ee 
There are  similar formulas for the second Symanzik polynomial, $\mathcal{A}$ which we will not review here, see for example \cite{Bogner:2010kv}. 

%%%%%%%%

%%%%%%%%%%%%%%%%%%%%%%%%%%%%%%%%%%%%%%
%%%%%%%%%%%%%%%%%%%%%%%%%%%%%%%%%%%%%%
%%%%%%%%%%%%%%%%%%%%%%%%%%%%%%%%%%%%%%
\section{Some examples of Symanzik polynomials and their zeros}\label{SymanzikExamples}
%%%%%%%%%%%%%%%%%%%%%%%%%%%%%%%%%%%%%%
%%%%%%%%%%%%%%%%%%%%%%%%%%%%%%%%%%%%%%
%%%%%%%%%%%%%%%%%%%%%%%%%%%%%%%%%%%%%%

Here we provide first some examples of Symanzik polynomials, which we use later to exemplify a method to compute their zeros. Consider first the following diagram $F_1$:
\begin{center}
\begin{tikzpicture}
  \begin{feynman}
    \vertex (a) ;
    \vertex [right=of a] (b);
    \vertex [above right=of b] (l1); 
    \vertex [below right=of b] (l2);
    \vertex [below right=of l1] (c);
    \vertex [right=of c] (d);

    \diagram* {
      (a) --  [momentum = $k_1$] (b) --  [quarter left, edge label=$a_2$](l1) --  [quarter left, edge label=$a_4$](c) --  [momentum = $k_2$] (d),
      (b) --   [quarter right, edge label'=$a_3$](l2),
      (l1) --  [ edge label=$a_1$](l2) --  [quarter right, edge label'=$a_5$](c) 
    };
  \end{feynman}  
\end{tikzpicture} 
\end{center}
The two Symanzik polynomials corresponding to this diagram are
%%%
\begin{align}
\begin{split}
\mathcal{U}_{F_1}(a_r) & = a_1(a_2+a_3+a_4+a_5)+ (a_2+a_3)(a_4+a_5)\, ,\\
\mathcal{A}_{F_1}(a_r) & = a_1(a_2a_3 +a_2a_5 + a_3 a_4 + a_4 a_5) + a_2(a_3a_4+a_3a_5 +a_4a_5) +a_3a_4a_5\, .
\end{split}
\end{align}
%%%
Now, another two-loop diagram is given by 
\begin{center}
\begin{tikzpicture}
  \begin{feynman}
    \vertex (a) ;
    \vertex [right=of a] (b);
    \vertex [right=of b] (l1);
    \vertex [right=of l1] (l2);
    \vertex [right=of l2] (c);
    \vertex [right=of c] (d);

    \diagram* {
      (a) --  [momentum = $k_1$] (b) --  [half left, edge label=$a_2$](l1) --  [edge label=$a_1$](l2) -- [half left, edge label=$a_4$] (c) -- [momentum = $k_2$] (d),
      (b) --   [half right, edge label'=$a_3$](l1),
      (l2) --  [ half right, edge label'=$a_5$](c)  
    };
  \end{feynman}
\end{tikzpicture}
\end{center}
In this case the graph polynomials are
%%%
\begin{align}
\begin{split}
\mathcal{U}_{F_2}(a_r) & = (a_2+a_3)(a_4+a_5)\, ,\\
\mathcal{A}_{F_2}(a_r) & = a_1(a_2+a_3)(a_4+a_5)\, .
\end{split}
\end{align}
%%%
A third diagram we can draw is
\begin{center} 
\begin{tikzpicture}
  \begin{feynman}
    \vertex (a) ;
    \vertex [right=of a] (b);
    \vertex [below right=of b] (l1);
    \vertex [right=of l1] (l2);
    \vertex [above right=of l2] (c);
    \vertex [right=of c] (d);

    \diagram* {
      (a) --  [momentum = $k_1$] (b) --  [quarter right, edge label'=$a_2$](l1) --  [half right, edge label'=$a_3$](l2) -- [quarter right, edge label'=$a_5$] (c) -- [momentum = $k_2$] (d),
      (b) --   [half left, edge label=$a_1$](c),
      (l1) --  [ half left, edge label=$a_4$](l2)  
    };
  \end{feynman}
\end{tikzpicture}
\end{center}
Its  polynomials are
%%%
\begin{align}
\begin{split}
\mathcal{U}_{F_3}(a_r) & =  (a_3+a_4)(a_1+a_2+a_5) +  a_3 a_4  \, ,\\
\mathcal{A}_{F_3}(a_r) & = a_1(a_2+a_5)(a_3+a_4) + a_1 a_3 a_4 \, .
\end{split}
\end{align}
%%%

The examples above allow us determine the zeros of Symanzik polynomials. Consider first the second Symanzik polynomial. What are the zeros of $\mathcal{A_F}$? Recall that $\mathcal{A}_F$ is obtained by removing internal lines from a diagram $F$ so that we are left with two disconnected components; this means that $\mathcal{A}_F$ will vanish if we set to zero those parameters that form a path from the incoming particle to the outgoing one. This is because the contributions to 
$\mathcal{A_F}$ i.e. two-trees can be classified by all possible single ``long cuts'' that divides the graph into two.  In the first example in Appendix \ref{SymanzikExamples} these long cuts are given by $\{a_2,a_3\}$, $\{a_4,a_5\}$, $\{a_2,a_1,a_5\}$ and $\{a_4,a_1,a_3\}$. Every term in  $\mathcal{A_F}$ should involve a product of the elements of these long cuts. Then $\mathcal{A_F}$ will clearly vanish if we set one element per long cut to zero. This defines nothing else but a path between the two external end points of the graph. 
In other words, we can characterize the zeros of $\mathcal{A}_F$ by looking at all possible connected paths in our diagram $F$. 

For example the paths connecting the two external insertions in the first example in Appendix \ref{SymanzikExamples} are given by $\{a_2,\, a_4\}$,  $\{a_3,\, a_5\}$,   $\{a_2,\, a_1, \, a_5\}$ and $\{a_3,\, a_1, \, a_4\}$. Hence the zeros of $\mathcal{A}_F$ in this case are given by $a_2=a_4=0$,  $a_3=a_5=0$,  $a_2=a_1=a_5=0$ and $a_3=a_1=a_4=0$. The same procedure determines the zeros of $\mathcal{A}_F$ in the second and third examples in Appendix \ref{SymanzikExamples} as $a_2=a_1=a_4=0$, $a_3=a_1=a_4=0$, $a_2=a_1=a_5=0$ and $a_3=a_1=a_5=0$ and $a_1=0$, $a_2=a_4=a_5=0$ and $a_2=a_3=a_5=0$ respectively. One can ask whether this procedure gives the full set of zeros. For example, one can generally try to solve the equation 
$\mathcal{A}_F=0$ for one of the $a_i$s in favor of others. But this will generally lead to $a_i<0$ as Symanzik polynomials only involve positive coefficients and $a_i$s are non-negative by definition.  

Similarly one can devise a method to obtain the zeros of the first Symanzik polynomial. This polynomial is obtained by removing $\ell$ internal lines such that all loops are broken and one is left with a tree graph. The zeros are then obtained by setting the Schwinger parameters in each independent loop of the graph to zero. This is because each element in $\mathcal{U}_F$ should involve at least one Schwinger parameter from an individual loop. If it does not then it would not correspond to a tree. Therefore setting all Schwinger parameters of a loop corresponds to a zero of $\mathcal{U}_F$. In the first example in Appendix \ref{SymanzikExamples} these loops are given by $a_2=a_1=a_3=0$, $a_4=a_1=a_5=0$ and $a_2=a_4=a_3=a_5=0$. Investigating the loops in the second example yield the zeros $a_2=a_3=0$ and  $a_4=a_5=0$. Similarly the zeros of the third example are again determined from the loops as $a_3=a_4=0$, $a_1=a_2=a_5=a_3=0$ and  $a_1=a_2=a_5=a_4=0$. 

We do not know whether this geometric determination of the zeros of Symanzik polynomials have been discussed elsewhere in the literature. It also remains to be seen whether this procedure can be extended to higher $n$-point functions or genus. 
 
\newpage

%{\color{red} is is clearly not a proof and I'm not sure this would work for more complicated diagrams. Also I'm not sure how to find the zeros of $\mathcal{U}$}

%%%%%%%%%%%%%%%%%%%%%%%%%%%%%%%
%%%%%%%%%%%%%%%%%%%%%%%%%%%%%%%
\bibliographystyle{JHEP}
\bibliography{ExactAdSCFT}
%%%%%%%%%%%%%%%%%%%%%%%%%%%%%%%
%%%%%%%%%%%%%%%%%%%%%%%%%%%%%%%

\end{document}